\documentclass{aa} 

 \usepackage{graphicx}
\usepackage{txfonts}
\usepackage{color}

\newcommand{\onvire}[1]{}

\newcommand{\beq}{\begin{equation}}
\newcommand{\eeq}{\end{equation}}
\newcommand{\excs}{\extracolsep{\fill}}

\usepackage{color}
\usepackage{epsfig}
\usepackage{amssymb}
\usepackage{graphicx,natbib}
\usepackage{pdflscape}
\usepackage{longtable}
\usepackage[figuresright]{rotating} 
\begin{document}
\title{C2D Spitzer-IRS spectra of disks around T~Tauri stars \\
V. Spectral decomposition}

   \author{J. Olofsson
          \inst{1,2}
          \and
          J.-C. Augereau
	  \inst{1}
          \and
	  E. F. van Dishoeck
	  \inst{3,4}
          \and
          B. Mer\'{\i}n
          \inst{5}	  
          \and
          N. Grosso
          \inst{6,7}
          \and
          F. M\'enard
          \inst{1}
          \and
          G. A. Blake
          \inst{8}
          \and
          J.-L. Monin
          \inst{1}
          }

   \offprints{olofsson@mpia.de}
   
   \institute{Universit\'e Joseph Fourier/CNRS, Laboratoire
     d'Astrophysique de Grenoble, UMR 5571, BP 53,
     38041 Grenoble Cedex 09, France \and
     Max Planck Institute for Astronomy, K\"onigstuhl 17, D-69117 
     Heidelberg, Germany \\
     \email{olofsson@mpia.de} \and
     Leiden Observatory, Leiden University, P.O. Box 9513, 2300 RA
     Leiden, The Netherlands \and
     Max Planck Institut f\"ur Extraterrestrische Physik,
     Giessenbachstrasse 1, 85748 Garching, Germany \and 
     Herschel Science Centre, SRE-SDH, ESA P.O. Box 78, 28691
     Villanueva de la Ca\~nada, Madrid, Spain \and
     Universit\'e de Strasbourg, Observatoire Astronomique de
     Strasbourg, 11 rue de l'universit\'e, 67000 Strasbourg, France \and
     CNRS, UMR 7550, 11 rue de l'universit\'e, 67000 Strasbourg, France \and
     Division of Geological and Planetary Sciences 150-21, California
     Institute of Technology, Pasadena, CA 91125, USA
  }

   \date{Received \today; accepted }

   \abstract
   {Dust particles evolve in size and lattice structure in protoplanetary
     disks, due to coagulation, fragmentation and crystallization, and
     are radially and vertically mixed in disks due to turbulent
     diffusion and wind/radiation pressure forces.}
   {This paper aims at determining the mineralogical composition and
     size distribution of the dust grains in planet forming regions of
     disks around a statistical sample of 58 T\,Tauri stars observed
     with Spitzer/IRS as part of the Cores to Disks (c2d) Legacy
     Program.}
   { We present a spectral decomposition model, named ``B2C'', that
     reproduces the IRS spectra over the full spectral range
     (5--35\,$\mu$m). The model assumes two dust populations: a warm
     component responsible for the 10\,$\mu$m emission arising from
     the disk inner regions ($\lesssim$\,1\,AU) and a colder component
     responsible for the 20--30\,$\mu$m emission, arising from more
     distant regions ($\lesssim$\,10\,AU). The fitting strategy relies
     on a random exploration of parameter space coupled with a
     Bayesian inference method.}
   { We show evidence for a significant \textit{size distribution
       flattening} in the atmospheres of disks compared to the typical
     MRN distribution, providing an explanation for the usual flat,
     boxy $10\,\mu$m feature profile generally observed in T~Tauri
     star spectra.  We reexamine the \textit{crystallinity paradox},
     observationally identified by Olofsson et al. (2009), and we find
     a simultaneous enrichment of the crystallinity in both the warm
     and cold regions, while grain sizes in both components are
     uncorrelated. We show that flat disks tend to have larger grains
     than flared disk. Finally our modeling results do not show
     evidence for any correlations between the crystallinity and
     either the star spectral type, or the X-ray luminosity (for a
     subset of the sample).}
   {The \textit{size distribution flattening} may suggests that grain
     coagulation is a slightly more effective process than
     fragmentation (helped by turbulent diffusion) in disk
     atmospheres, and that this imbalance may last over most of the
     TTauri phase. This result may also point toward small grain
     depletion via strong stellar winds or radiation pressure in the
     upper layers of disk. The non negligible cold crystallinity
     fractions suggests efficient radial mixing processes in order to
     distribute crystalline grains at large distances from the central
     object, along with possible nebular shocks in outer regions of
     disks that can thermally anneal amorphous grains.}

   \keywords{Stars: pre-main sequence -- planetary systems:
   protoplanetary disks -- circumstellar matter -- Infrared: stars --
   Methods: statistical -- Techniques: spectroscopic}
\authorrunning{Olofsson et al.}
\titlerunning{Spectral decomposition of C2D/IRS spectra of T\,Tauri
  disks}

   \maketitle
%
\section{Introduction}

The mid-infrared spectral regime probes the warm dust grains located
in the planet forming region (typically 1--10\,AU for a classical
T~Tauri disk). At these wavelengths, the young disks are optically opaque to
the stellar light, and the thermal emission arises from the disk
surface, well above the disk midplane. Single-aperture imaging of
disks in the mid-IR suffers from both a relatively low spatial
resolution compared to optical/near-IR telescopes, and from poorly
extended emission zones. On the other hand, silicates
have features due to stretching and bending resonance modes which make
spectroscopy at mid-IR wavelengths of very high interest and feasible
with current instrumentation. Silicates are indeed the most abundant
sort of solids in disks, and therefore constitute a very important
ingredient in any planet formation theory.

The dust grains that are originally incorporated into protoplanetary circumstellar disks are essentially of interstellar nature. They are thought to be particles much smaller than $1\,\mu$m, and mostly composed of silicates or organic refractories. \citet{Kemper2005} placed an upper limit of 2.2\% on the amount of crystalline silicates in the interstellar medium (hereafter ISM), which suggests an amorphous lattice structure for the pristine silicates in forming protoplanetary disks. In the very early stages of the disk evolution, the tiny dust grains are so coupled with the gas that grain-grain collisions occur at sufficiently low relative velocity to allow the grains to coagulate and grow. This results in fractal aggregates that will tend to settle toward the disk midplane as their mass increases. From simple theoretical arguments, considering only the gravitational and drag forces on the grains in a laminar disk, one can show that $\mu$m-sized grains at 1\,AU should settle in the midplane in less than $10^5$ years in a classical T~Tauri disk (\citealp{Weidenschilling1980}). In fact, the grains are expected to settle even faster as their mass increases in the course of their journey to the disk midplane. Because the T~Tauri disks are a few million years old, this would suggest that the inner disk upper layers should be devoided of $\mu$m-sized or larger grains in the absence of turbulence and grain fragmentation. This is a prediction that can be tested observationnaly, especially in the infrared where spectroscopic signatures of silicate grains are present.

The Spitzer Space Telescope, launched in August 2003, had a sensitivity that surpassed previous mid-IR space missions by orders of magnitudes until the cryogenic mission ended in May 2009. As part of the ``Core to Disks'' (c2d) Legacy survey (Evans et al., 2003), more than a hundred T~Tauri stars were spectroscopically observed, to confirm or invalidate some of the predictions concerning dust processing and grain dynamics in protoplanetary disks. In \citet{Kessler-Silacci2006,Kessler-Silacci2007}, and \citet{Olofsson2009}, we showed that most of the objects display silicate emission features arising from within 10\,AU from the star.  This allowed both a classical study of grain coagulation and a comprehensive statistical analysis of dust crystallization in planet forming regions of disks around young solar analogs.

In \citet{Olofsson2009}, we showed that not only the warm amorphous silicates had grown, but so had the colder crystalline silicates as probed by their $23\,\mu$m emission feature. In fact, the emission features in IRS spectra are very much dominated by micron-sized grains in the upper layers of disks, pointing toward vertical (turbulent?) mixing of the dust grains to compensate for gravitational settling, together with grain-grain destructive fragmentation in the innermost regions of most protoplanetary disks to compensate for grain growth. This equilibrium seems to last over several millions of years and be to independent of the specific star forming region (\citealp{Oliveira2010}). Winds and/or radiation pressure, in complement to these processes, can act to remove a fraction of the submicron-sized grains from disk atmospheres, and may thus contribute to the transport of crystalline silicate grains toward the outermost disk regions \citep[see discussion in][]{Olofsson2009}.

Crystalline silicates appear to be very frequent in disks around T~Tauri disks and in regions much colder than their presumed formation regions, suggesting efficient outward radial transport mechanisms in disks \citep{Bouwman2008,Olofsson2009, Watson2009}. Therefore, the determination of the composition and size distribution of the dust grains in circumstellar disks is one of the keys in understanding the first steps of planet formation as it can trace the dust fluxes in planet-forming disks. Detailed modeling of the $10\,\mu$m silicate emission feature has already been performed for HAEBE disks by \citet{Bouwman2001} and \citet{2005}. Mineralogical studies of dust in disks around very low mass stars and brown dwarfs were also led by \citet{Apai2005}, \citet{Riaz2009}, \citet{Mer'in2007} and \citet{Bouy2008}. The two last studies introduced a novel method to fit the spectra over the entire IRS spectral range, which allows to decompose the spectra into two main contributions at different temperatures. This in turn allowed to compare the degrees of crystallinity and the grain sizes in two different disk regions. The latter compositional fitting approach is further supported by the analysis by \citet{Olofsson2009} who showed that the energy and frequencies at which crystalline silicate features are seen at wavelengths larger than $20\,\mu$m are largely uncorrelated to the amorphous $10\,\mu$m feature observational properties.

We present in this paper an improved version of the compositional fitting method used in \citet{Mer'in2007} and \citet{Bouy2008}. We apply the model to a subsample (58 stars) of high SNR spectra presented in \citet{Olofsson2009} to derive the dust content in disks about young solar analogs. The method relies on the fact that the IRS spectral range is sufficiently broad so that the regions probed at around $10\,\mu$m and between $20$ and $30\,\mu$m do overlap only partially. This is illustrated for instance in \citet{Kessler-Silacci2006} who show that for a typical T\,Tauri star, a factor of 2 increase in wavelength ($10\,\mu$m~$\rightarrow$~$20\,\mu$m), translates into a factor of 10 ($1\,$AU~$\rightarrow$~$10\,$AU) into the regions from which most of the observed emission arises.  An additional goal of our model is therefore to search for differential effects in crystallinity and grain sizes between the warm and slightly cooler disks regions which may be indicative of radial and/or vertical dependent chemical composition and grain size, due for instance to differential grain evolution and/or grain transport.

We develop in Sec.\,\ref{secomp} our procedure to model IRS spectra of Class~II objects, and present the tests for robustness of this procedure in Appendix\,\ref{sec:tests}. The results for 58 objects are presented in Sec.\,\ref{sec:res}, and they are discussed in terms of dust coagulation and dust crystallization in Sec.\,\ref{sec:disc}.  Sec. \ref{sec:sum} summarizes the implications of our results on disks dynamics at this stage of evolution and critically discuss the use of the shape and strength of the amorphous $10\,\mu$m silicate feature as a grain proxy.

\section{Spectral decomposition with the B2C method}
\label{secomp}

We elaborate in this section a modeling procedure whose goal is to
reproduce the observed IRS spectra, from about $5$ to $35\,\mu$m, in
order to infer the composition and size of the emitting dust
grains. This is achevied by using two dust grain populations at two
different temperatures. We will refer to these two populations as two
different ``components''. The first component aims at reproducing the
features at around $10\,\mu$m, its temperature is generally around
$T_{\mathrm{w}}\sim 300$\,K (warm component hereafter). The second
component aims at reproducing the residuals, over the full spectral
range and its temperature is colder, $T_{\mathrm{c}} \sim 100$\,K
(cold component hereafter). Each of these two components is described
by several grain compositions, including amorphous and crystalline
silicates, and sizes as detailed below. Bayesian inference is used to
best fit the IRS spectra and to derive uncertainties on the
parameters, hence the name of the procedure, ``B2C'', which stands for
{\it Bayesian inference with 2 Components}.

\subsection{Theoretical opacities and grain sizes}

To reproduce the observed spectra, we consider five different dust
species. The amorphous species include silicates of olivine
stoichiometry (glassy MgFeSiO$_{4}$, density of 3.71 g.cm$^{-3}$, optical
indexes from \citealp{Dorschner1995}), silicates of pyroxene
stoichiometry (glassy MgFeSiO$_{6}$, density of 3.2 g.cm$^{-3}$,
\citealp{Dorschner1995}), and silica (amorphous quartz, density of
3.33 g.cm$^{-3}$, \citealp{Henning1997}). For the Mg-rich
\citep[see][]{Olofsson2009} crystalline species, we consider
enstatite (MgSiO$_{3}$, density of 2.8 g.cm$^{-3}$, \citealp{Jaeger1998})
and forsterite (Mg$_{2}$SiO$_{4}$, density of 2.6 g.cm$^{-3}$,
\citealp{Servoin1973}).

\begin{figure*}
\begin{center}
\resizebox{\hsize}{!}{\includegraphics{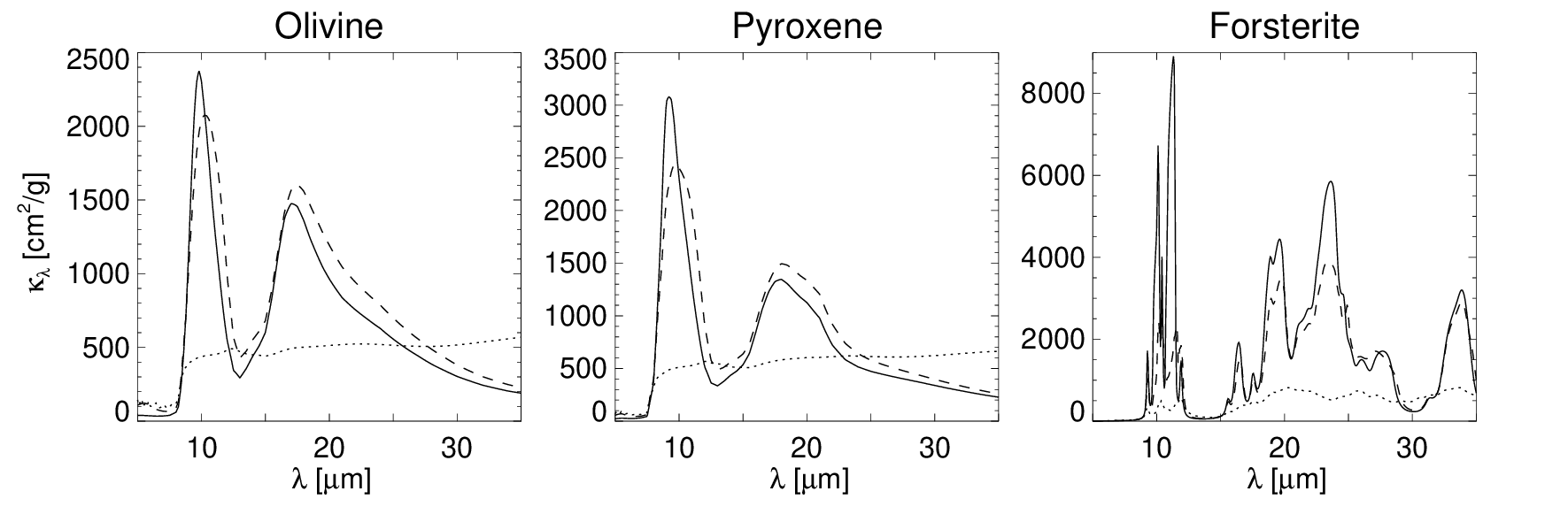}}
\resizebox{0.7\hsize}{!}{\includegraphics{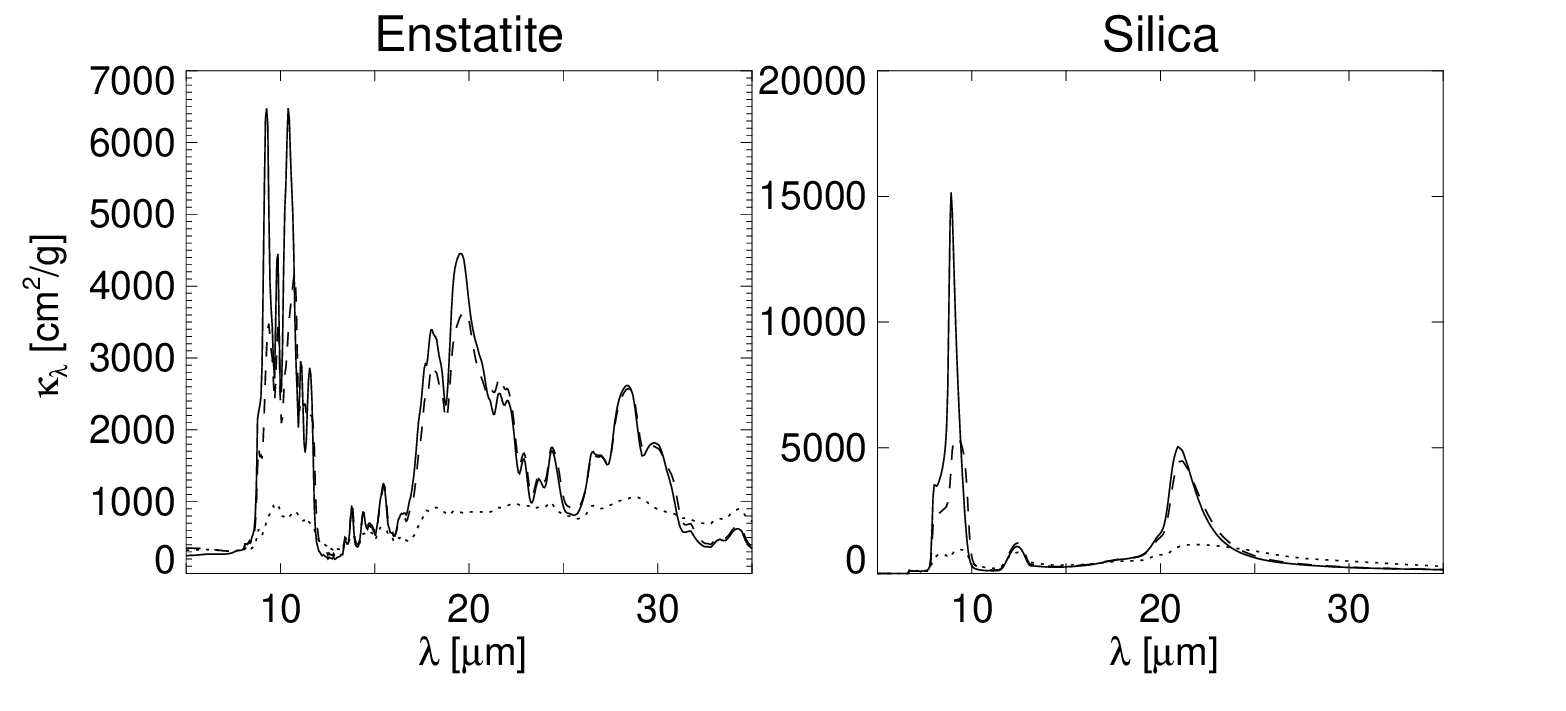}}
\caption{\label{blowopac}Blowup of the opacities (in units of
  cm$^2$.g$^{-1}$) used to model the IRS spectra. The solid line is for
  0.1\,$\mu$m grains, dashed line for 1.5\,$\mu$m grains and dotted
  line for 6.0\,$\mu$m grains.}
\end{center}
\end{figure*}

Following previous papers (e.g. \citealp{Bouwman2001,Bouwman2008},
\citealp{Juh'asz2009}), the theoretical opacities $\kappa_{\lambda}$
of amorphous species are computed assuming homogeneous sphere (Mie
theory), while the DHS theory (Distribution of Hollow Spheres,
\citealp{Min2005}) is employed for the crystalline silicates in order
to simulate irregularly-shaped dust particles. To limit the number of
free parameters in the model, we consider three spectroscopically
representative grain sizes radii, which are $0.1\,\mu$m, $1.5\,\mu$m
and $6.0\,\mu$m. These three grain sizes are supposed to best mimic
the behaviour of very small grains, intermediate-sized grains and
large grains \citep[e.g.][]{Bouwman2001,Bouwman2008}. We limit
ourselves to the two smallest sizes, 0.1 and 1.5\,$\mu$m, for the
crystalline species. The first reason for this choice is that large
crystalline grains present a high degeneracy with large amorphous
grains, as 6.0\,$\mu$m-sized grains are mostly featureless. Therefore
large crystals can be used by the procedure instead of large amorphous
grains, thereby introducing a bias toward high crystallinity
fractions. The second reason is that, according to crystallisation
models (e.g. \citealp{Gail2004a}), one does not expect to produce such
large pure crystals via thermal annealing. Grains more likely grow via
collisional aggregation of both small crystalline and amorphous
material (e.g. \citealp{Min2008}).  Following these two
considerations, we decided not to include 6.0\,$\mu$m-sized
crystals. This choice is in line with previous works from
\citet{Sargent2009a} or \citet{Riaz2009}. The fifteen opacity curves
(5 grain compositions, 3 grain sizes) used in this paper are displayed
in Fig.\,\ref{blowopac}, including 6.0\,$\mu$m-sized crystals to show
the degeneracy with large amorphous grains.

\subsection{The B2C model}

The B2C model elaborated in this paper assumes that the IRS spectrum
can be fitted by considering a continuum emission, and two main
components, a warm and a cold one, essentially responsible for the
10$\,\mu$m and 20--30\,$\mu$m emissions, respectively. The two
component approach is supported by the work of \citet{Olofsson2009}
who show that disks usually have inhomogeneous compositions with
respect to the dust temperature (crystalline features being more
  frequently detected at long than at short wavelengths, the
so-called {\it crystallinity paradox}), and that the emission features
at around 10$\,\mu$m are essentially uncorrelated with those
  appearing at wavelengths larger than 20$\,\mu$m.  This approach is
also supported by the T~Tauri model described in
\citet{Kessler-Silacci2006} where they show that the emission at
10\,$\mu$m is arising from regions within $\sim$\,1\,AU while the flux
at 20\,$\mu$m is arising from within $\sim$\,10\,AU. Furthermore, we
first used the simpliest solution, with only one component, without
successful results. Any compositional method that aims at fitting IRS
spectra over the full spectral range should therefore be able to
handle inhomogeneous disk compositions. The two component approach
developed in this paper is a simple attempt to go in that
direction. This modeling strategy has already proven sufficient to
provide adequate fits to IRS spectra, from 5 to 35$\,\mu$m
\citep{Mer'in2007,Bouy2008}. In this paper, we improve both the model
and the fitting strategy originally developed in \citet{Mer'in2007}
and \citet{Bouy2008} in order to apply the decompositional method to a
large number of objects.

The first step of the modeling approach is the estimate of the
continuum to be subtracted to the observed IRS spectrum before
performing a compositional fit.  This could be done using a radiative
transfer code \citep[e.g.][]{Mer'in2007,Bouy2008}, but given the
number of objects (58) to be analyzed, and given the objectives of the
paper which is oriented toward searching for trends thanks to the
analysis of a large sample, obtaining a satisfying model for every
object is not a manageable task. We instead adopt a continuum built by
using a power-law ($\lambda^{\alpha}$) plus a black-body at
temperature $T_{\rm cont}$, to make our problem more tractable:
\begin{eqnarray}
  F_{\mathrm{\nu,cont}} = K_{\rm pl} \lambda^{\alpha} + K_{\rm bb} B_{\nu}(T_{\rm
    cont})
\end{eqnarray}
where $K_{\rm pl}$ and $K_{\rm bb}$ are two positive constants. The
power law index, $\alpha$, is determined independently from the
compositional procedure. It represents the mid-IR tail of the emission
from both the star itself and from the inner rim of the disk. Its
value is estimated by fitting the IRS spectrum from its shortest
wavelength (about 5$\,\mu$m) up to blue foot of the 10\,$\mu$m
amorphous silicate feature. The black-body is aimed at contributing at
wavelengths larger than $20\,\mu$m and $T_{\rm cont}$ therefore
constrained to be less than 150\,K. A typical value for the
temperature $T_{\rm cont}$ is found to be about 110\,K.

In our fitting approach, only two free parameters characterize the
continuum: the black-body temperature $T_{\rm cont}$ and a
normalization offset $O_{\nu_2}$ at $\lambda_2 = c/\nu_2 \sim
13.5\,\mu$m. The implementation of a variable offset comes from the
fact that the 10 and 18$\,\mu$m amorphous features partly overlap at
$\lambda\sim$13--15\,$\mu$m (Fig.\,\ref{blowopac}), and any realistic
continuum should therefore pass below the observed flux at these
wavelengths. For a given set of $\alpha$, $T_{\rm cont}$ and offset
value $O_{\nu_2}$, the synthetic continuum is obtained by solving the
following two-equation system with respect to the normalization
coefficients $K_{\rm pl}$ and $K_{\rm bb}$:
\begin{eqnarray}
  \left\{
      \begin{array} {ccl}
        K_{\rm pl} \times \lambda_{1}^{\alpha} + K_{\rm bb} \times
        B_{\nu_1}(T_{\mathrm{cont}})  & = &  F_{\mathrm{\nu_1,\mathrm{obs}}} \\
        K_{\rm pl} \times  \lambda_{2}^{\alpha} + K_{\rm bb} \times
        B_{\nu_2}(T_{\mathrm{cont}}) & = &
        F_{\mathrm{\nu_2,\mathrm{obs}}} \times (1 -
        O_{\nu_2})
      \end{array}
    \right .
\end{eqnarray}
where $F_{\nu,\mathrm{obs}}$ is the observed spectrum, in units of
Jansky. The constraints on the normalization coefficients ($K_{\rm
  pl}>0$ and $K_{\rm bb}\geq 0$) imply that for some objects, the best
continua require $K_{\rm bb}=0$, corresponding to continua
described by a pure power-law. The normalization wavelengths are
chosen to bracket the amorphous 10$\,\mu$m silicate feature, with
$\lambda_{1}=c/\nu_1 \sim 7.5$\,$\mu$m, and $\lambda_{2}=c/\nu_2 \sim
13.5$\,$\mu$m.

The fit to the continuum-subtracted IRS spectrum is performed in two
steps.  First, a fit to the 10\,$\mu$m feature is obtained between
$\lambda_{1}$ and $\lambda_{2}$ (warm component), then a second fit to
the residual spectrum is obtained for wavelengths between $\lambda_1$
and $\lambda_3\sim$35\,$\mu$m (cold component).  The
continuum-substracted IRS spectrum ($F_{\nu,{\rm obs}} - F_{\nu,{\rm
    cont}}$) around $10\,\mu$m is reproduced within the range
[$\lambda_{1}$, $\lambda_{2}$], by summing up the thirteen mass
absorption coefficients $\kappa_{i}^{j}$ ($N_{\rm species}=5$ dust
species, $i$ index, and $N_{\rm sizes}=3$ or 2 grain sizes, $j$ index
depending on the lattice structure), multiplied by a blackbody
$B_{\nu}(T_{\mathrm{w}}$) at a given warm temperature
$T_{\mathrm{w}}$, and weighted with relative masses $M_{{\rm w},
  i}^{j}$:
\begin{eqnarray}
  F_{\mathrm{\nu,warm}} = 
  B_{\mathrm{\nu}}(T_{\mathrm{w}}) \times 
  \sum_{j=1}^{\mathrm{N_{\rm sizes}}} 
  \sum_{i=1}^{\mathrm{N_{\rm species}}}  
  \kappa_{i}^{j} \times M_{{\rm w},i}^{j}.
  \label{eq:warmcomp}
\end{eqnarray}
The residuals ($F_{\nu,{\rm obs}} - F_{\nu,{\rm cont}} -
F_{\mathrm{\nu,warm}}$) are then similarly fitted between $\lambda_1$
and $\lambda_3$ with a synthetic cold component spectrum that writes:
\begin{eqnarray}
  F_{\mathrm{\nu,cold}} = 
  B_{\mathrm{\nu}}(T_{\mathrm{c}}) \times 
  \sum_{j=1}^{\mathrm{N_{\rm sizes}}} 
  \sum_{i=1}^{\mathrm{N_{\rm species}}}  
  \kappa_{i}^{j} \times M_{{\rm c},i}^{j}.
  \label{eq:coldcomp}
\end{eqnarray}
The final synthetic spectrum, obtained in two steps, then reads:
\begin{eqnarray}
  F_{\nu,{\rm synt}} = F_{\nu,{\rm cont}} + F_{\nu,{\rm warm}} + F_{\mathrm{\nu,cold}}.
\end{eqnarray}
It depends on $2 \times 13$ parameters for the relative mass
abundances of the warm and cold components (the $M_{{\rm w},i}^{j}$
and $M_{{\rm c},i}^{j}$), two temperatures ($T_{\rm warm}$ and $T_{\rm
  cold}$), and two parameters for the continuum ($T_{\rm cont}$ and
$O_{\nu_2}$), which leads to 30 free parameters in total.

\begin{figure}
  \begin{center}
    \resizebox{\hsize}{!}{\includegraphics{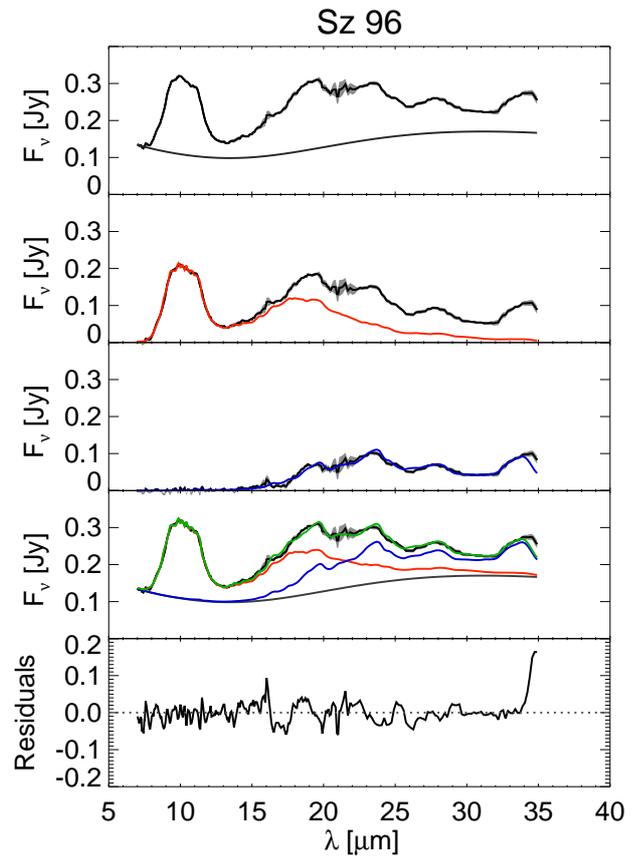}}
    \caption{\label{fig:didact} Detailed example of B2C modeling steps
      for the object \object{Sz\,96}. From top to bottom: the original
      spectrum and the estimated continuum (top panel), the continuum
      substracted spectrum and the fit to the warm component (in
      orange, 2$^{\rm nd}$ panel), the residuals from the fit to the
      warm component and the fit to this cold component (in blue,
      3$^{\rm rd}$ panel), the final fit to the entire spectrum (in
      green) and the various contributions (4$^{\rm th}$ panel), the
      relative residuals between the original spectrum and the final
      fit (last panel).}
  \end{center}
\end{figure}
The successive steps of the B2C modeling approach are decomposed in
Figure\,\ref{fig:didact} for the illustrative case of \object{Sz\,96}.
The original spectrum ($F_{\nu, \rm obs}$) and the estimated continuum
($F_{\nu,\rm cont}$) are displayed in the top panel. Then, the fit to
the 10\,$\mu$m feature is performed ($F_{\nu, \rm warm}$, orange curve
on second panel) and subtracted from the continuum-subtracted spectrum
(black line on the third panel). It shows that, although the
contribution of the first fit is not negligible up to
$\sim$\,25\,$\mu$m, a second, colder component is required to account
for the large wavelength spectrum. The fit to the residuals ($F_{\nu,
  \rm cold}$, in blue on the third panel) is then computed, leading to
an overall fit ($F_{\nu, \rm synt}$, green line) to the entire IRS
spectrum as drawn in the fourth panel together whith all the
contributions (continuum, warm and cold components). The last panel of
Fig.\,\ref{fig:didact} displays the final, relative residuals for the
full fit of the IRS spectrum, calculated as follows: $( F_{\nu,
  \mathrm{obs}} - F_{\nu, \mathrm{synt}}) / F_{\nu, \mathrm{synt}}$.

\subsection{Fitting process}

\begin{figure*}
  \begin{center}
\hspace*{-0.5cm}\includegraphics[angle=0,width=1.7\columnwidth,origin=bl]{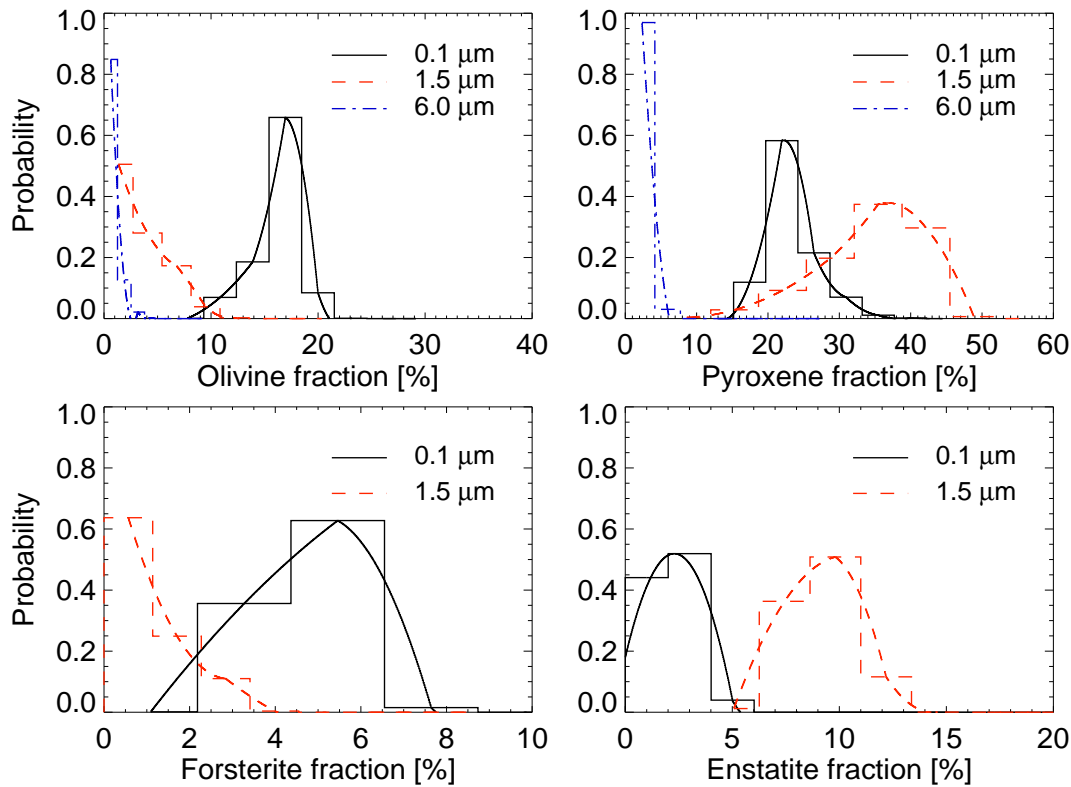}
\caption{\label{fig:proba}Examples of probabilities distributions for
  the fit to the Sz\,96 spectrum. From the top left panel, clockwise,
  are displayed the probability distributions for the warm olivine,
  pyroxene, enstatite and forsterite grains. On each panel are
  represented 0.1\,$\mu$m grains (black plain histogram), 1.5\,$\mu$m
  grains (red dashed histogram) and 6.0\,$\mu$m grains (blue
  dot-dashed histogram).}
\end{center}
\end{figure*}

The parameter space has a high dimensionality in our problem and
imposes a specific fitting approach to appreciate the reliability of
the results. We develop in this paper a method based on a Bayesian
analysis, combined with a Monte Carlo Markov chain (MCMC)--like
approach to explore the parameter space.

Our procedure is built in order to randomly explore the space of free
parameters ($2 \times$ 13 dust compositions, 2 dust temperatures and 2
parameters for the continuum). We start with a randomly chosen initial
set of parameters, then one of these parameters is randomly modified,
while all the others remain unchanged. When the chosen parameter is
related to the abundance of a grain species (the $M_{{\rm w},i}^{j}$
and $M_{{\rm c},i}^{j}$), the maximal value for the increment is 1\%
of the previous total mass for the considered component (warm or
cold). When the chosen parameter is a temperature or the offset
$O_{\nu , 2}$, we allow increments of at most 4\% of the previous
temperature or the previous offset, respectively. These maximum values
were chosen to obtain small enough increments and therefore explore
continuously the parameter space.

For each set of parameters, a synthetic spectrum $F_{\nu,{\rm synt}}$
is calculated and the goodness of the fit to the observed spectrum is
evalutated with a reduced $\chi^{2}_{\mathrm{r}}$. The parameter space
exploration is therefore very much alike a MCMC approach, with two
specificities: we only perform jumps for one parameter at a time, and
second, the jumps are always accepted\footnote{the implementation of a
  Metropolis-Hastings rule to decide on whether a jump should be
  accepted or rejected with a certain probability did not improve the
  fitting process, while increasing the calculation time}. After $n$
iterations (typically $n = 800$), we set all the parameters to those
that gave the lowest $\chi^{2}_{\mathrm{r}}$ among the $n$ previous
iterations. This loop is done $m$ times (usually $m$ is also set to
800) and all the compositions, temperatures, offset and their
associated $\chi^{2}_{\mathrm{r}}$ values are stored.
\begin{figure*}
  \begin{center}
    \includegraphics[width=7cm]{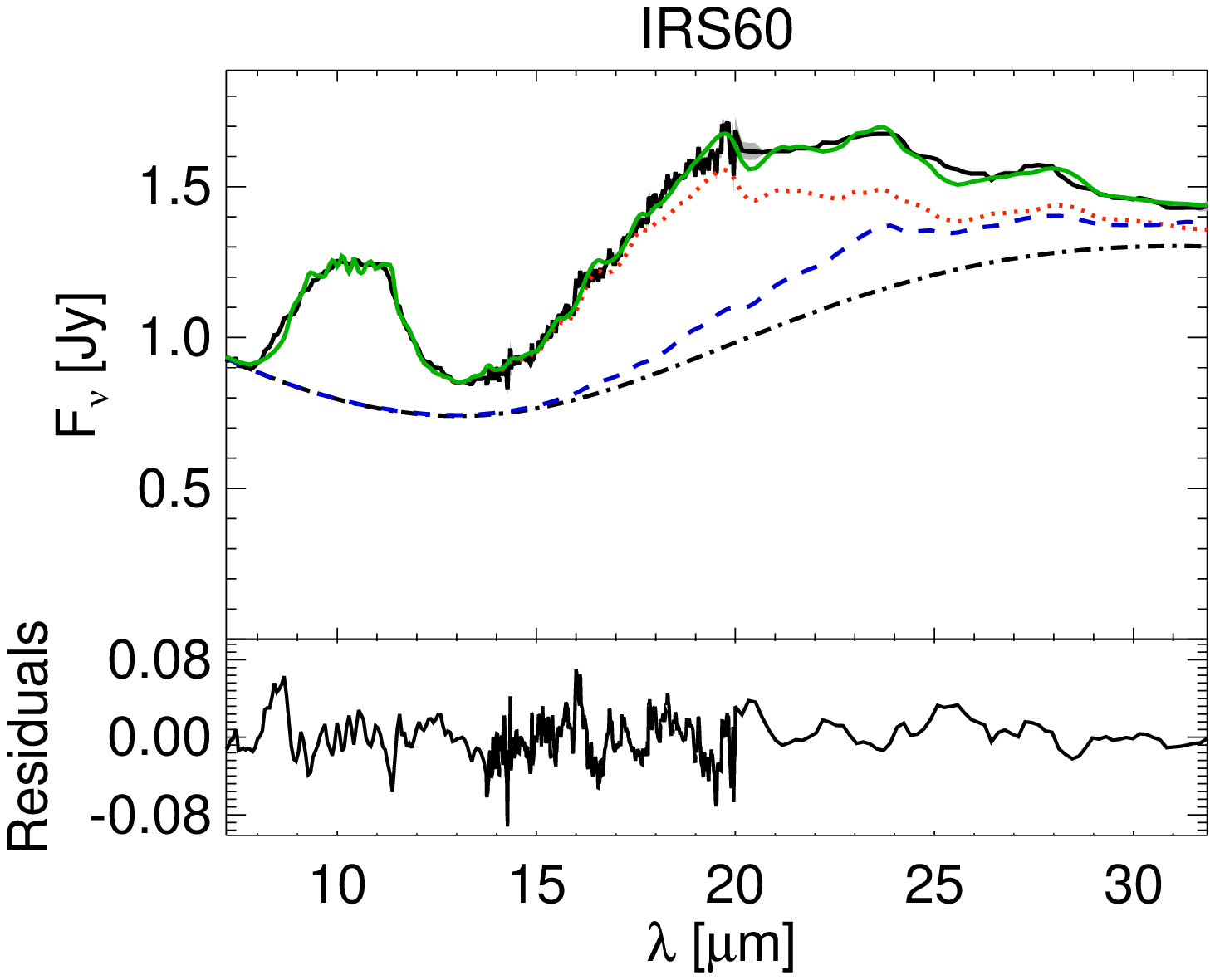}
    \includegraphics[width=7cm]{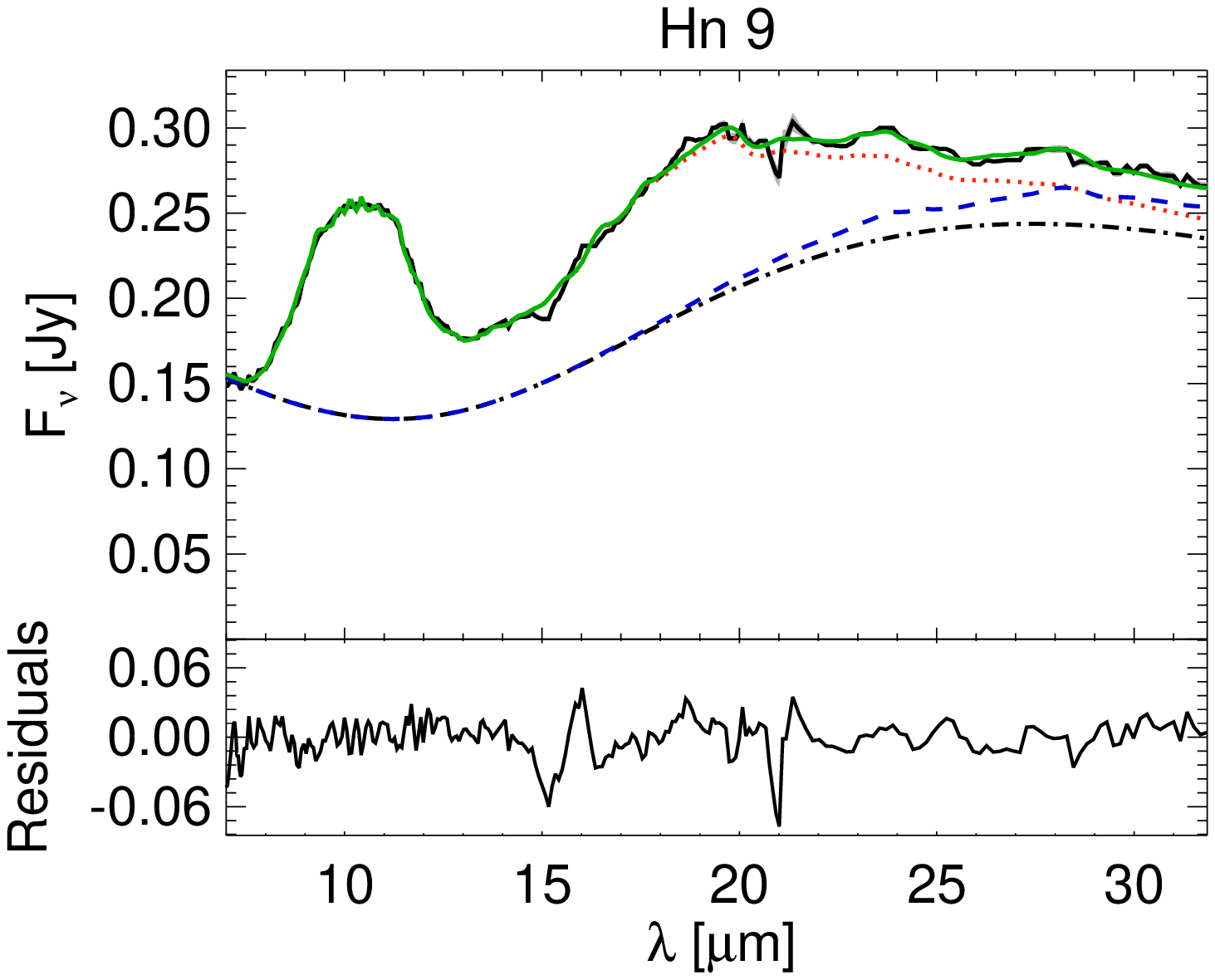}
    \includegraphics[width=7cm]{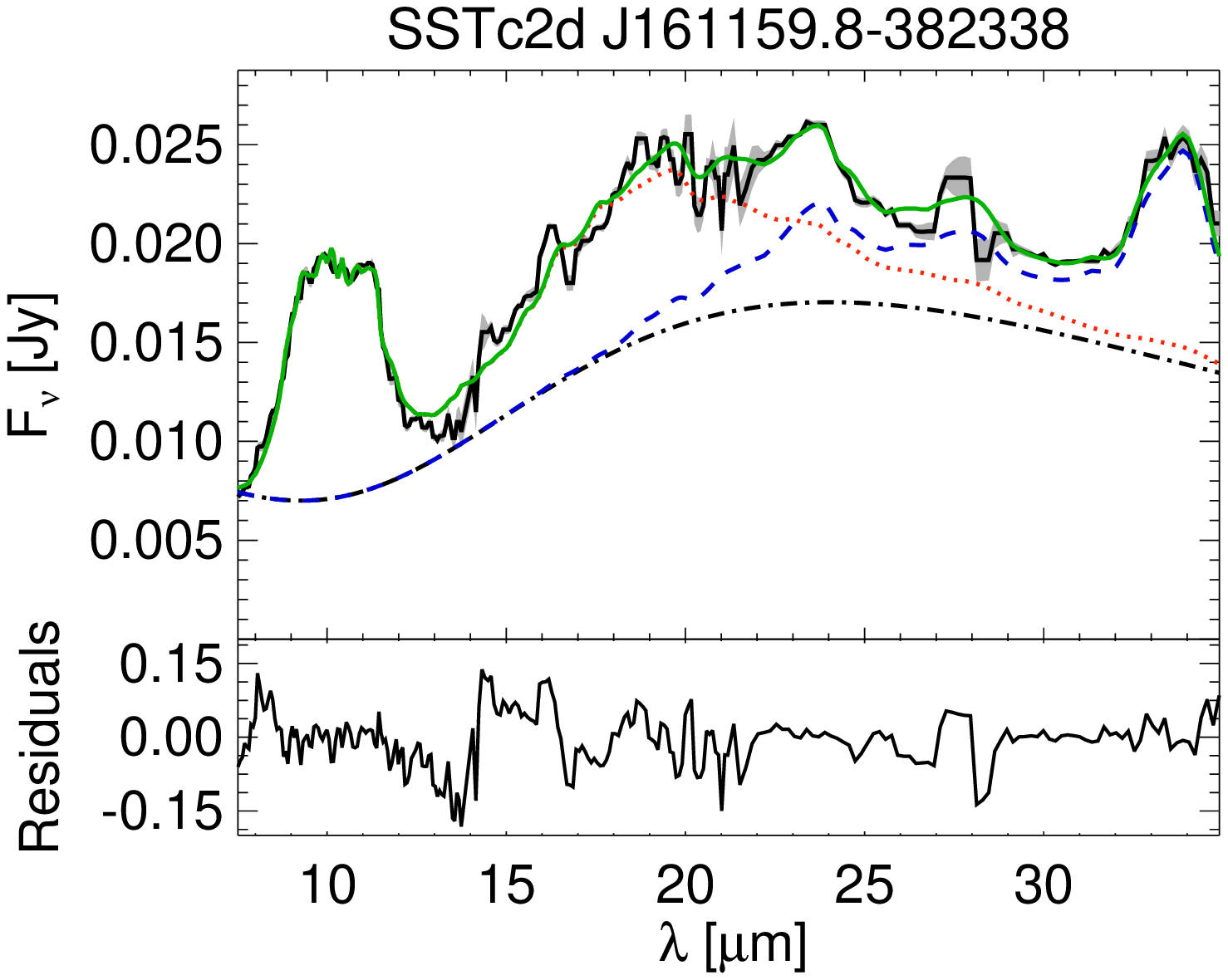}
    \includegraphics[width=7cm]{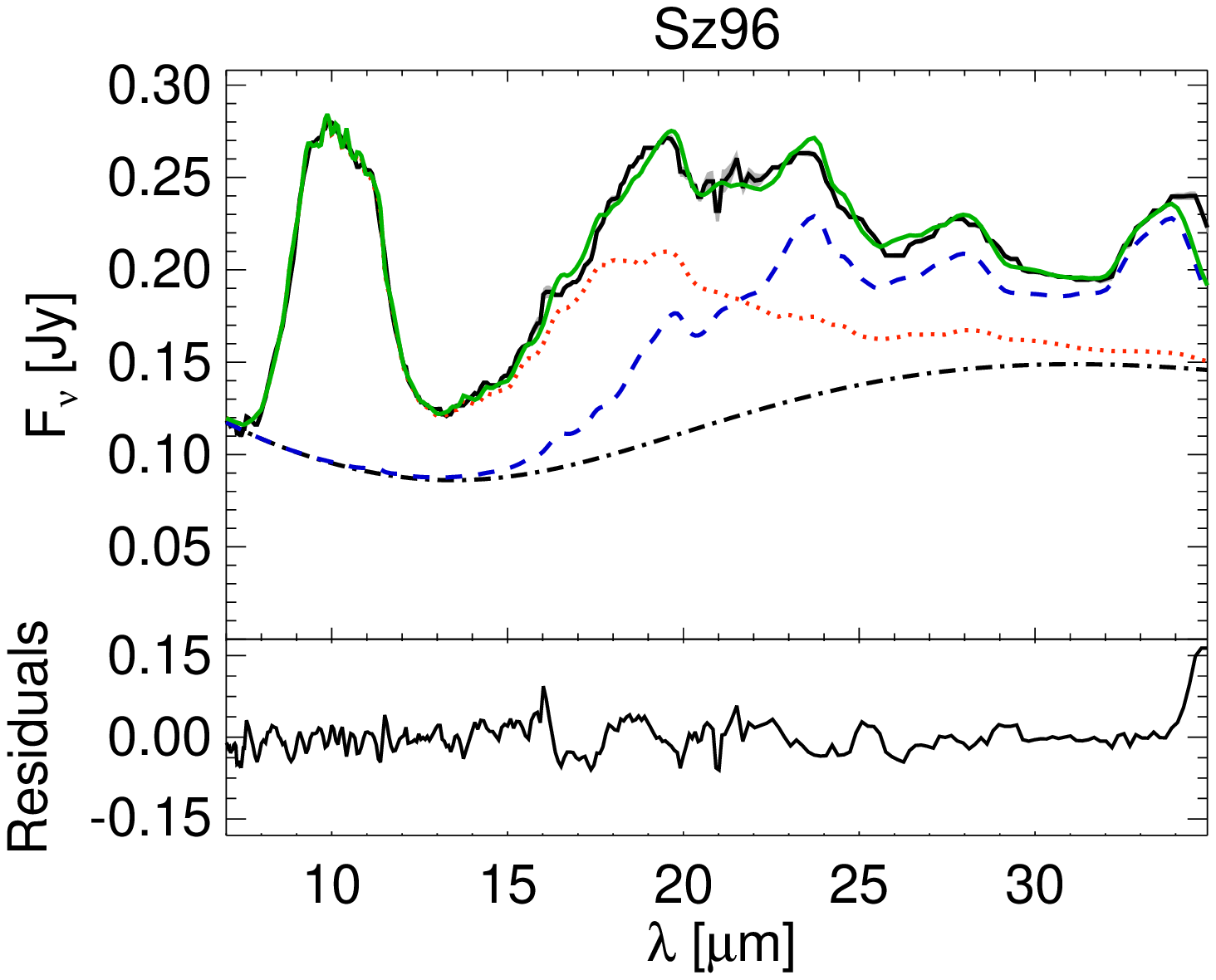}
    \caption{\label{fitcomp}Examples of B2C fits over the entire IRS
      spectral range for four objects: \object{IRS 60},
      \object{Hn\,9}, \object{SSTc2d\,J161159.8-382338} and
      \object{Sz\,96}. The warm component is displayed in dotted
      orange line, the cold component in dashed blue line, and final
      fit to the entire spectrum in green. The dot-dashed grey line
      represents the continuum. The light grey enveloppe represents
      the 3-$\sigma$ uncertainty on the observed spectrum. For each
      star, the bottom panel shows the relative residuals.}
  \end{center}
\end{figure*}

The $n \times m$ reduced $\chi^{2}_{\mathrm{r}}$ values are
transformed into probabilities assuming a Gaussian likelihood function
($\propto\exp(- \chi^{2}_{\mathrm{r}}/2)$) for Bayesian
analysis. Marginalized probability distribution for each free
parameter are then obtained by projection of these probabilities onto
each dimension of the parameter space. The best fit to the observed
spectrum among all the simulations (i.e., the one with the lowest
$\chi^2 _{\mathrm{r}}$ value), yields relative masses for all the dust
species and grain sizes ($M_{{\rm w},i}^{j}$ and $M_{{\rm c},i}^{j}$
parameters in Eqs.\,\ref{eq:warmcomp} and \ref{eq:coldcomp}) for both
the warm and cold components, as well as best temperatures
$T_{\mathrm{w}}$ and $T_{\mathrm{c}}$, which typically range between
$200\,K$ and $300\,$K for the warm component, and between $70\,K$ and
$150\,K$ for the cold one. The 1-$\sigma$ uncertainties on the
parameters are derived from the probability distributions for each
parameter. We over-sample each probability distribution to compute
half-width at half maximum for both sides of the distribution, and
derive minimum and maximum 1-$\sigma$
uncertainties. Figure\,\ref{fig:proba} displays ten examples of
well-peaked probability distributions, with their respective
over-sampled distributions overplotted, for some parameters of the fit
to the \object{Sz\,96} spectrum, indicating the parameters are
constrained by the B2C fitting approach. Figure\,\ref{fitcomp}
displays four example spectra with their best fits (and residuals)
over the entire IRS spectral range, showing that good fits can be
obtained for spectra with different shapes. Spectral regions with high
residuals mostly correspond to regions with low signal-to-noise ratio.

The robustness of the B2C procedure has been intensively tested and
this work is reported in Appendix\,\ref{sec:tests}. Using theoretical
spectra we search for any possible deviations to the input dust
mineralogy, that could either affect the inferred grain sizes or
crystallinity fractions. The main result is that even if there is a
slight deviation for a few individual cases, in a statistical point of
view for a large sample, the B2C procedure is robust when determining
the dust mineralogy.

\section{Spectral decomposition of 58 T\,Tauri IRS spectra with the
  B2C model \label{sec:res}}

We run the B2C compositional fitting procedure on 58 different stars
(most of them being T\,Tauri stars, except \object{BD+31 634} which is
an Herbig Ae star), for which we obtained Spitzer/IRS spectra as part
of the c2d Legacy program. The spectra are presented in
\citet{Olofsson2009} and we refer to that paper for details about data
reduction. The selection of the 58 objects out of 96 in
\citet{Olofsson2009} is based on several criteria. First, some objects
do not have Short-Low data therefore the amorphous 10\,$\mu$m feature
is not complete. Second, as the goal is to determine the dust
mineralogy, we do not run the procedure for objects that do not show
clear silicate emission features, or with peculiar spectra
\citep[e.g. cold disks like \object{LkHa\,330} or
\object{CoKu\,Tau\,/4},][]{Brown2007}. Finally, objects for which
continuum estimation in the 5--7.5\,$\mu$m spectral region was not
possible using a power-law were eliminated.

The spectral range used for the fits is always limited to a maximum
wavelength of 35\,$\mu$m. The first reason of this choice is that the
products of the c2d extraction pipeline are, for our sample, in
average limited to 36.6\,$\mu$m. In addition to this, the degrading
quality of the end of the spectra, likely caused by the quality of the
relative spectral response function used at these wavelengths, may
lead to an over-prediction of the crystalline content, for the cold
component.  In a few cases, the spectra are rising for wavelengths
larger than 35\,$\mu$m, and the fitting procedure tries to match this
rise with crystalline features (which are the only features strong
enough in this spectral range). Finally, longer wavelengths may probe
an even colder dust content and this would require the implementation
of a third dust component in order to reproduce the entire spectral
range. For all these reasons, we choose to limit the modeling to
wavelengths smaller than 35\,$\mu$m.

The source list and relative abundances for every object can be found
in Table\,\ref{tbl:silicate}. Because of some degeneracy between
amorphous olivine and pyroxene opacities, we sum their respective
abundances to a single amorphous component. The final fits to the 58
IRS spectra are displayed in Figs\,\ref{all:fit}--\ref{all:fit3}.

\subsection{Grain size properties\label{sec:coag}}

\begin{figure}
  \hspace*{-0.5cm}\includegraphics[angle=0,width=\columnwidth,origin=bl]{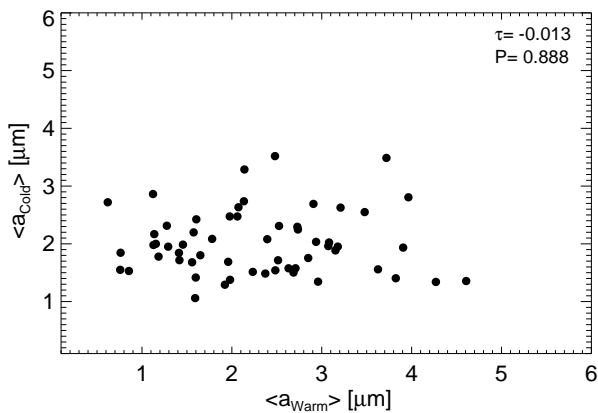}
  \caption{Mean grain sizes for the cold component as a function of
    the mean grain size for the warm component\label{fig:gs}}
\end{figure}

\subsubsection{Mean mass-averaged grain sizes}
With the outputs of the ``B2C'' procedure for the 58 objects processed
we have statistical trends on typical grain sizes necessary to
reproduce the spectra. The mean mass-averaged grain sizes for the warm
and cold components, $\langle a_{\rm warm} \rangle$ and $\langle
a_{\rm cold} \rangle$, respectively, are calculated as follows:
\begin{eqnarray}
  \langle a_{\rm warm/cold} \rangle = 
  \left(\sum_{j=1}^{\mathrm{N_{\rm sizes}}}  a_{j} 
    \sum_{i=1}^{\mathrm{N_{\rm species}}}  
    M_{{\rm w/c},i}^{j} \right)
  \times \left( \sum_{j=1}^{\mathrm{N_{\rm sizes}}} 
    \sum_{i=1}^{\mathrm{N_{\rm species}}}  
    M_{{\rm w/c},i}^{j} \right)^{-1}
  \label{eq:meana}
\end{eqnarray}
where $a_1 = 0.1\,\mu$m (small grains), $a_2 = 1.5\,\mu$m
(intermediate-sized grains) and $a_3 = 6\,\mu$m (large grains), with
the masses as defined in Eqs.\,\ref{eq:warmcomp} and
\ref{eq:coldcomp}. We further define $\langle a_{\rm warm/cold}^{\rm
  amo}\rangle$ and $\langle a_{\rm warm/cold}^{\rm cry}\rangle$ the
mass-averaged grains sizes for amorphous and crystalline grains,
respectively, in the warm or cold component.

We obtain a mean mass-averaged size of $\langle a_{\rm warm} \rangle
=$ 2.28\,$\mu$m for the warm component, and a comparable value of
$\langle a_{\rm cold} \rangle =$ 2.02\,$\mu$m for the cold
component. Figure\,\ref{fig:gs} shows that mass-averaged grain sizes
for both components are uncorrelated with each other. In order to
quantify the strength of the correlation, we compute the Kendall
$\tau$ correlation coefficient and its associated probability $P$. The
$\tau$ value denotes if there is any correlation or anti-correlation
($\tau = 1$ or $-1$, respectively), and the $P$ probability
corresponds to the significance probability (from 0 to 1, from the
most to the less significant). Regarding the latter two quantities we
obtain a Kendall $\tau$ value of -0.013 and a significance probability
$P =$ 0.888. This overall suggests that the warm and cold disk regions
considered in this study are independent, {\it as both components show
  uncorrelated grain sizes in the inner and outer regions}. This
suggests that differential grain growth is not the sole process
explaining the observed variations from object to object. This result
is in line with the conclusions of \citet{Olofsson2009} and supporting
the B2C model assumptions. In their study of 65 TTauri stars,
\citet{Sargent2009a} find that grains are larger in the inner regions
compared to outer regions, and argue this difference can be explained
by faster grain coagulation in the inner regions, where dynamical
timescales are shorter. In our study, we find no strong evidence of
such a difference (2.28 versus 2.02\,$\mu$m). However
Fig.\,\ref{fig:gs} shows a larger dispersion in grain sizes for the
warm component compared to the cold component. This could be a
consequence of shorter dynamical timescales in the inner regions where
grains are not frozen and will be strongly submitted to both
coagulation and fragmentation processes, compared to the outer
regions.

Additionally, we investigate the different mass-averaged grain sizes,
for both the crystalline and amorphous grains. For the warm component,
crystalline grains have a mean mass-averaged grain size of $\langle
a_{\rm warm}^{\rm cry} \rangle =$ 1.14\,$\mu$m while amorphous grains
have $\langle a_{\rm warm}^{\rm amo}\rangle =$ 2.50\,$\mu$m. For the
cold component, we find $\langle a_{\rm cold}^{\rm cry} \rangle =$
0.79\,$\mu$m and $\langle a_{\rm cold}^{\rm amo} \rangle =$
2.40\,$\mu$m. As we did not include large crystalline grains in the
fitting process, it is not surprising to obtain smaller mass-averaged
sizes for the crystalline grains compared to amorphous grains. However
this trend is supported by the results from \citet{Bouwman2008}, for
seven T\,Tauri stars.

\begin{figure}
  \begin{center}
    \hspace*{-0.5cm}\includegraphics[angle=0,width=\columnwidth,origin=bl]{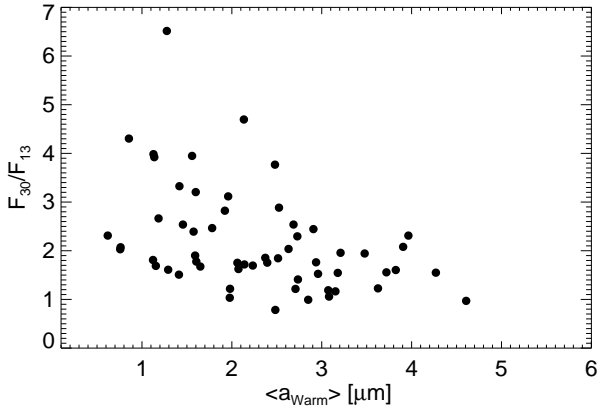}
    \caption{Flaring indices $F_{30}/F_{13}$ as a function of the warm
      mass-averaged mean grain size.\label{fig:C23_F30}}
\end{center} 
\end{figure}

\subsubsection{Mass-averaged grain size versus disk flaring}

As in \citet{Olofsson2009}, we find a trend between $\langle a_{\rm
  warm} \rangle$ and disk flaring as measured by the flux ratio
$F_{30}/F_{13}$ (fluxes in Jy integrated between $13 \pm 0.5$\,$\mu$m
for $F_{13}$, and $30 \pm 1$\,$\mu$m for $F_{30}$). As illustrated in
Figure\,\ref{fig:C23_F30}, we find that {\it large grains in the warm
  component are mostly present in flat disks, while smaller grains can
  be found in both flared or flat disks}. This anti-correlation for
the warm component has a Kendall correlation coefficient $\tau$ of
-0.31 with a significance probability of 5.0$ \times
10^{-4}$. This anti-correlation is still present, when considering
only the warm amorphous grain sizes ($\tau =$ -0.32 with $P =$
3.5 $\times 10^{-4}$). A similar anti-correlation has also been
found by \citet{Bouwman2008}, and by \citet{Watson2009} in their
Taurus-Auriga association sample.

Considering $\langle a_{\rm cold}^{\rm cry} \rangle$ as a function of
the flaring degree ($F_{30}/F_{13}$), we search for a similar result
as in \citet{Olofsson2009} where we found that small crystalline
grains are preferentially seen in flattened disks, while large
crystalline grains can be found in a variety of flat or flared
disks. We did not find a similar trend from the outputs of the
modeling, the correlation coefficient being $\tau =$ -0.040, with a
significance probability of 0.66. According to this model, the size of
the cold crystalline material seems to be strongly unrelated with the
flaring degree of disks.

\subsubsection{Flattened grain size distributions\label{sec:strength}}

\begin{figure}
  \hspace*{-0.cm}\includegraphics[angle=0,width=1.\columnwidth,origin=bl]{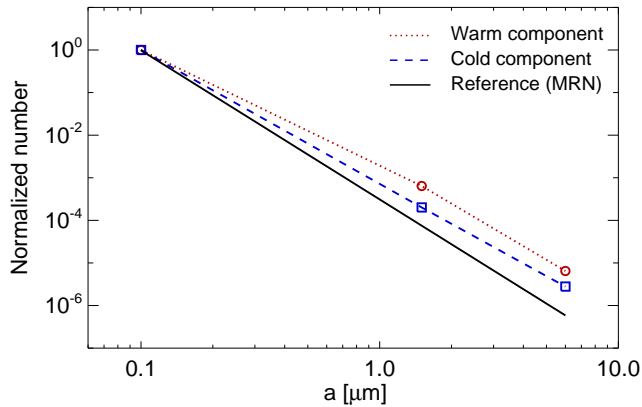}
  \caption{Grain sizes distribution, for warm (red circles and dotted
    line) and cold (blue squares and dashed line) component. The MRN
    reference distribution (-3.5 index) is the black plain
    line.\label{fig:pl}}
\end{figure}

Since our spectral decomposition includes three grain sizes, for the
amorphous species, we can evaluate the shape of the size distribution
in the atmospheres of disks. Assuming a differential size distribution
$\mathrm{d}n(a) = K_a a^{p}\mathrm{d}a$ where $K_a$ is a normalization
constant, we approximate the number $N_a$ of grains of size $a$ with:
\begin{eqnarray}
  N_a& \simeq &  \int_{a-\delta a/2}^{a+\delta a/2} \mathrm{d}n(a) 
  =  \frac{K_a}{1+p} \left[\left(a+\frac{\delta a}{2}\right)^{1+p}
    - \left(a-\frac{\delta a}{2}\right)^{1+p}\right] \label{eq:taylor}\\
  & \simeq & K_a a^{p}\delta a
\end{eqnarray}
after a Taylor expansion to the first order of Eq.\,\ref{eq:taylor}
about $a$ assuming $\delta a \ll a$. This shows that $N_{a_i}/N_{a_k}
\simeq (a_i/a_k)^p$ for two grain sizes $a_i$ and $a_k$, and therefore
that relative numbers of grains can directly be used to evaluate the
slope of the size distribution $p$. For each grain size, we compute
the mean mass obtained from the B2C simulations, and then divide it by
the corresponding volume ($\propto a^3$). We then normalize these
relative grain numbers so that the total number of 0.1\,$\mu$m-sized
grain equals 1.

Fig.\,\ref{fig:pl} shows the mean differential grain size distribution
in normalized number of grains obtained following this procedure. On
Fig.\,\ref{fig:pl}, the warm component is represented in filled
circles, the cold component in open circles, and a reference MRN
differential size distribution ($p=-3.5$) with a dashed line. Assuming
power-law size distributions, we find $p$ indexes of $p_{\rm
  warm}=$-2.89 and $p_{\rm cold}=$-3.13 for the warm and cold
components, respectively, indicating much flatter size distributions
compared to the MRN size distribution. Two additional runs of the B2C
procedure for the 58 objects allowed to confirm this trend, and to
estimate uncertainties on the slopes. For the warm component, we
typically find $p = -2.90 \pm 0.1$, and $p = -3.15 \pm 0.15$ for the
cold component.

Because the $p$ indexes are larger or close to $-3$, an immediate
consequence of this result is that the emission in disk upper layers
is statistically dominated by the $\mu$m-sized grains in our stellar
sample, especially for the warm component, and is largely independent
of the minimum grain size of the size distribution as long as its
value is small enough (submicronic, see discussion in Sec. 5.3 of
\citealp{Olofsson2009}). This suggests that the flat, boxy $10\,\mu$m
feature profile for most T\,Tauri stars discussed in terms of a
depletion of small grains in \citet{Olofsson2009}, is more precisely
revealing a {\it significant flattening of the size distribution,
  i.e. a relative lack of submicron-sized grains with respect to
  micron-sized grains, but not a complete depletion}. This {\it size
  distribution flattening} is further discussed in
Sec.\,\ref{sec:disc}.

\subsection{Silicate crystals}
\label{sec:crystals}

\subsubsection{The crystallinity paradox reexamined}
\begin{figure}
  \hspace*{-0.cm}\includegraphics[angle=0,width=1\columnwidth,origin=bl]{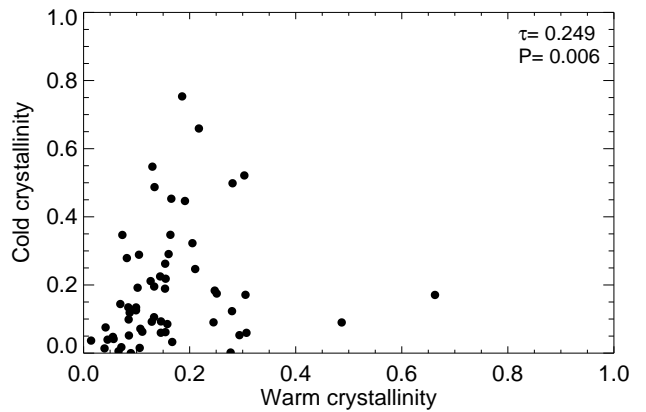}
  \caption{Crystalline fraction of the cold component ($C_{\rm cold}$)
    as a function of the warm crystalline fraction ($C_{\rm
      warm}$).\label{fig:cryst}}
\end{figure}

\begin{figure*}
  \begin{center}
  \hspace*{-0.5cm}\includegraphics[angle=0,width=1\columnwidth,origin=bl]{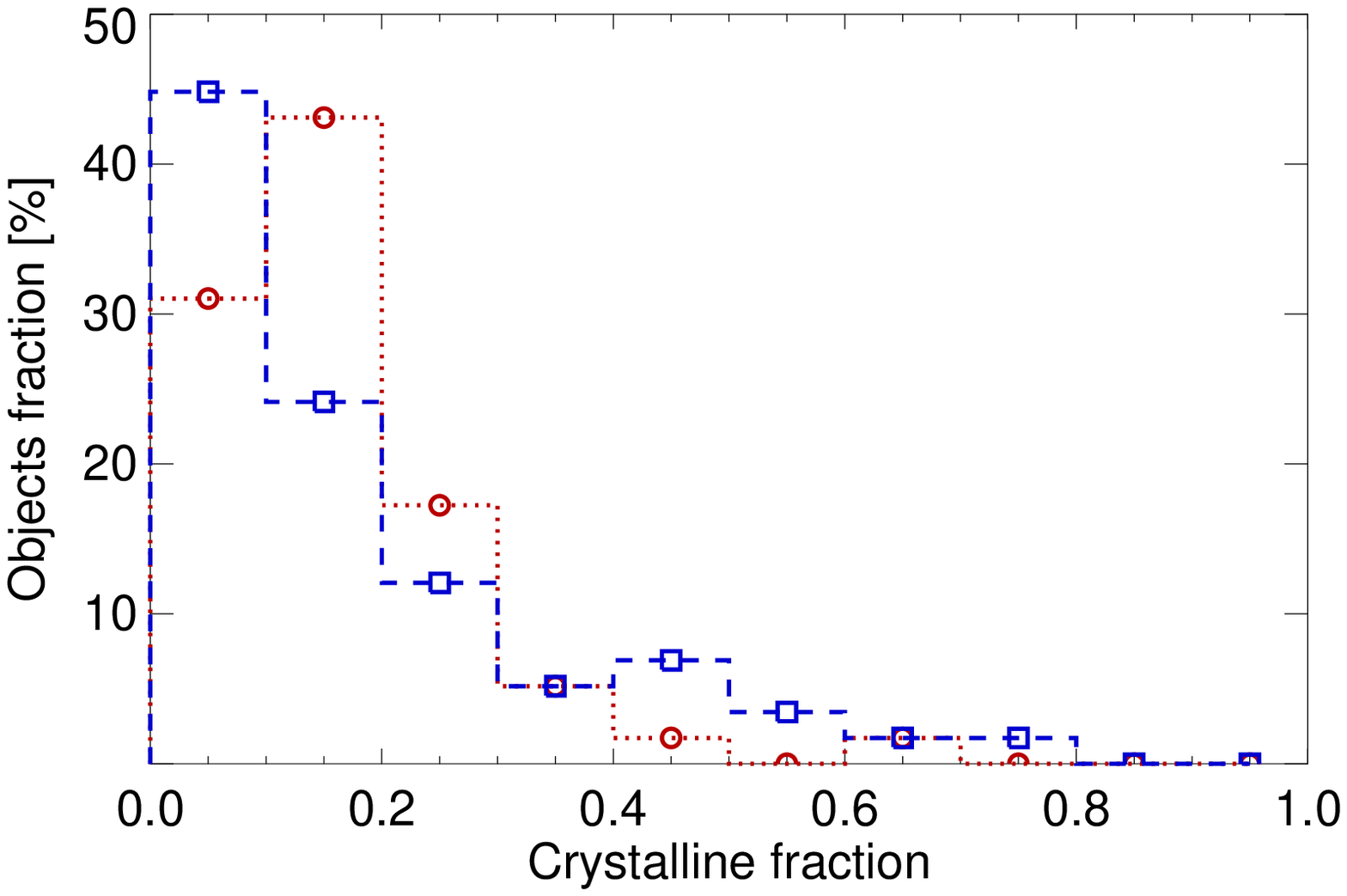}
  \hspace*{-0.5cm}\includegraphics[angle=0,width=1\columnwidth,origin=bl]{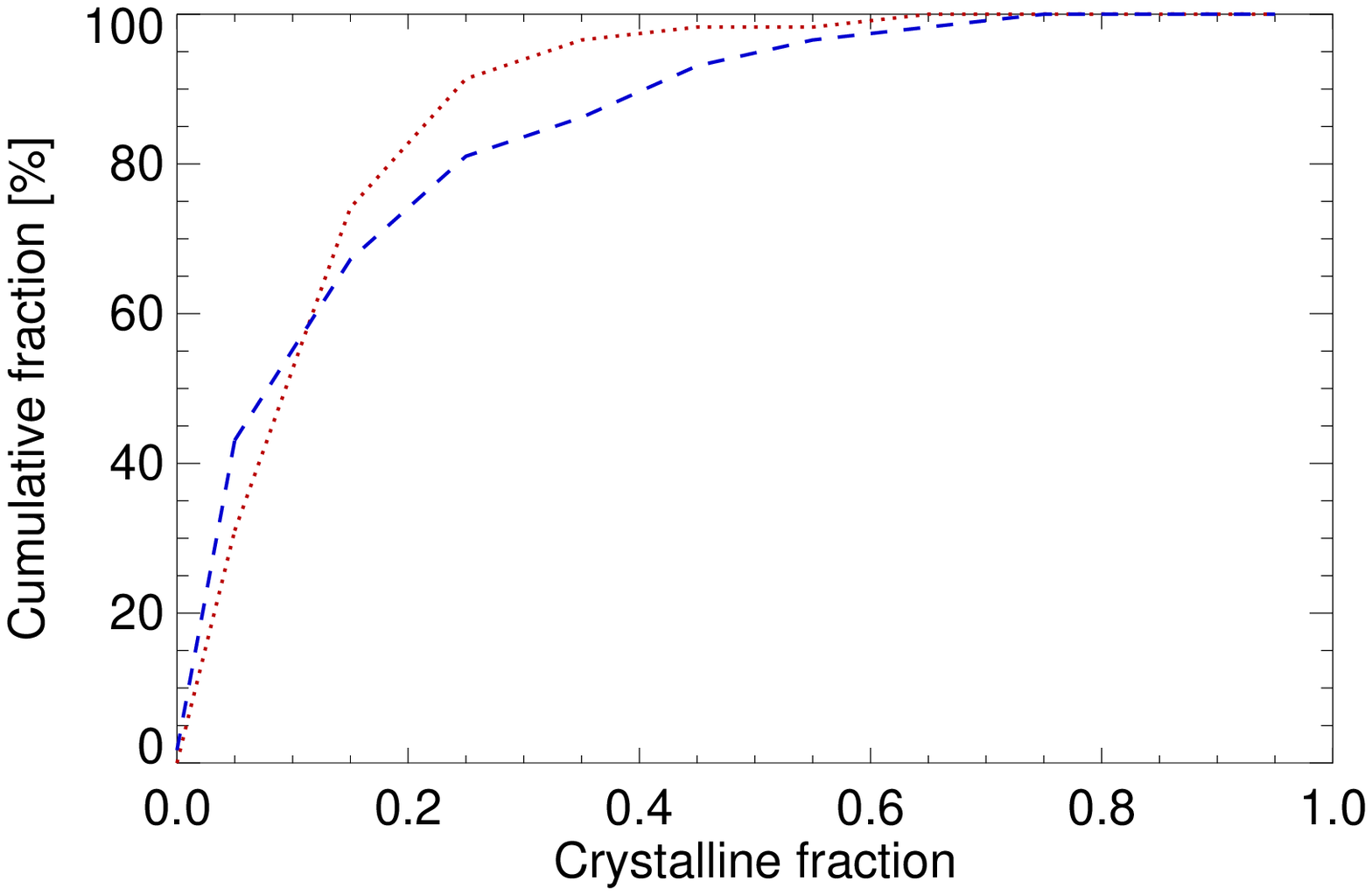}
  \caption{{\it Left panel:} crystalline distributions for warm
    component ($C_{\rm warm}$, red circles and dotted histogram) and
    cold component ($C_{\rm cold}$, blue squares and dashed
    hsitogram). {\it Right panel: }cumulative fractions of the
    crystallinity fractions, for the warm (red dotted line) and cold
    (blue dashed line) components.\label{fig:cdist}}
  \end{center}
\end{figure*}

Figure\,\ref{fig:cryst} shows the correlation between the warm and
cold crystalline fractions (which we denote as $C_{\rm warm}$ and
$C_{\rm cold}$, respectively, in the following), with Kendall's $\tau
= $0.25 and a significance probability $P =$ 5.8 $\times
10^{-3}$. {\it While slightly dispersed, there is a tendency for a
  simultaneous increase of the crystallinity in both the warm and cold
  components}. The crystalline distributions for the warm and cold
components are displayed on the left panel of Fig.\,\ref{fig:cdist}.
For the warm component, the mean crystalline fraction is $\langle
C_{\rm warm}\rangle\simeq$ 16\%, while this fraction shifts up to
$\langle C_{\rm cold}\rangle\simeq$ 19\% for the cold
component. Overall, these results show no significant difference
between the crystalline fractions in both components.

These modeling results give us new insights concerning the
crystallinity paradox identified observationally in
\citet{Olofsson2009}. The crystallinity paradox expresses the fact
that crystalline features at long wavelengths are $\sim$\,3.5 times
more frequently detected than those at shorter wavelengths. Using
simple models of dust opacities, we concluded that a contrast effect
(i.e. the strong 10\,$\mu$m amorphous feature masking smaller
crystalline features) could not be accounted for the low detection
frequency of crystalline features in the 10\,$\mu$m range, a result
valid for crystallinity fraction larger than 15\% for 1.5\,$\mu$m
grains (\citealp{Olofsson2009}). For lower crystallinity values, we
indeed showed that the synthetic spectra computed in
\citet{Olofsson2009} were not representative of the observations. This
contrast effect could therefore not be constrained for
low-crystallinity fractions, where most of our objects lie for
  the warm component, according to Fig.\,\ref{fig:cdist}. With the
outputs of the modeling procedure, we can investigate this issue with
a new look. Right panel of Fig.\,\ref{fig:cdist} shows the cumulative
fractions as a function of the crystallinity fractions. Even if there
are a few differences between the warm (red dotted) and cold (blue
dashed) components, the two cumulative fractions display a very
similar behavior that cannot solely explain the crystallinity paradox
that we derived from the observations. This therefore means that, with
this modeling procedure, a contrast issue around 10\,$\mu$m is
required to match the observations described in
\citet{Olofsson2009}. In other words, for some objects, few
crystalline grains are required to reproduce the 10\,$\mu$m feature,
but these grains do not produce strong features (on top of the
amorphous feature) that can easily be detected in the spectra. Still,
the left panel of Fig.\,\ref{fig:cdist} shows that the cold
crystalline distribution is wider than the warm distribution. For the
warm component, 3.5\% of the objects have a crystallinity fractions
larger than 40\%, while this fraction shifts up to 13.8\% for the cold
component. This overall means that in a few cases, we see more
crystalline grains in the cold component compared to the warm
component, as result also found by \citet{Sargent2009a}

\subsubsection{Crystallinity versus disk and stellar properties}
We also search for correlations between disk flaring proxies and
crystallinity. We find no striking correlations regarding the disk
flaring indexes $F_{30}/F_{13}$ and warm crystalline fraction $C_{\rm
  warm}$ ($\tau =$ -0.11 with a significance probability $P =$
0.23). Flared or flat disks present a wide range of crystalline
fractions, with a strong dispersion. We find similar results, with an
important dispersion, for the flaring index versus cold crystalline
fraction $C_{\rm cold}$ ($\tau =$ -0.16 with a significance
probability $P =$ 0.08), meaning that crystallinity fraction of the
cold component does not strongly depend on the disk flaring.

\begin{figure}
\begin{center} \hspace*{-0.5cm}\includegraphics[angle=0,width=1.\columnwidth,origin=bl]{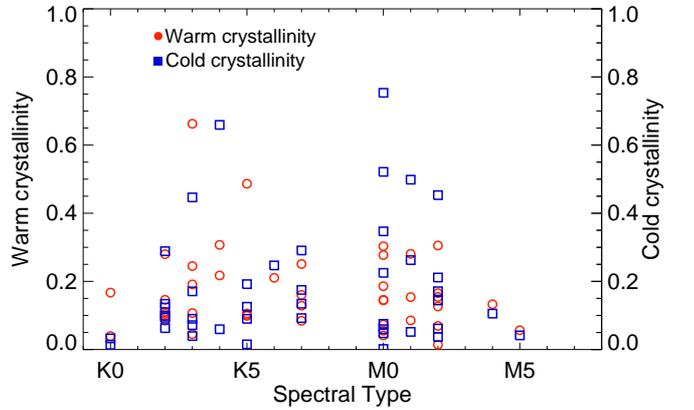}
  \caption{Crystalline fraction for the warm ($C_{\rm warm}$, open red
    circles) and the cold ($C_{\rm cold}$, open blue squares)
    components, as a function of the spectral
    types.\label{fig:spt_cr}}
\end{center}
\end{figure}
 
Figure\,\ref{fig:spt_cr} displays the dependence of crystalline
fractions for both the warm ($C_{\rm warm}$, red open circles) and
cold components ($C_{\rm cold}$, blue open squares) as a function of
the spectral type, for stars between K0 and M5. We find no correlation
for the warm component with the spectral type ($\tau =$ 0.13 with $P
=$ 0.20), suggesting that the degree of crystallization does not
depend upon the spectral type (in the explored range), and that
crystallization processes are very general for TTs. This result is in
good agreement with \citet{Watson2009}, who find no correlation
between crystallinity and stellar luminosity or stellar mass. On the
other hand, \citet{Watson2009} only studied the cold disk regions
crystallinity via the presence of the 33.6\,$\mu$m forsterite
feature. Using the outputs of our B2C procedure, we are able to better
quantify the crystalline fraction of the cold component, and we do not
find any correlation of $C_{\rm cold}$ with spectral type, the
dispersion being too large ($\tau =$ 0.12 with $P =$ 0.24), thereby
confirming the result by \citet{Watson2009}.

\subsubsection{Dust evolution: coagulation versus crystallization}

\begin{figure}
 \hspace*{-0.5cm}\includegraphics[angle=0,width=1.1\columnwidth,origin=bl]{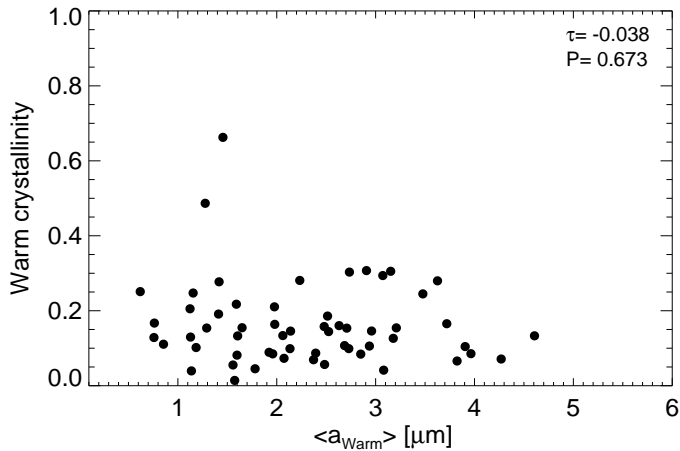}
  \caption{Warm crystalline fraction $C_{\rm warm}$ versus warm
    mass-averaged grain size $\langle a_{\rm
      warm}\rangle$.\label{fig:wgc}}
\end{figure}

Dust is evolving inside circumstellar disks, either in size or in
lattice structure, in particular through coagulation and
crystallization. To investigate possible links between these two
phenomena, we search for correlations between grain sizes, for
amorphous and crystalline dust contents ($\langle a_{\rm
  warm/cold}\rangle$), as a function of crystalline fractions, in both
the warm and cold components ($C_{\rm warm/cold}$). 

We find no evidence that grains are growing and crystallizing at the
same time in the warm component. Considering $\langle a_{\rm
  warm}\rangle$ and $C_{\rm warm}$ values, we obtain the following
coefficients: $\tau=$ -0.038, with a significance probability equal to
0.67. This result is displayed in Fig.\,\ref{fig:wgc}. We obtain an
even less favorable correlation coefficients for the cold component
($\tau$ = 0.013 with $P =$ 0.89). This overall indicates that {\it the
  processes that govern the mean size of the grains and their
  crystallinity are independent phenomena in disks}.

We also search for correlations between the amorphous and crystalline
dust contents for the two components, to see how they evolve with
respect to each others. Based on our B2C spectral decomposition, we
find no significant correlations between the several amorphous or
crystalline dust populations.

\subsubsection{Enstatite, forsterite and silica}

\begin{figure}
  \begin{center}
    \resizebox{\hsize}{!}{\includegraphics{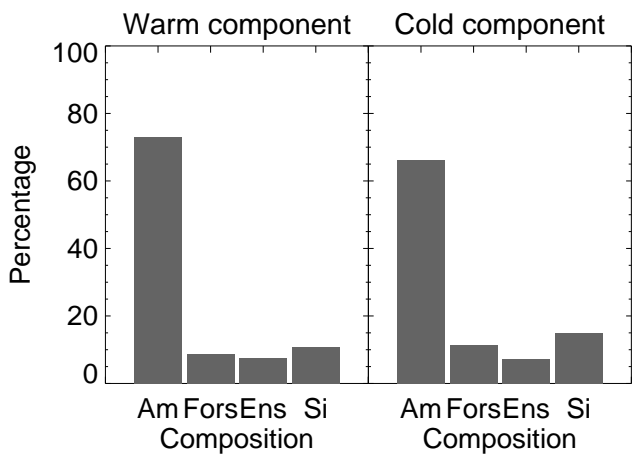}}
    \caption{\label{fig:cps}Mean composition for warm and cold
      components according to the results of the 58 objects
      processed.}
  \end{center}
\end{figure}

\begin{figure*}
  \begin{center}
    \hspace*{-0.5cm}\includegraphics[angle=0,width=1.\columnwidth,origin=bl]{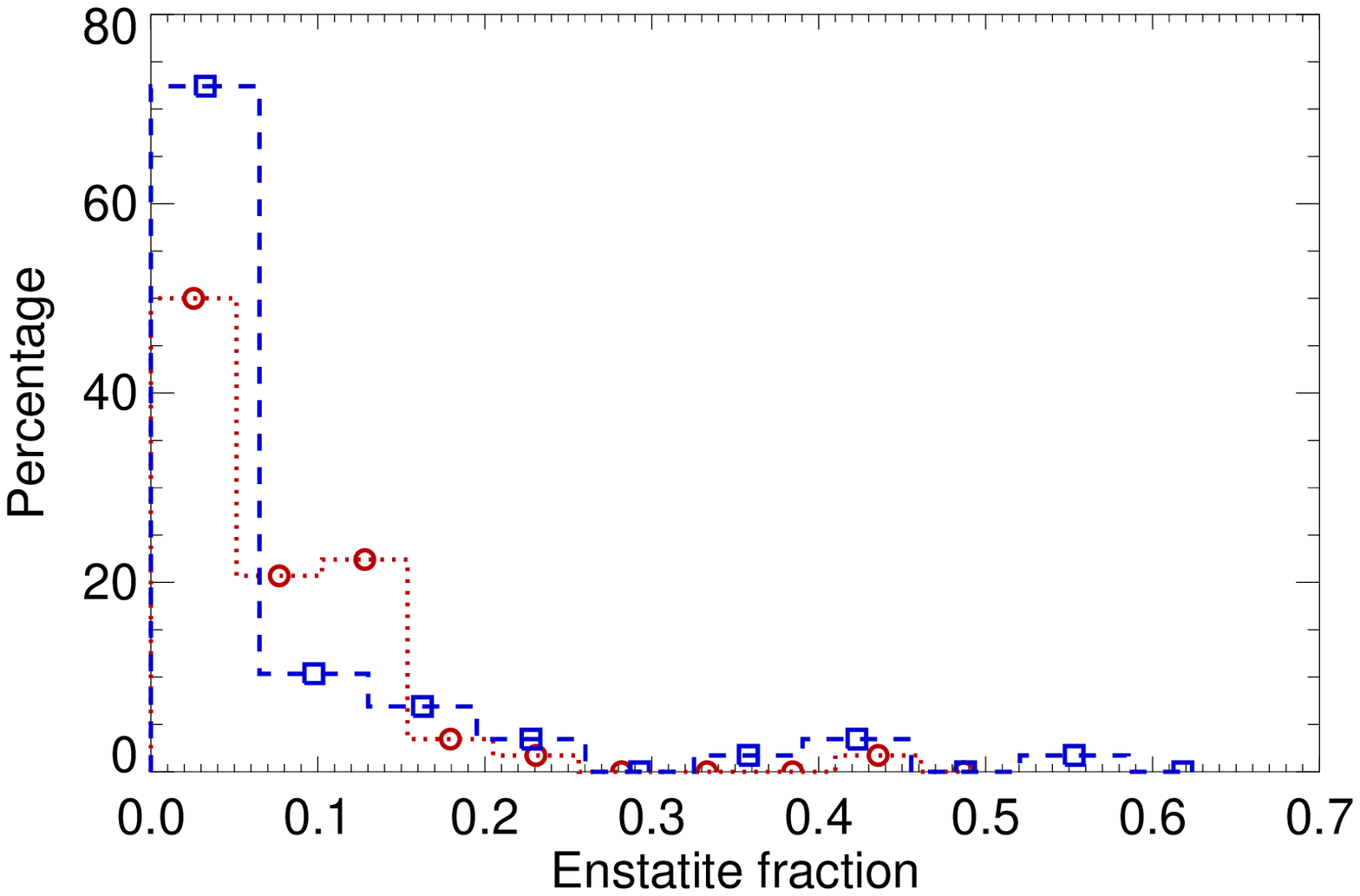}
    \hspace*{-0.5cm}\includegraphics[angle=0,width=1.\columnwidth,origin=bl]{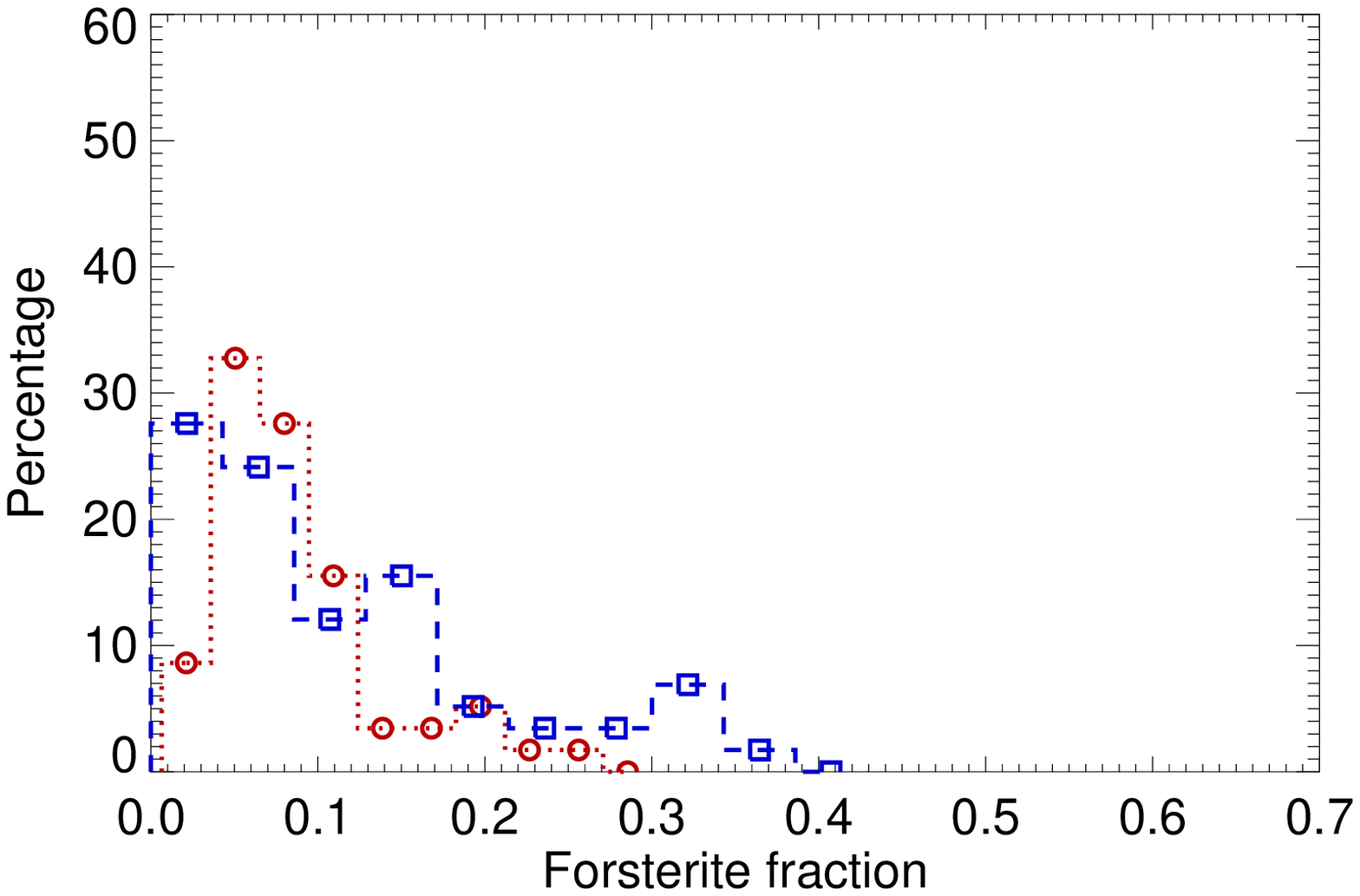}
    \caption{Distribution for the warm (red circles and dotted
      histogram), cold (blue squares and dashed histogram) for
      enstatite ({\it left panel}) and forsterite ({\it right panel})
      crystalline abundances.\label{fig:ens_for}}
  \end{center}
\end{figure*}

\cite{Bouwman2008} find that enstatite dominates in the innermost warm
regions over forsterite, and on the other side that forsterite is the
dominant silicate crystal in the cooler outer
regions. Fig.\,\ref{fig:cps} shows the mean composition we obtain for
the 58 objects, for both components (warm and cold). As amorphous
pyroxene and amorphous olivine spectral signatures are very similar,
we merge together their relative contributions. Regarding the
crystalline content, even if there is no clear predominance of
forsterite or enstatite masses for the warm component (8.7\% vs. 7.4\%
respectively), for the cold component forsterite appears more abundant
than enstatite (11.4\% vs. 7.4\% in mass). In order to have more
details on the presence or absence of these two crystals,
Figure\,\ref{fig:ens_for} shows the distribution for the enstatite and
forsterite abundances (left and right panel, respectively), for the
warm and cold components (filled circles and open circles,
respectively). {\it No predominance of either enstatite or forsterite
  is found in the warm component, while forsterite seems to be more
  frequent than enstatite in the cold component}. Similar
distributions are found by \citet{Sargent2009a} regarding the warm and
cold forsterite and enstatite.

We search for correlations between the masses of the warm or cold
enstatite populations, and warm or cold forsterite populations. We
find a correlation between warm enstatite and warm forsterite grains,
with a Kendall's $\tau$ value of 0.33 and with a significance
probability $P$ equal to 2.7$ \times 10^{-4}$. This trend tends to
confirm the results displayed in Fig.\,\ref{fig:cps}: the
crystallisation processes in the warm component do not favor the
formation of one or the other crystals. We also find a correlation
between the relative masses of warm enstatite grains and cold
forsterite grains, with $\tau = $ 0.31 and $P$ equal to 7.0$ \times
10^{-4}$.  {\it This tends to indicate that the enrichment in
  crystalline grains in the disk is global within $\sim$10\,AU, and
  not local}.

Knowing that forsterite and silica can interact with each other to
form enstatite ($\mathrm{Mg_2SiO_4} + \mathrm{SiO_2} \to
2 \mathrm{MgSiO_3}$), we search for correlations between the relatives
masses of forsterite, enstatite and silica in both components. We may
expect to find an anti-correlation between the presence of enstatite
and silica, as silica disappears from the disk medium in the above
reaction. We did not find any striking trend between all the masses,
even when considering only small and intermediate-sized grains (0.1
and 1.5\,$\mu$m). This suggests that {\it the reaction between
  forsterite and silica is not the main path for enstatite
  production}.

\subsection{Sample homogeneity}

\begin{table}
  \begin{center}
    \caption{\label{tbl:clouds}Cloud-to-cloud variations for several
      parameters from the spectral decomposition.}
    \begin{tabular}{l|ccccc}
      \hline\hline
      Cloud & Number & $\alpha_{\rm warm}$ & $\alpha_{\rm cold}$ & $\langle C_{\rm warm}\rangle$
      & $\langle C_{\rm cold}\rangle$ \\
      \hline
      Chamaleon & 16 & -2.97 & -3.05 & 18.4 & 19.9 \\
      Ophiuchus & 14 & -3.02 & -3.10& 16.0 & 19.0 \\
      Lupus  & 13 & -2.96 & -3.06 & 16.6 & 20.8 \\
      Perseus  & 8 & -2.60 & -3.27 & 13.2 & 12.3 \\
      Serpens  & 5 & -2.80 & -3.11 & 11.6 & 14.3 \\
      Taurus  & 1 & -2.84 & -3.00 & 30.3 & 52.1 \\
      \hline
   \end{tabular}
  \end{center}
\end{table}

The studied stellar sample is drawn from several star forming
regions. It may therefore be interesting to see if any cloud-to-cloud
variations can be observed. The 58 Class\,II objects are distributed
among 6 different clouds: Chamaleon, Ophiuchus, Lupus, Perseus,
Serpens and Taurus (except for \object{IRAS~08267-3336} which is an
isolated star). Table\,\ref{tbl:clouds} shows several results as a
function of the corresponding clouds: the number of objects, the
grain-size distribution slopes for the warm and cold components, and
both the warm and cold crystallinity fractions. The results for the
Perseus, Serpens and Taurus clouds are not statistically significant,
given the low-statistics number of stars for these regions. On the
other side, the three other clouds show very similar behavior for all
the considered results. The grain size distributions and crystallinity
fractions are very close to each other (with a number of objects
between 13 and 16 per cloud). 

Recently, \citet{Oliveira2010} studied the amorphous silicate features
in a large sample of 147 sources in the Serpens, and compared their
results with the Taurus young stars. That study, along with the
results from \citet{Olofsson2009}, show that the 10\,$\mu$m feature
appears to have a similar distribution of shape and strength,
regardless of the star forming region or stellar ages. In terms of
silicate features, the stellar sample analyzed in this study is
therefore representative of a typical population of TTauri stars.


\section{Discussion\label{sec:disc}}

\subsection{Implications for the dust dynamics in the atmospheres of
  young disks}

In \citet{Olofsson2009}, we discussed the impact of a size
distribution on the shape of the $10\,\mu$m feature, and could not
disentangle between two possibilities to explain the flat, boxy
$10\,\mu$m feature profile that is characteristic of micron-sized
grains. One possibility was that the grain size distribution is close
to a MRN-like distribution ($p = -3.5$), extending down to sizes of at
least several micrometers, with the consequence that the Spitzer/IRS
observations would then probe a truncation of the size distribution at
the minimum grain size (of the order of a micrometer). A second
possibility implied a differential size distribution
$\mathrm{d}n(a)\propto a^p \mathrm{d}a$ that departs from a MRN-like
distribution, being much flatter and with an index $p$ close to, or
larger than $-3$ to get the dust absorption/emission cross sections
dominated by grains with size parameters $2\pi a /\lambda$ close to
unity.

By model fitting 58 IRS spectra of T\,Tauri stars, we find a general
flattening of the grain size distribution ($p_{\rm warm} \simeq$ -2.9
$p_{\rm cold} \simeq $ -3.15) in the atmospheres of disks around
T~Tauri stars with respect to the grain size distribution in the ISM
up to about 10\,AU (this effect being stronger for the inner regions
of disks). Our results suggest that the frequently observed boxy shape
of the $10\,\mu$m feature is rather due to the slope of the size
distribution, than to a minimum grain size in the disk atmospheres
close to a micrometer. This finding very much relaxes the need for a
sharp, and quite generic truncation of the size distribution at around
one micrometer as proposed in \citet{Olofsson2009}. Nevertheless,
radiation pressure and/or stellar winds which had been proposed to
explain such a cut-off are likely to operate anyway, and may thus
contribute to the flattening of the size distribution by removing a
fraction of the submicron-sized grains. This would furthermore help
transporting crystalline grains outwards.

An alternative explanation for the {\it size distribution flattening}
involves an overabundance of micrometer-sized grains with respect to
the submicron-sized grains. At first glance, this situation may sound
counterintuitive as micrometer-sized grains are more prone to settle
toward the disk midplane which would, in turn, tend to steepen the
size distribution in disk atmospheres. This ignores that both
turbulent diffusion and collisions between pebbles may complicate this
picture, and may even possibly revert this trend by supplying the disk
atmospheres with micrometer-sized (and larger) grains. Therefore, the
IRS observations may in fact indicate that coagulation and vertical
mixing are slightly more efficient processes to supply grains in disk
atmospheres than fragmentation, resulting in an overabundance of
micrometer-sized grains compared to an ISM-like distribution. This
situation may last over most of the TTauri phase.

Interestingly, \citet{Lommen2010} also found a flattening of the grain
size distributions for T~Tauri stars, probed by the millimeter slope
of the SED. They found a tentative correlation between the shape of
the 10\,$\mu$m feature and the mm slope for a large sample of sources,
with flatter 10\,$\mu$m feature being associated with shallower mm
slopes. Similarly, \citet{Ricci2010} also find grain size
distributions flatter than in the ISM, with 3\,mm PdBI observations of
T~Tauri stars in the Taurus-Auriga star forming region. Therefore, the
size distribution flattening in disk atmospheres may extend to deeper
and more distant regions, and to larger grains sizes than those probed
by mid-IR spectroscopy.

\subsection{Silicate crystallization and amorphization}

In this study, we have been able to quantify the fraction of crystals
for both the warm and cold components. In \citet{Olofsson2009}, we
identify a {\it crystallinity paradox} expressing the fact that
crystalline features are much more frequently detected at long
wavelengths compared to short wavelengths. By building crystallinity
distributions with the results from the spectral decomposition, we
find that even if the cold crystalline distribution is slightly
broader than the warm distribution, they both show a similar behavior
(see right panel of Fig.\ref{fig:cdist}). Therefore, this may suggest
that the observational {\it crystallinity paradox} could be the
consequence of a combined effect: relatively low crystalline fractions
($\leq$15-20\%) for a great number of the TTauri stars, and a contrast
issue at short wavelengths preventing the direct identification of
weak crystalline features on top of the 10\,$\mu$m amorphous
feature. Nevertheless, it is noteworthy that correcting for the
contrast issue through B2C spectral decomposition does not revert the
trend, but instead makes the distribution of crystals in the warm and
cold disk regions quite similar.

In order to explain the non negligible crystallinity fractions
obtained for the cold component (up to 70\% in a few cases,
e.g. Fig.\,\ref{fig:cryst}), there is a strong need for radial outward
transport, assuming that the only source of production of crystals is
thermal annealing or gas-phase condensation/annealing, taking place in
the inner regions of disks. Several models have shown that such
mechanism is likely to happen in circumstellar disks around active
stars (see \citealp{Keller2004} or \citealp{Ciesla2009}). According to
\citet{Ciesla2009}, for an accretion rate larger than
10$^{-7}$\,M$_\odot / yr$, grains can be transported to distances out
to 20\,AU in about $\sim$\,10$^5$ years. A result in line with the 2D
time-dependent disk model described in \citet{Visser2010}. However
given the relatively young ages of our stellar sample ($\sim$ few
Myrs), and the rather high accretion rate required
(10$^{-7}$\,M$_\odot / yr$), we cannot assess that radial transport is
the only mechanism responsible for the strong similarity of both warm
and cold crystallinity distributions. This instead suggests that other
crystallization processes are necessary, that must take place in the
outer regions of disks. A theoretical model, detailed in
\citet{Tanaka2010}, shows that depending on the gas density,
crystallization can take place with typical temperature of a few
hundred Kelvin. Such process can be triggered in the outer regions by
nebular shocks (\citealp{Desch2002}) which lead to graphitization of
the carbon mantle providing the latent heat for silicate
crystallization.

However, crystallization by thermal annealing is likely to happen on
very short timescales (few hours at $T \sim 1000$\,K, according to
laboratory measurements, see \citealp{Hallenbeck1998}). Therefore, one
can expect to observe higher crystallinity fractions in the inner warm
regions compared to the outer cold regions. One possible explanation
to this issue may rely on the importance of accretion activity and on
amorphization processes of crystals that are preferentially taking
place in the inner regions of disks, as tentatively proposed by
\citet{Kessler-Silacci2006} and \citet{Olofsson2009}, and recently
discussed in more detail in \citet{Glauser2009}. This possibility is
further supported by the results of \citet{'Abrah'am2009} who studied
the TTauri star EX\,Lup with IRS, at two different epochs separated by
a three-year time span, first in a quiescent phase (see spectrum in
\citealp{Kessler-Silacci2006}), and then two months after an optical
outburst driven by an increase of the mass accretion rate. They find
an increase of crystallinity caused by this outburst in the inner disk
regions, supposedly due to thermal annealing of amorphous silicates.
The authors did not find any colder crystals in the outer component
after the outburst, and therefore excluded shock heating at larger
distances. This study shows that, obviously, the accretion activity
may have a strong impact on the dust chemistry in circumstellar
disks. Recently, \citet{Glauser2009} identified an anti-correlation
between X-ray luminosity of the stars and crystalline fraction for the
warm disk component ($C_{\rm warm}$ in Sec.\ref{sec:crystals}). The
authors performed a decomposition fit to the $10\,\mu$m feature for a
stellar sample of 42 TTauri stars, and find an anti-correlation
between $C_{\rm warm}$ and X-ray luminosities for a subsample of 20
objects with ages between 1 and 4.5\,Myr. This anti-correlation
suggests that the stellar activity and energetic ions from stellar
winds, traced by X-ray flux, can amorphize crystalline grains in disk
atmospheres due to high-energy particles.

\begin{figure}
\begin{center}
\resizebox{\hsize}{!}{\includegraphics{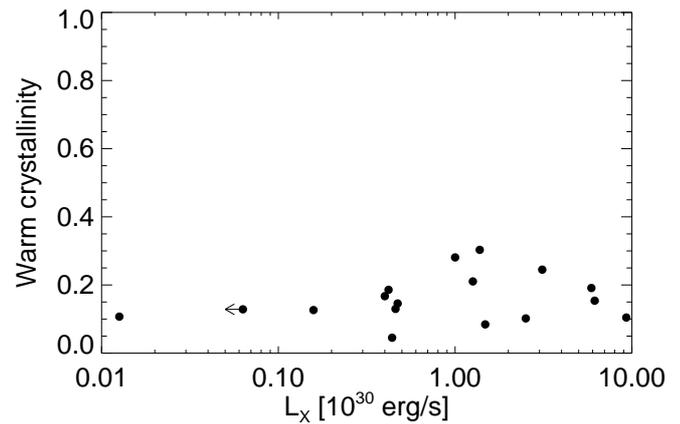}}
\caption{\label{fig:xray}Crystalline fraction for the warm component
  ($C_{\rm warm}$) as a function of the X-ray luminosities.}
\end{center}
\end{figure}

In a similar approach as \citet{Glauser2009}, we attempt to
investigate the impact of high-energy particles, assumed to be traced
by X-ray luminosity, on the disk mineralogy. Table\,\ref{tbl:xray}
displays the X-ray luminosities (in units of erg.s$^{-1}$) available
in the literature or obtained by fitting the spectra from the Second
{\sl XMM-Newton} serendipitous source catalogue
(\citealp{Watson2009a}) with the X-ray emission from one or
two-temperature plasma combined with photoelectric absorption, for a
subset of 17 objects among our stellar sample analyzed with the B2C
compositional fitting model. The low overlap between X-ray
observations with the c2d sample can be explained by the different
observation strategies: X-ray campaigns are mostly targetting dense
cores while our observations are located at the periphery of the
clouds. Fig.\,\ref{fig:xray} displays the crystalline fraction $C_{\rm
  warm}$ for the warm component inferred with the B2C procedure, as a
function of the X-ray luminosity. The associated Kendall $\tau$ value
is 0.16 with a significance probability $P =$ 0.36, suggesting a
dispersed weak trend, if any. To confirm any weak trend, we would need
to have more measurements below about 2 $\times$ 10$^{29}$
erg.s$^{-1}$. Above this luminosity value no trend is visible. Still,
we observe that objects with high X-ray luminosities seem to have
higher crystalline fractions, which seems at first in contradiction
with the conclusion by \citet{Glauser2009}. Given the limited number
of objects for which we have X-ray data in our sample, however, we
were not able to perform any age selection as in \citet{Glauser2009}.

\begin{table*}
  \begin{center}
    \caption{\label{tbl:xray}X-ray luminosities for a subset of our
      stellar sample.}
    \begin{tabular}{l|ccc}
      \hline\hline
      Star Name & Satellite & L$_X$ [erg/s] & Ref \\ \hline
      HM\,27 & ROSAT & $<$ 6.3$\times 10^{28}$ & \cite{Feigelson1993}  \\
      LkHA\,271 & CXO & 1.3$\times 10^{28}$ & \cite{Getman2002}  \\
      Hn\,9 & XMM & 1.6$\times 10^{29}$ & \cite{Robrade2007} \\
      TW\,Cha & ROSAT & 4.0$\times 10^{29}$ & \cite{Feigelson1993}  \\
      IM\,Lup & XMM & 4.2$\times 10^{29}$ & this study \\
      CK4 & CXO & 4.4$\times 10^{29}$ & \cite{Giardino2007} \\
      SSTc2d\,J162715.1-245139 & XMM & 4.6$\times 10^{29}$ & \cite{Ozawa2005} \\
      SX\,Cha & XMM & 4.7$\times 10^{29}$ & this study \\
      XX\,Cha & XMM & 1.0$\times 10^{30}$ & \cite{Robrade2007}  \\
      VZ\,Cha & XMM & 1.3$\times 10^{30}$  & \cite{Robrade2007}  \\
      V710\,Tau & XMM & 1.4$\times 10^{30}$ & \cite{Gudel2007} \\
      WX\,Cha & XMM & 1.5$\times 10^{30}$ & \cite{Stelzer2004} \\
      SR\,9 & ROSAT & 2.5$\times 10^{30}$ & \cite{Casanova1995} \\
      RU\,Lup & XMM & 3.1$\times 10^{30}$ & this study \\
      C7-11 & XMM & 5.9$\times 10^{30}$ & \cite{Stelzer2004} \\
      Sz96 & XMM & 6.2$\times 10^{30}$ & this study \\
      VW\,Cha & XMM & 9.3$\times 10^{30}$ & \cite{Stelzer2004} \\
      \hline
    \end{tabular}
  \end{center}
\end{table*}

\subsection{Strength of the 10\,$\mu$m feature, a proxy for grain size ?}

\begin{figure*}
  \hspace*{-0.5cm}\includegraphics[angle=0,width=1.1\columnwidth,origin=bl]{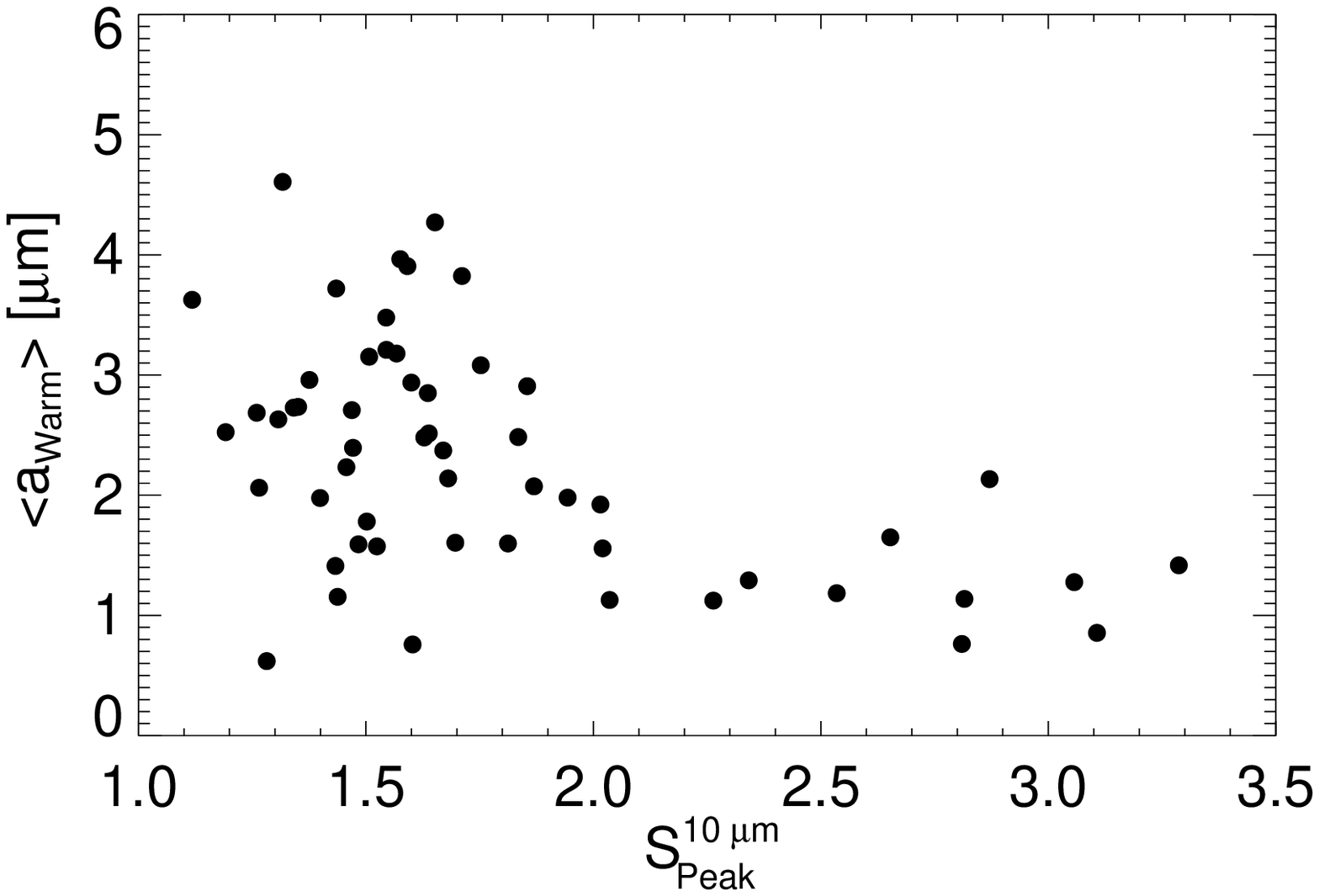}
  \hspace*{-0.5cm}\includegraphics[angle=0,width=1.1\columnwidth,origin=bl]{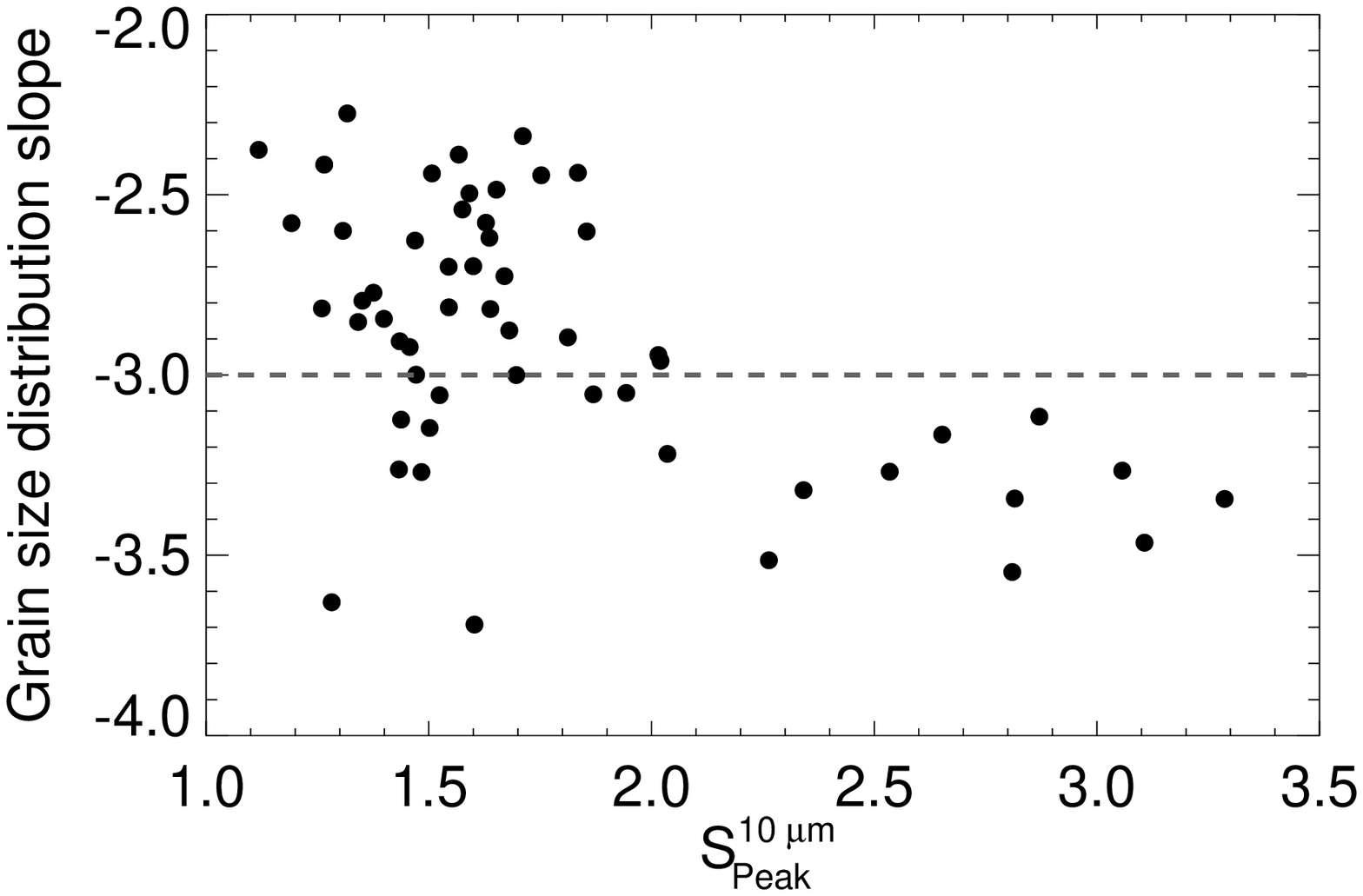}
  \caption{{\it Left panel: } Mass-averaged grain sizes for the warm component, as a function of the
    strength of the 10\,$\mu$m feature ($S_{\mathrm{Peak}}^{10 \mu
      m}$).  {\it Right panel: } power-law index $p$ of the grain size
    distribution as a function of $S_{\mathrm{Peak}}^{10 \mu
      m}$.\label{fig:sh_si}}
\end{figure*}

The strength $S_{\mathrm{Peak}}^{10 \mu m}$ of the 10\,$\mu$m feature
has been commonly used in previous studies (e.g. \citealp{2005},
\citealp{Kessler-Silacci2006}, \citealp{Olofsson2009}) as a proxy for
grain size, small $S_{\mathrm{Peak}}^{10 \mu m}$ values tracing large
(micrometer-sized) grains, while larger $S_{\mathrm{Peak}}^{10 \mu m}$
would be tracing smaller (submicron-sized) grains. In a recent paper,
\citet{Watson2009} instead attribute the diversity of strengths of the
10\,$\mu$m feature to a combination of sedimentation and an increase
of the crystallinity degree (as in \citealp{Sargent2009a}). With the
outputs of our B2C spectral decompositional procedure, we revisit the
interpretation of the strength of the 10\,$\mu$m feature as a proxy
for grain size.

The left panel of Fig.\,\ref{fig:sh_si} displays the correlation
between the mass-averaged grain size for the warm grains ($\langle
a_{\rm warm}\rangle$) and the strength of the 10\,$\mu$m feature
($S_{\mathrm{Peak}}^{10 \mu m}$). We find a Kendall $\tau$ correlation
coefficient of about -0.30 with a significance $P$ probability smaller
than 1.1 $\times 10^{-3}$, showing that grain size is well traced by
the strength of the feature. It is noteworthy to point out that
\citet{Sargent2009a} find a different result regarding these two
quantities, in their spectral decomposition analysis of 65 T~Tauri
stars. As shown in the review by \citet{Watson2009b}, based on
\citet{Sargent2009a} results, they find a correlation coefficient of
0.08 with a significant probability of 0.55. This means, according to
their model, that the strength of the 10\,$\mu$m feature is not
correlated with the presence of large grains. More precisely, they
find that seven objects among their sample with the smallest
10\,$\mu$m feature strength, can be reproduced with low abundances of
large grains. However, our opposite result is supported by several
points: we successfully checked for all dependencies that could lead
to a possible overestimate of the grain sizes (e.g the continuum
offset $O_{\nu_2}$) and we show in App.\,\ref{sec:chi}, based on the
analysis of synthetic spectra, that we may eventually underestimate
the grain sizes in the warm component rather than over-estimate it
(underestimated by $7\pm 11$\% with respect to the input value). This
deviation ($\langle a \rangle / \langle a_{\mathrm{inp}} \rangle $) is
furthermore independent of the strength of the 10\,$\mu$m feature
($S_{\mathrm{Peak}}^{10 \mu \mathrm{m}}$), with a $\tau$ value of
-0.06 and a significance probability of 0.52.

\begin{figure}
  \hspace*{-0.5cm}\includegraphics[angle=0,width=1.1\columnwidth,origin=bl]{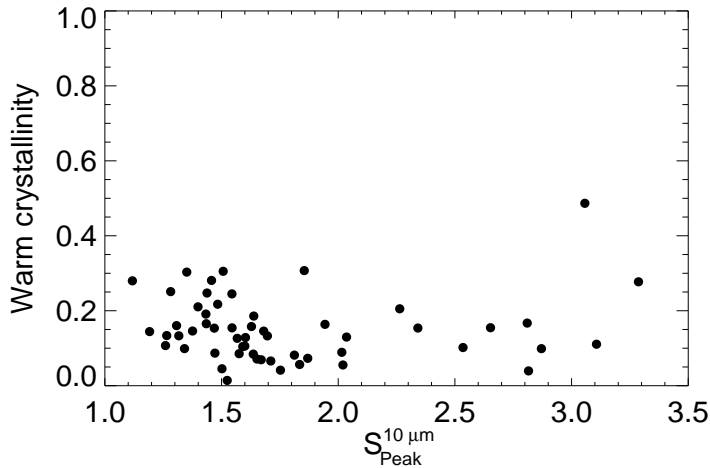}
  \caption{Crystalline fraction for the warm component ($C_{\rm
      warm}$), as a function of the strength of the 10\,$\mu$m feature
    ($S_{\mathrm{Peak}}^{10 \mu m}$).\label{fig:sh_cr}}
\end{figure}

We also search for a correlation between the warm crystalline fraction
$C_{\rm warm}$ and the strength of the 10\,$\mu$m feature
($S_{\mathrm{Peak}}^{10 \mu m}$), and find a highly dispersed relation
between the two quantities. Fig.\,\ref{fig:sh_cr} shows that the
distribution is flat ($\tau =$ -0.15 with a significance probability
of 0.10), which suggests that crystallization cannot be the sole
explanation for the diversity of 10$\,\mu$m feature strengths. This is
in line with the results by \citet{Bouwman2008} who do not find any
correlation between these two quantities in their seven T\,Tauri star
sample.

As mentioned in \citet{Olofsson2009} and in this paper, the grain size
distribution in the upper layers of disks has a strong impact on the
10\,$\mu$m feature strength. The right panel of Fig.\,\ref{fig:sh_si}
shows the power-law indexes $p$ of the size distributions (calculated
as in Sec.\,\ref{sec:strength}), as a function of the
$S_{\mathrm{Peak}}^{10 \mu m}$ values. We find an anti-correlation
($\tau =$ -0.28 with $P =$ 2.3 $\times 10^{-3}$), indicating the shape
of the 10$\,\mu$m feature is indeed related to the slope of the grain
size distribution. In fact we see two different regimes appearing: the
small $S_{\mathrm{Peak}}^{10 \mu m}$ values mostly correspond to
objects with slopes $p$ larger than $-3$ (dashed horizontal line on
the figure), while objects with large $S_{\mathrm{Peak}}^{10 \mu m}$
values essentially have slopes $p$ smaller than $-3$. This points in
the same direction as previously discussed, namely that the variation
in $S_{\mathrm{Peak}}^{10 \mu m}$ may to a large extent trace the
variation in the slope of the size distribution. Not surprisingly, we
find a very strong correlation between the mean mass-averaged grain
size and the slope of the grain size distribution, with $\tau =$ 0.79
and a significance probability below $10^{-38}$.

\section{Summary and conclusion\label{sec:sum}}

In this paper, we present a large statistical study of the dust
mineralogy in 58 proto-planetary (Class~II) disks around young solar
analogs.  We develop a robust routine that reproduces IRS spectra over
their entire spectral range (5--35\,$\mu$m), with a MCMC-like approach
coupled with a Bayesian inference method. This procedure explores
randomly the parameter space and reproduces IRS spectra using two
independent dust populations: a warm and a cold component, arising
from inner and outer regions from disks, respectively. This
2-component approach is supported by our previous statistical analysis
of the silicate features in \citet{Olofsson2009}. This B2C procedure
has been tested over synthetic spectra giving, statistically speaking,
coherent and reproducible results. In its current form, the B2C model
allows to derive relative abundances for 5 different dust species, 3
different grain sizes for the amorphous grains, 2 sizes for the
crystalline grains, for both inner regions ($\leq 1$\,AU) and outer
regions ($\sim 10$\,AU) of disks. Based on the modeling of the 58 IRS
spectra of TTauris stars, we find the following results:
\begin{enumerate}

\item The grain sizes in the inner and outer regions of disks are
  found to be uncorrelated, reinforcing the idea of two independent
  components probed by IRS. However, crystallinity fractions in warm
  and cold components are correlated to each other, indicative of a
  simultaneous enrichment of crystals within the first 10\,AU of
  disks. This suggests that dynamical processes affecting the size
  distribution are essentially local rather than global, while
  crystallization may rather trace larger scale phenomena (e.g. radial
  diffusion). Finally, we find that grain size and crystallinity
  fraction are two independent quantities with respect to each other
  in both the warm and cold components.

\item We quantify the growth of grains compared to the ISM. We find a
  significant {\it size distribution flattening} in the upper
  atmospheres of disks probed by IRS, compared to the MRN grain size
  distribution. While the MRN size distribution has a power law index
  $p = -3.5$, we find flatter distributions for both the inner regions
  ($\langle p \rangle = -2.90\pm0.1$), and the outer regions ($\langle
  p \rangle = -3.15\pm0.15$). This explains why the IR emission of
  T\,Tauri stars studied using Spitzer-IRS is dominated by
  $\mu$m-sized grains, despite the possible presence of significant
  amounts of submicron-sized grains. The $S_{\rm peak}^{10\mu m}$
  value turns out to be a proxy of the slope of the grain size
  distribution. This finding, combined with recent similar results
  from observations at millimeter wavelengths, suggests that the size
  distribution flattening is not confined to the disk upper layers but
  may extend to deeper and more distance disk regions, and larger
  grains.

\item We reexamined the {\it crystallinity paradox} identified in
  \citet{Olofsson2009}, by building crystallinity distributions for
  the warm and cold components. The mean crystallinity fractions are
  16 and 19\% for the warm and cold components, respectively.  Even if
  the crystalline distribution for the cold component is slightly
  wider compared to the warm component, both distributions show a very
  similar behavior, suggesting the silicate crystals are well mixed
  within the first 10\,AU of disks around T~Tauri stars. According to
  the B2C model, the 3.5 times more frequently detected crystalline
  features at long (20-30\,$\mu$m) than at short ($\sim$10\,$\mu$m)
  wavelengths arises from the combination of rather low crystalline
  fractions ($\leq$15-20\%) for many T~Tauri stars, and a contrast
  effect that makes the crystalline features more difficult to
  identify at around 10\,$\mu$m where the strong amorphous feature can
  hide smaller crystalline features.

\item We see a trend where flared disks (high $F_{30}/F_{13}$ values)
  tend to preferentially show small warm grains while flat disks
  (small $F_{30}/F_{13}$) show a large diversity of grain sizes. We do
  not find any correlations between the disk flaring and the
  crystallinity, meaning that the presence of such crystals is not
  related to the shape of the disks.

\item We find no predominance of any crystal species in the warm
  component, while forsterite seems to be more frequent compared to
  enstatite in the cold component. Regarding the different dust
  compositions, we find no evidence of correlations between them,
  except for the warm enstatite and forsterite grains, suggesting that
  the crystallisation processes in the inner regions of disks do not
  favor any of the two crystals. We find no link between silica and
  enstatite as one may expect if the main path for enstatite
  production would be the reaction between silica plus forsterite.

\item We do not find any striking correlation between spectral type
  and warm or cold crystallinity fraction. We do not either find any
  correlation between X-ray activity and crystallinity. As suggested
  by \citet{Watson2009}, the time variability in young objects may be
  responsible for erasing most of the correlations regarding dust
  mineralogy.

\end{enumerate}


\begin{acknowledgements}
  The authors thank Christophe Pinte for his help on the bayesian
  study to derive uncertainties on the output parameters of the
  compositionnal fitting procedure. We thank the referee,
  Dan~M.~Watson, for his very constructive comments, that helped
  improving both the modeling procedure, especially on the question of
  large 6.0\,$\mu$m crystalline grains, and the general quality of the
  paper. We also thank N.J.~Evans II for his very useful comments that
  helped improving this study. We finally thank the {\it Programme
    National de Physique Stellaire} (PNPS) and ANR (contract
  ANR-07-BLAN-0221) for supporting part of this research. This
  research is also based on observations obtained with {\sl
    XMM-Newton}, an ESA science mission with instruments and
  contributions directly funded by ESA Member States and
  NASA \end{acknowledgements}

\bibliography{13909_bib}
\appendix

\section{Validation}
\label{sec:tests}
\subsection{Procedure validation\label{sec:chi}}

We evaluate the robustness of our procedure by fitting synthetic
spectra with known abundances, temperatures and continua. This allows
to check whether these quantities could be recovered when the spectra
are processed with the B2C procedure. To this end, we used the best
fit, synthetic spectra for the 58 objects analyzed in this paper
(and presented in Sec.\,\ref{sec:res}), as fake, but representative
observations for serving as inputs to our B2C procedure. We
  reconstructed these synthetic spectra using the outputs of the
  fitting process on the original data. Namely, the same continuum
  emission, the same relative abundances for the dust species and
  their corresponding temperatures. The uncertainties chosen for the
synthetic spectra are those of the original observed spectra.

\begin{figure}
\begin{center}
\resizebox{\hsize}{!}{\includegraphics{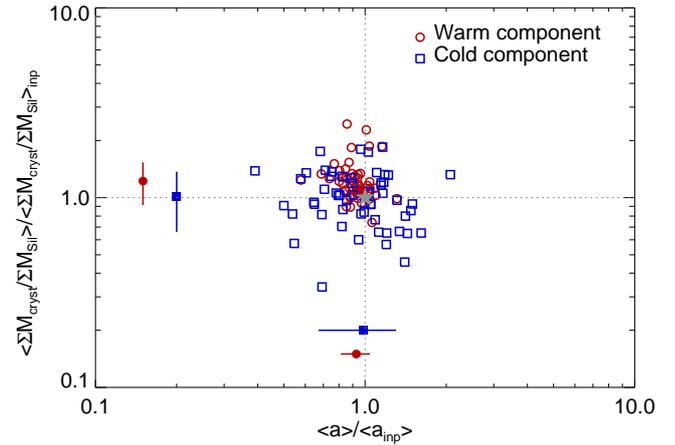}}
\caption{\label{diver}Results for the fits to the 58 synthetic spectra
  to test our ``B2C'' procedure. The $x$-axis shows the dispersion in
  grain size (ratio of inferred over input mean mass-averaged grain
  sizes) and the $y$-axis the ratio between inferred and input
  crystallinity. Red open circles correspond to the warm component,
  and blue open squares to the cold component. The filled circles are
  mean values, with error bars corresponding to standard deviations.}
\end{center}
\end{figure}

Fig.\,\ref{diver} shows the ratio between the output and input
crystallinity, as a function of the ratio between the output and input
grain sizes, for both the warm and cold components. The plot shows
ratios of about 1 on both axis, especially for the cold component
(blue squares on the figure), suggesting some best fits obtained with
the procedure may not be unique. But from a statistical point view,
the B2C procedure produces reproducible results as the mean ratios
between output and input quantities remain close to 1 (vertical and
horizontal bars on the figure).

More precisely, the warm component crystallinity is slightly
overestimated by 22\% with a dispersion (i.e. the standard deviation)
about this value of 30 \%. The inferred cold component crystallinity
is satisfactorily reproduced at the 2\% level with respect to the
input value, and the dispersion, 35\%, is rather similar to that for
the crystallinity of the warm grains. The mean mass-averaged size of
the warm grains is slightly underestimated by 7$\pm$11\% with respect
to the input value, while the mean mass-averaged size of the cold
grains is well reproduced but with a larger dispersion ( 1$\pm$32\%).

\begin{figure} \begin{center}
    \resizebox{\hsize}{!}{\includegraphics{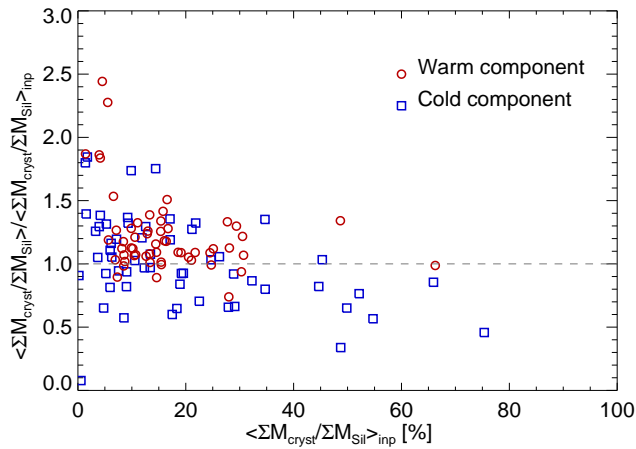}} 
    \caption{\label{fig:v_cryst}Ratios between output over input
      crystallinity fractions as a function of the input crystallinity
      fractions, for the fits to the 58 synthetic spectra. Red open
      circles are results for the warm component and blue open squares
      are for the cold component.}
\end{center}
\end{figure}

A closer look at the dispersion of the calculated crystalline fraction
as a function of the input crystallinity is shown in
Figure\,\ref{fig:v_cryst} for the 58 fits to synthetic spectra, for
both the warm (red open circles) and cold (blue open squares)
components. The over-estimation of the warm component crystalline
fractions seen in Fig.\,\ref{diver} is mostly visible for objects with
low-crystalline fractions (below 20\%). For the cold component, it
actually tends to be rather underestimated for large crystalline
fractions (above $\sim$40\%). The overall trend for both the warm and
cold components is that the dispersion raises with the decreasing
crystallinity. This finds an explanation in the fact that objects
showing high crystalline fractions display strong, high-contrast
crystalline features that are less ambiguously matched by theoretical
opacities than objects with lower crystalline fractions.

\subsection{Influence of the continuum on the cold component}

The continuum estimation is a critical step when modeling spectra,
especially for the cold component, and may contribute to the larger
uncertainties on the inferred cold component crystallinities and sizes
discussed in previous subsection. The continuum may affect the model
outputs in the following way: a high-flux continuum in the
20--30$\,\mu$m spectral range leaves very little flux to be fitted
under the spectrum, possibly leading to a composition with few large
and/or featureless grains. A low-flux continuum could, on the other
hand, lean toward large/featureless grains to fill the flux left. We
therefore examine if (whether) such trends are present in the results
of our B2C procedure (or not) for the cold component.

\begin{figure*}
  \begin{center}
    \hspace*{-0.5cm}\includegraphics[angle=0,width=1.\columnwidth,origin=bl]{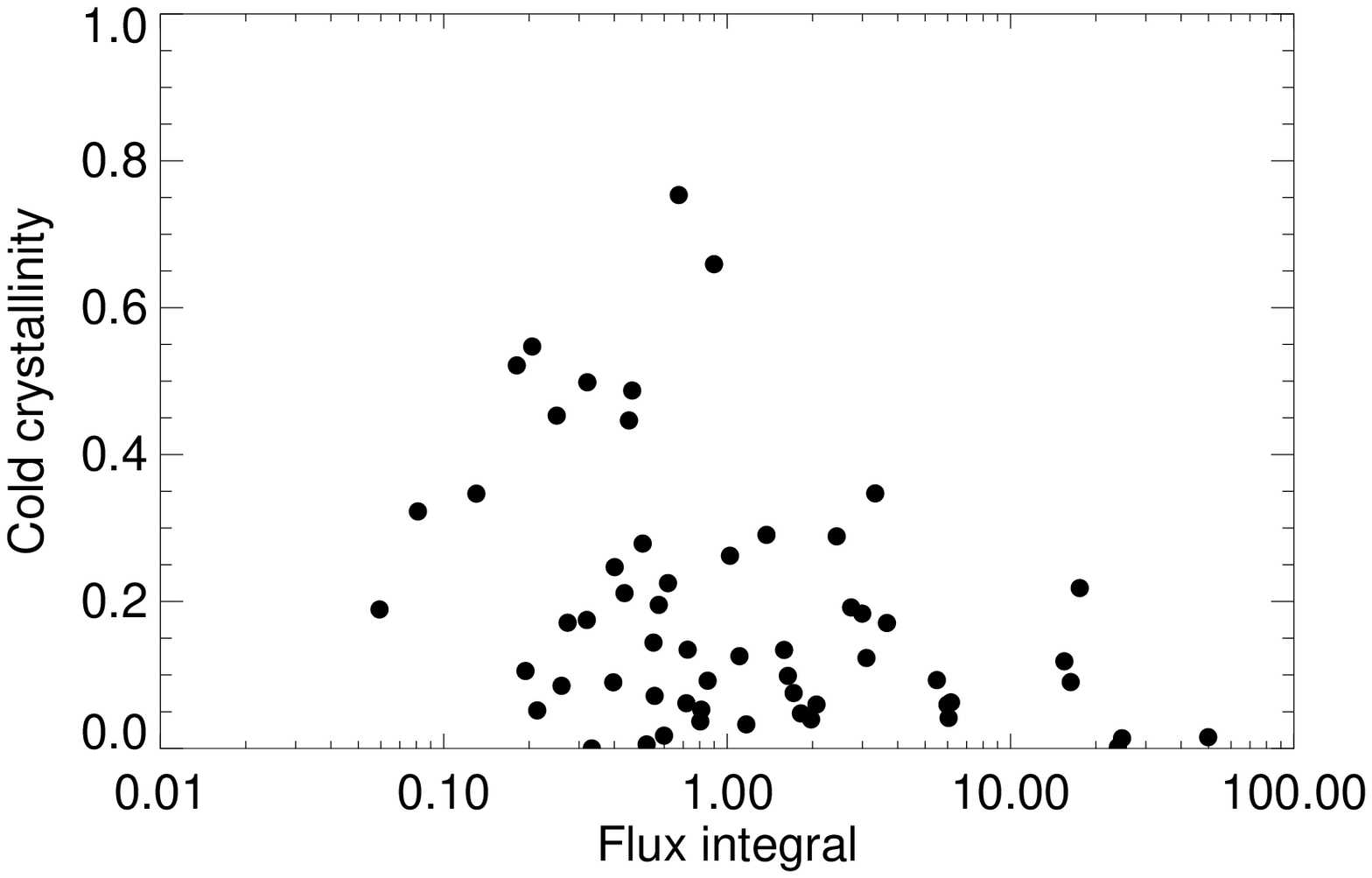}
    \hspace*{-0.5cm}\includegraphics[angle=0,width=1.\columnwidth,origin=bl]{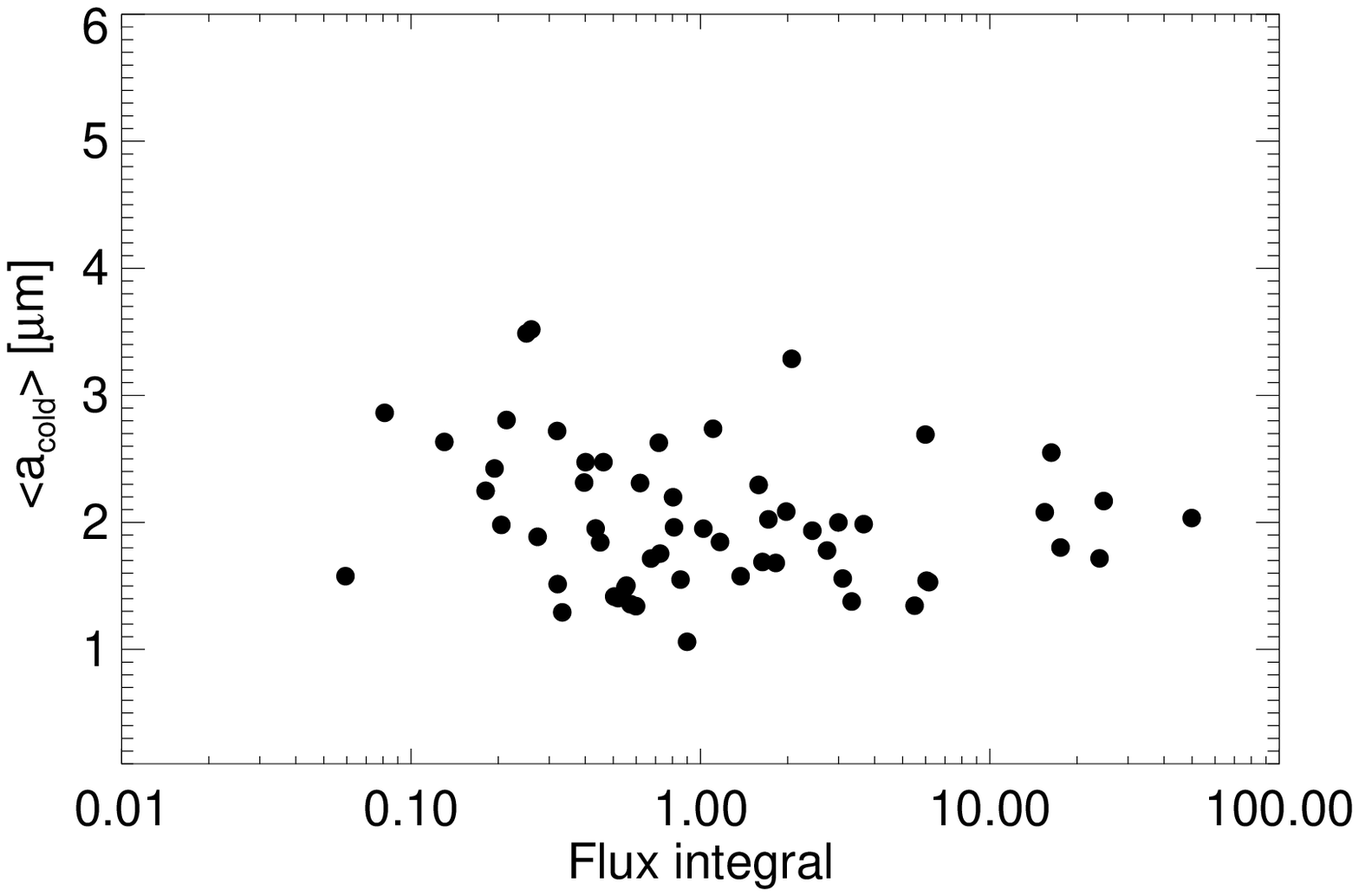}
    \caption{{\it Left panel: }Crystalline fraction for the cold
      component as a function of the integrated continuum substracted
      flux in the range 22-35\,$\mu$m (see text for details). {\it
        Right panel: }Same for the cold mean mass-averaged grain
      size. \label{fig:cont_check}}
\end{center}  
\end{figure*}

We quantify the level of continuum by integrating the
continuum-substracted spectra between 22\,$\mu$m and 35\,$\mu$m
($x$-axis on Fig.\,\ref{fig:cont_check}). Low-flux continua have high
integrated fluxes and are located on the right side of the plot in
Fig.\,\ref{fig:cont_check}, the high-flux continua being on the left
side. The left panel of Fig.\,\ref{fig:cont_check} shows the
crystalline fraction for the cold component, as a function of the
integrated flux left once the continuum is substracted. The trend low
continuum -- low crystalline fraction is visible for the highest
integrated flux values. But the large dispersion for low integrated
flux values indicates that no strong bias is introduced as we obtain
cases with very high-flux continua but still with very low crystalline
fractions. It remains that for the low-flux continua cases, we cannot
possibly obtain very high crystalline fractions as a lot of flux needs
to be filled to match the spectra. Indeed, amorphous grains are the
best choice to fulfill this requirement, therefore diminishing the
cold component crystalline fraction.

The influence of the continuum on the estimated mass-averaged grain
size is shown on the right panel of Fig.\,\ref{fig:cont_check}. A very
weak and dispersed anti-correlation is found between the two
parameters. We obtain a $\tau $ value of -0.12 with a significance
probability $P = 0.17$, meaning that the adopted shape for the
continuum is not strongly influencing the inferred mean grain
size. The trend is mostly caused by a few low-flux continua objects
modeled with large grains.

To conclude, the continuum estimation is a challenging problem for the
cold component and its adopted shape will always have an impact on the
results for the crystallinity and grain size at the same time. Still,
we have checked that, statistically speaking, we are not introducing a
strong and systematic bias with our simple, two free parameter
continuum.

\subsection{Importance of silica and necessity for large grains}

\begin{figure}
  \begin{center}
    \resizebox{\hsize}{!}{\includegraphics{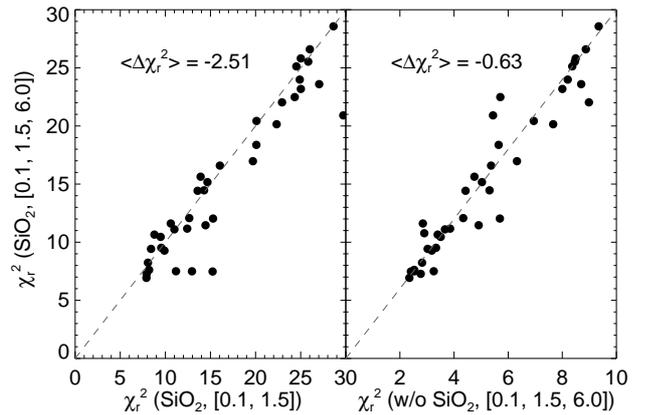}}
    \caption{\label{fig:chi}Change in reduced $\chi^2_{\mathrm{r}}$
      values for fits with and without silica ({\it right panel}) and
      fits with two grain sizes (0.1 and 1.5\,$\mu$m, {\it left
        panel}). The dashed line corresponds to $y = x$. The $<\Delta
      \chi^2_{\mathrm{r}}>$ values correspond to the mean difference of
      reduced $\chi^2_{\mathrm{r}}$ between x-axis and y-axis simulations.}
  \end{center}
\end{figure}

\begin{figure}
\begin{center}
\resizebox{\hsize}{!}{\includegraphics{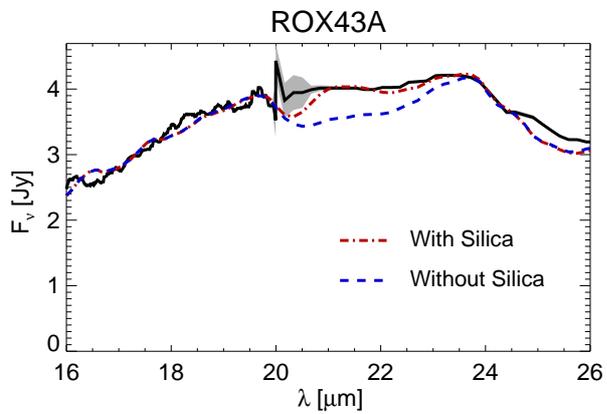}}
\caption{\label{sio}Blowup on the fit to the ROX43A spectrum,
  performed with silica (red dot-dashed line) and without (blue dashed line).}
\end{center}
\end{figure}

Usually, Mg-rich silicates are considered for the dust mineralogy in
proto-planetary disks (e.g. \citealp{Henning2009}). But both
\citet{Olofsson2009} and \citet{Sargent2009a} attribute some features
in IRS spectra of young stars to silica (composition SiO$_2$). To
gauge the importance of silica in our B2C compositional approach, we
run the B2C model with and without silica. Many fits were improved
adding silica in the dust population (as shown for one example in
Fig.\,\ref{sio}, the 20--22\,$\mu$m range being the spectral range
where the improvement is the more noticeable). As can be seen on the
right panel of Fig.\,\ref{fig:chi}, showing the reduced
$\chi^2_\mathrm{r}$ for simulations with and without silica, many fits
are improved when silica is included, demonstrating the non-negligible
importance of silica for our B2C model.

We have also critically examined the need for large, amorphous,
6.0\,$\mu$m-sized grains in the B2C model as they are almost
featureless contrary to the 0.1 and 1.5\,$\mu$m-sized grains. We
therefore run the B2C procedure on the 58 same objects with only 0.1
and 1.5\,$\mu$m-sized grains. Left panel of Fig.\,\ref{fig:chi} shows
that the mean value of the reduced $\chi^2_{\mathrm{r}}$ for all the
simulations is augmented by 2.50 when using only two grain sizes,
showing that the majority of the fits are improved using three grain
sizes instead of two.

One can also worry about the influence of the offset $O_{\nu,2}$
values on the inferred grain sizes. As a large offset value leads to
larger flux below the continuum--substracted spectrum, it may indeed
favor larger grains. We therefore examined if there was any
correlation between the fraction of large 6.0\,$\mu$m-sized grains and
the $O_{\nu,2}$ values, and we quantified these relations with
correlation coefficients. Considering the warm large grains and the
$O_{\nu,2}$ offsets, we find a $\tau$ value of 0.04 with a
significance probability of $P = 0.69$. For the cold large grain
fraction and the $O_{\nu,2}$ values, we obtain $\tau = -0.03$ and $P =
0.70$. This means that there is no significant influence of the
offsets $O_{\nu,2}$ on the inferred grain sizes.

\onecolumn
\begin{landscape}
\begin{longtable}{@{\excs}lcccc|cccc|cccc}
\caption{Dust composition derived using the "B2C" procedure for 58 objects from our sample. For each star, first line corresponds to the warm component abundances and second line corresponds to the cold component abundances.\label{tbl:silicate}}\\
\hline \hline
 & \multicolumn{4}{c|}{0.1\,$\mu$m} & \multicolumn{4}{c|}{1.5\,$\mu$m} & \multicolumn{4}{c|}{6.0\,$\mu$m} \\
\hline
Starname &  Am & Ens & For & Sil & Am & Ens & For & Sil & Am & Ens & For & Sil \\
   & \% & \% & \% & \% & \% & \% & \% & \% & \% & \% & \% & \%  \\
\hline
\endfirsthead
\caption{continued.}\\
\hline\hline
 & \multicolumn{4}{c|}{0.1\,$\mu$m} & \multicolumn{4}{c|}{1.5\,$\mu$m} & \multicolumn{4}{c|}{6.0\,$\mu$m} \\
\hline
Starname &  Am & Ens & For & Sil & Am & Ens & For & Sil & Am & Ens & For & Sil \\
   & \% & \% & \% & \% & \% & \% & \% & \% & \% & \% & \% & \%  \\
\hline
\endhead
\hline
\endfoot
AS\,205 &    7.0$^{+   1.4}_{-   0.9}$ &    1.5$^{+   0.8}_{-   0.7}$ &    0.8$^{+   0.7}_{-   0.0}$ &    0.0$^{+   0.3}_{-   0.0}$ &   47.6$^{+   3.0}_{-   3.6}$ &    5.1$^{+   2.0}_{-   1.2}$ &    3.3$^{+   2.0}_{-   1.8}$ &    0.0$^{+   0.6}_{-   0.0}$ &   16.3$^{+   2.2}_{-   2.6}$ &    - &    - &   18.5$^{+   1.9}_{-   2.9}$ \\
... &   27.3$^{+   2.1}_{-   3.2}$ &    1.5$^{+   1.0}_{-   0.9}$ &    0.0$^{+   0.7}_{-   0.0}$ &    0.1$^{+   0.7}_{-   0.0}$ &   50.2$^{+   4.5}_{-   5.0}$ &    0.0$^{+   0.9}_{-   0.0}$ &    0.0$^{+   0.9}_{-   0.0}$ &    0.0$^{+   1.3}_{-   0.0}$ &    9.3$^{+   2.5}_{-   1.6}$ &    - &    - &   11.6$^{+   5.2}_{-   3.6}$ \\
B35 &   32.4$^{+   7.5}_{-   7.1}$ &    0.7$^{+   1.3}_{-   0.0}$ &    0.3$^{+   1.8}_{-   0.0}$ &    0.0$^{+   2.4}_{-   0.0}$ &   54.1$^{+   4.5}_{-   4.5}$ &    0.0$^{+   2.5}_{-   0.0}$ &    0.5$^{+   1.3}_{-   0.0}$ &    0.0$^{+   1.7}_{-   0.0}$ &    8.8$^{+   1.7}_{-   0.0}$ &    - &    - &    3.3$^{+   1.7}_{-   0.0}$ \\
... &   23.5$^{+   2.7}_{-   2.6}$ &    0.0$^{+   0.7}_{-   0.0}$ &    0.0$^{+   0.9}_{-   0.0}$ &    0.0$^{+   1.3}_{-   0.0}$ &   50.0$^{+   6.6}_{-  11.1}$ &    3.7$^{+   1.8}_{-   2.1}$ &    0.0$^{+   1.2}_{-   0.0}$ &    0.0$^{+   0.9}_{-   0.0}$ &    5.3$^{+   1.3}_{-   1.1}$ &    - &    - &   17.5$^{+   4.2}_{-   4.5}$ \\
BD+31\,634 &   16.6$^{+   2.2}_{-   1.2}$ &    2.0$^{+   1.5}_{-   0.0}$ &    0.0$^{+   0.4}_{-   0.0}$ &    0.0$^{+   0.6}_{-   0.0}$ &   56.9$^{+   4.6}_{-   6.2}$ &    6.9$^{+   3.6}_{-   3.2}$ &    0.0$^{+   1.3}_{-   0.0}$ &    2.4$^{+   1.5}_{-   1.4}$ &    7.0$^{+   1.5}_{-   0.0}$ &    - &    - &    8.2$^{+   1.4}_{-   2.0}$ \\
... &   56.1$^{+   4.3}_{-   4.4}$ &    0.0$^{+   0.8}_{-   0.0}$ &    0.0$^{+   0.7}_{-   0.0}$ &    0.0$^{+   1.3}_{-   0.0}$ &   31.1$^{+   3.0}_{-   2.4}$ &    0.0$^{+   0.6}_{-   0.0}$ &    0.0$^{+   0.6}_{-   0.0}$ &    0.0$^{+   0.6}_{-   0.0}$ &    7.2$^{+   1.2}_{-   0.5}$ &    - &    - &    5.6$^{+   1.4}_{-   0.8}$ \\
\end{longtable}
\end{landscape}

\onecolumn
\begin{longtable}{@{\excs}lcc|cc|cc}
\caption{Dust temperatures, continuum parameters (blackbody temperature and offset $O_{\nu , 2}$) derived using the "B2C" procedure for 60 objects from our sample.\label{tbl:silicate2}}\\
\hline \hline
 & \multicolumn{2}{c|}{Dust temperatures} & \multicolumn{2}{c|}{Continuum} & \multicolumn{2}{c|}{Goodness} \\
\hline
Starname & $T_{\mathrm{warm}}$ & $T_{\mathrm{cold}}$ & $T_{\mathrm{cont}}$ & $O_{\nu ,2}$ & $\chi^2_{\mathrm{r}}$ & Dof \\
  & (K) & (K) & (K) & ($\times$ 100) & - & - \\
\hline
\endfirsthead
\caption{continued.}\\
\hline\hline
 & \multicolumn{2}{c|}{Dust temperatures} & \multicolumn{2}{c|}{Continuum} & \multicolumn{2}{c|}{Goodness} \\
\hline
Starname & $T_{\mathrm{warm}}$ & $T_{\mathrm{cold}}$ & $T_{\mathrm{cont}}$ & $O_{\nu ,2}$ & $\chi^2_{\mathrm{r}}$ & Dof \\
  & (K) & (K) & (K) & ($\times$ 100) & - & - \\
\hline
\endhead
\hline
\endfoot
\object{AS\,205} &  228.9$^{+   2.7}_{-   2.1}$ &   58.2$^{+   1.3}_{-   0.9}$ &  121.6$^{+   1.2}_{-   1.8}$ &   20.1$^{+   0.3}_{-   0.5}$ &   22.6 &         1119 \\
\object{B35} &  207.3$^{+   5.0}_{-   3.7}$ &   79.5$^{+   1.7}_{-   0.0}$ &   97.4$^{+   3.0}_{-   1.1}$ &   16.0$^{+   0.7}_{-   0.8}$ &   16.9 &          232 \\
\object{BD+31\,634} &  596.2$^{+  27.1}_{-  16.2}$ &   98.3$^{+   2.3}_{-   2.6}$ &  107.6$^{+   0.9}_{-   0.7}$ &    6.9$^{+   0.3}_{-   0.2}$ &   61.0 &          231 \\
\object{C7-11} &  188.2$^{+   2.9}_{-   1.8}$ &   75.2$^{+   4.9}_{-   1.1}$ &   95.1$^{+   0.9}_{-   1.2}$ &    6.1$^{+   0.3}_{-   0.1}$ &   15.9 &          238 \\
\object{CK4} &  238.0$^{+   9.3}_{-   8.5}$ &   80.9$^{+   1.3}_{-   0.8}$ &   79.3$^{+   1.3}_{-   0.9}$ &   12.9$^{+   0.2}_{-   0.1}$ &   25.6 &         1151 \\
\object{EC82} &  491.6$^{+  25.7}_{-  19.9}$ &   88.8$^{+   1.8}_{-   1.0}$ &   72.0$^{+   3.4}_{-   0.0}$ &   27.3$^{+   0.5}_{-   0.4}$ &  157.7 &         1159 \\
\object{GQ\,Lup} &  192.8$^{+   2.5}_{-   1.6}$ &   70.7$^{+   2.2}_{-   3.7}$ &   85.2$^{+   0.7}_{-   0.5}$ &   12.9$^{+   0.2}_{-   0.1}$ &   23.8 &          611 \\
\object{GW\,Lup} &  341.4$^{+  10.7}_{-  13.1}$ &  102.9$^{+   3.7}_{-   5.5}$ &  119.9$^{+   1.6}_{-   1.3}$ &    8.4$^{+   0.2}_{-   0.3}$ &   46.3 &          639 \\
\end{longtable}

\begin{figure}
\begin{center}
\caption{\label{all:fit}Fits to the 60 objects using the B2C procedure.}
\includegraphics[width=.2\columnwidth,origin=bl]{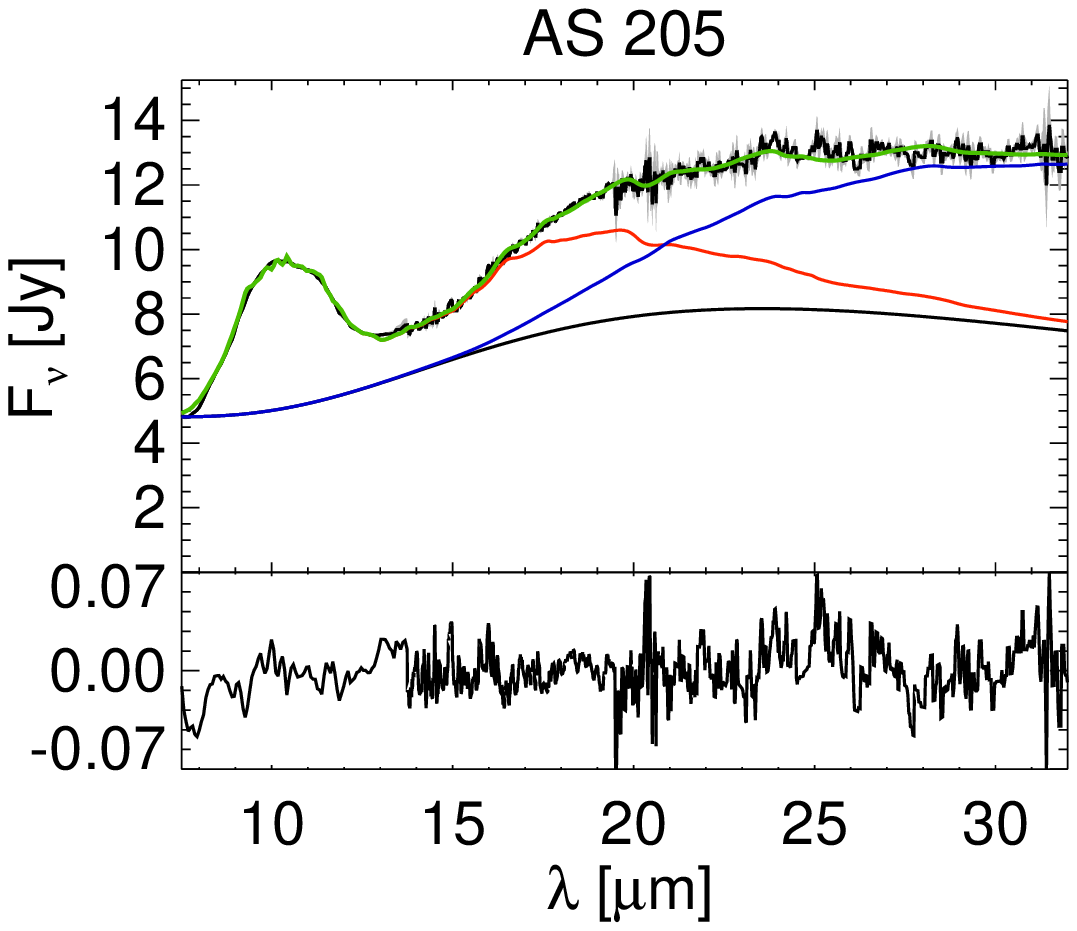}
\includegraphics[width=.2\columnwidth,origin=bl]{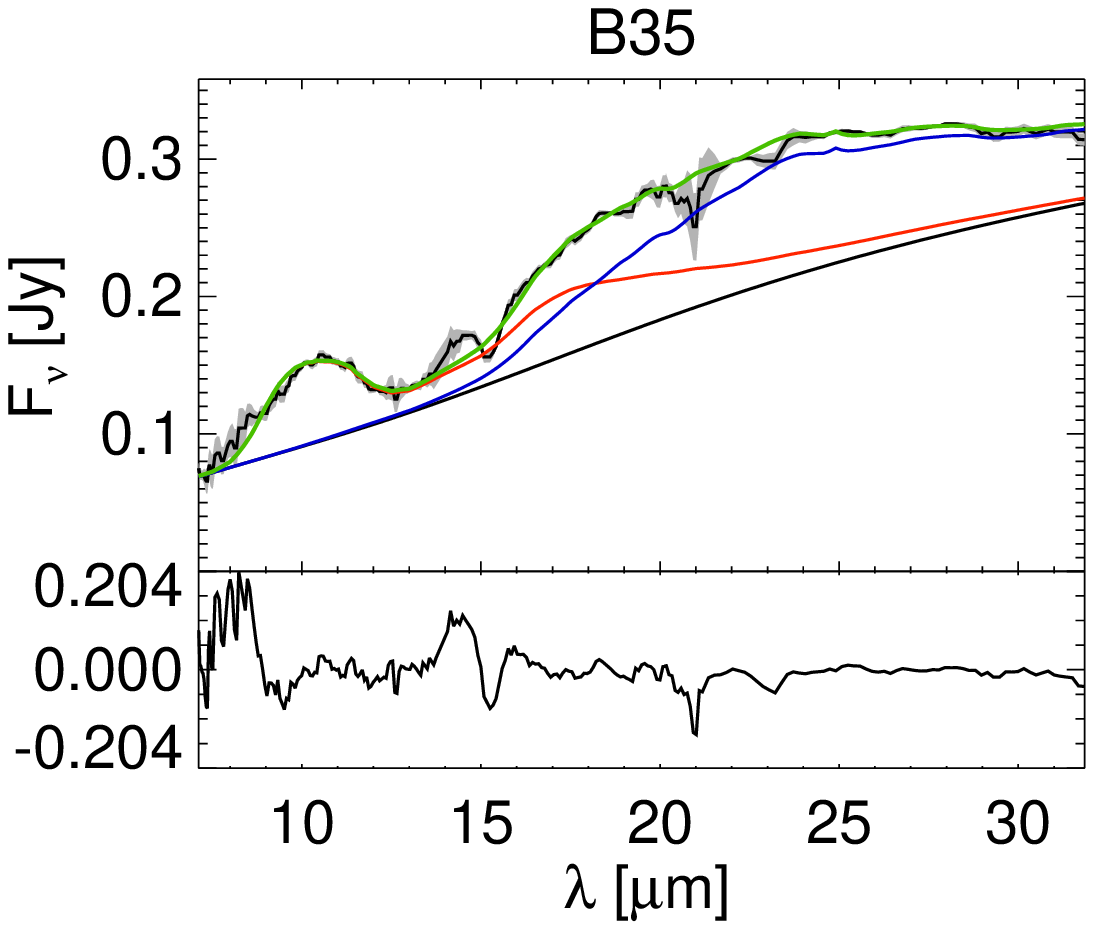}
\includegraphics[width=.2\columnwidth,origin=bl]{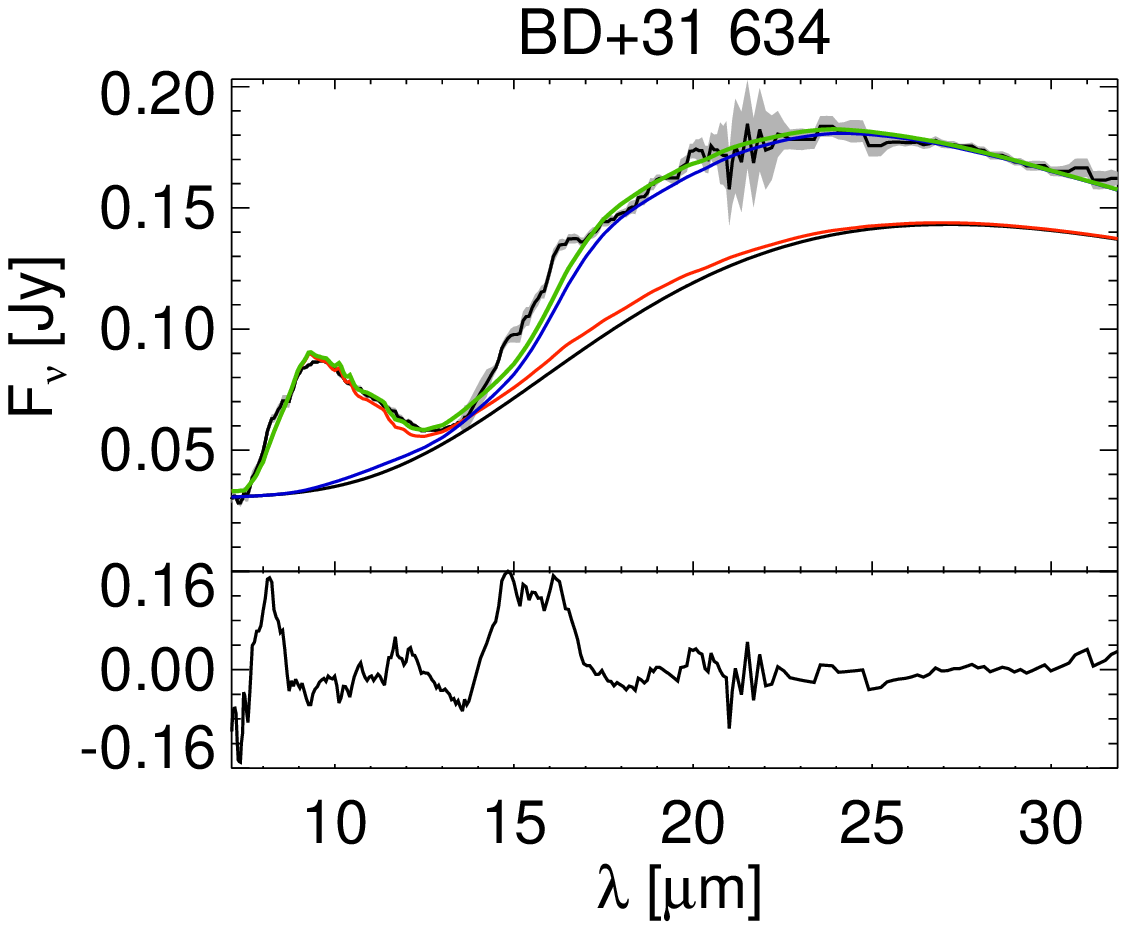}
\includegraphics[width=.2\columnwidth,origin=bl]{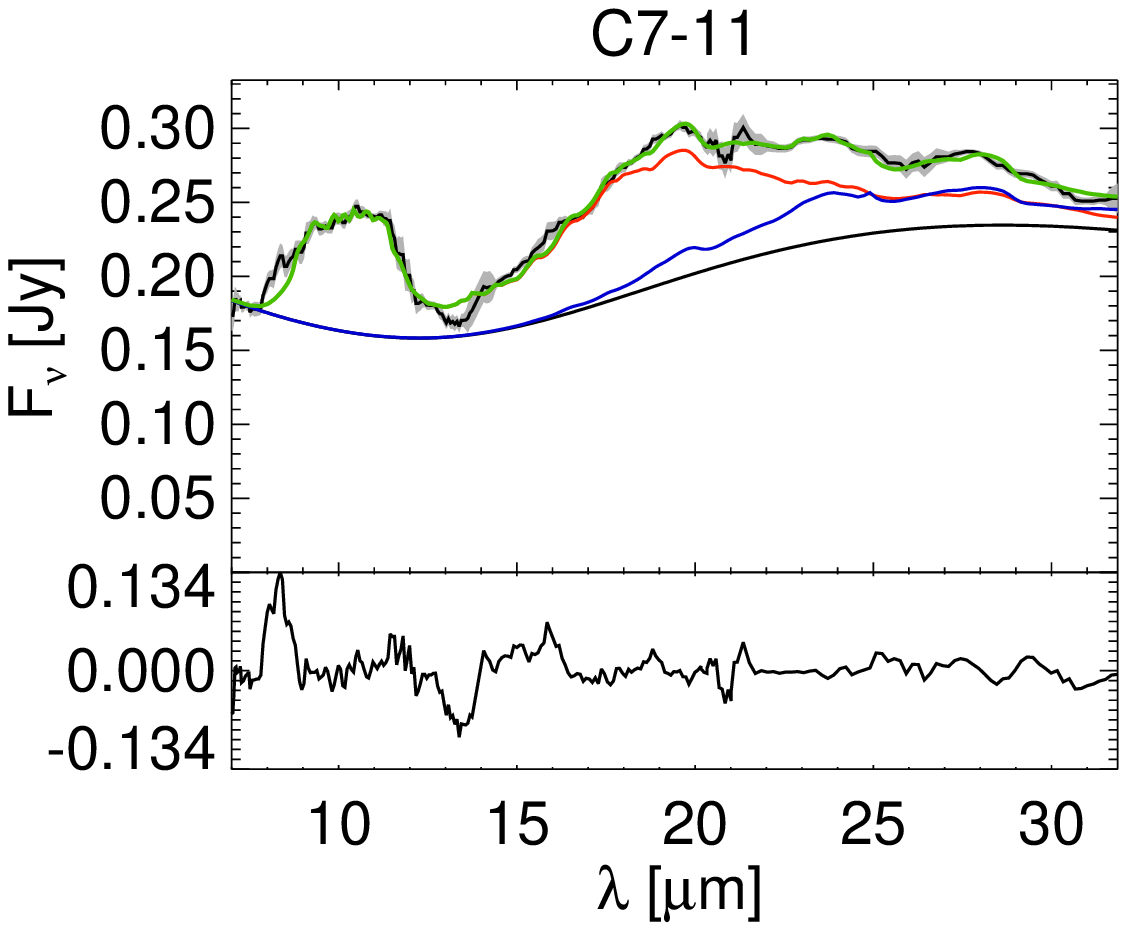}
\includegraphics[width=.2\columnwidth,origin=bl]{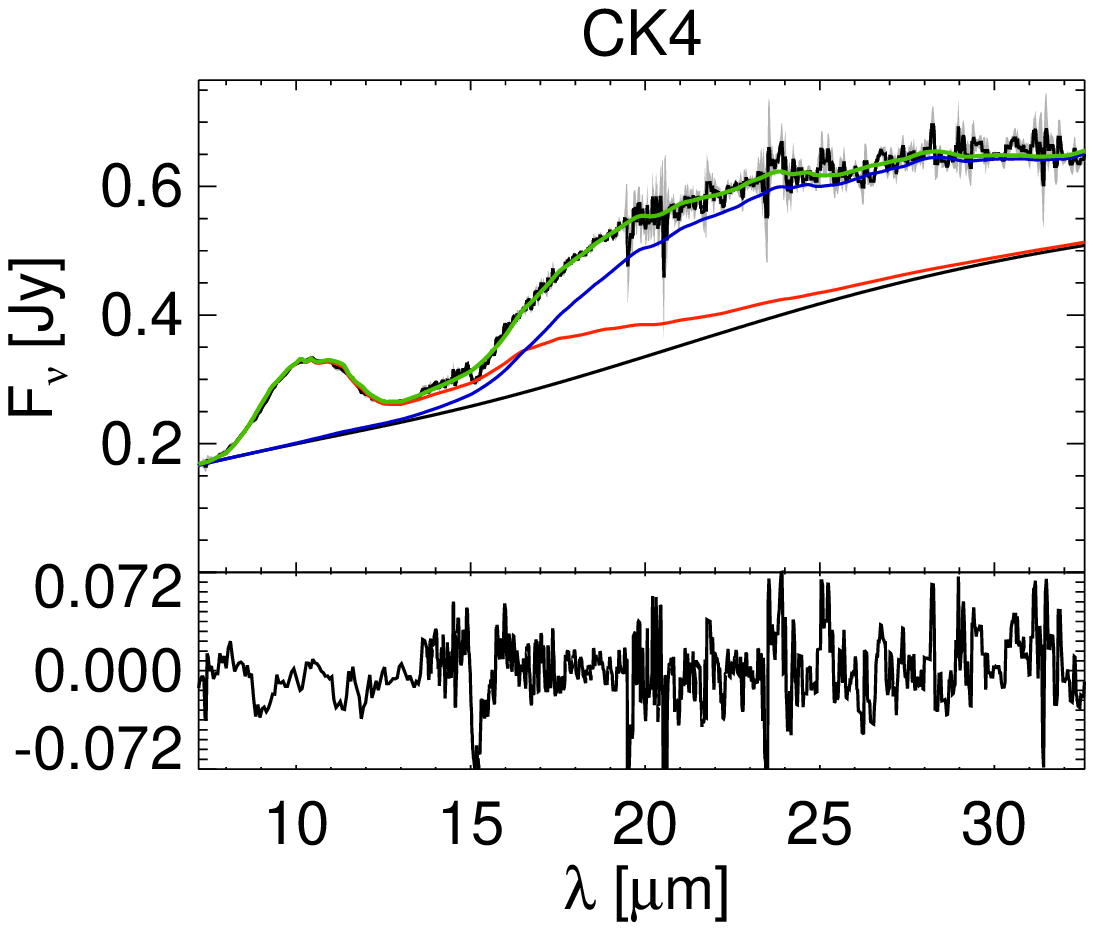}
\includegraphics[width=.2\columnwidth,origin=bl]{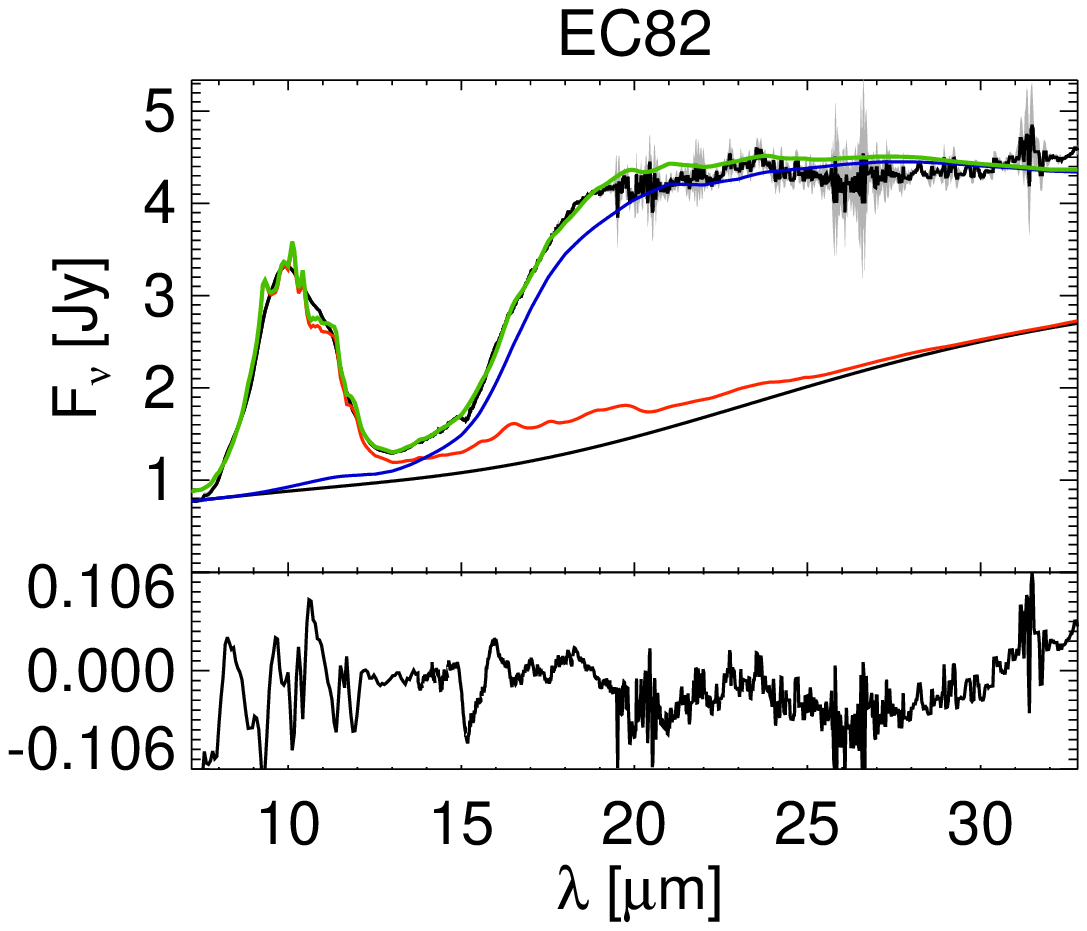}
\includegraphics[width=.2\columnwidth,origin=bl]{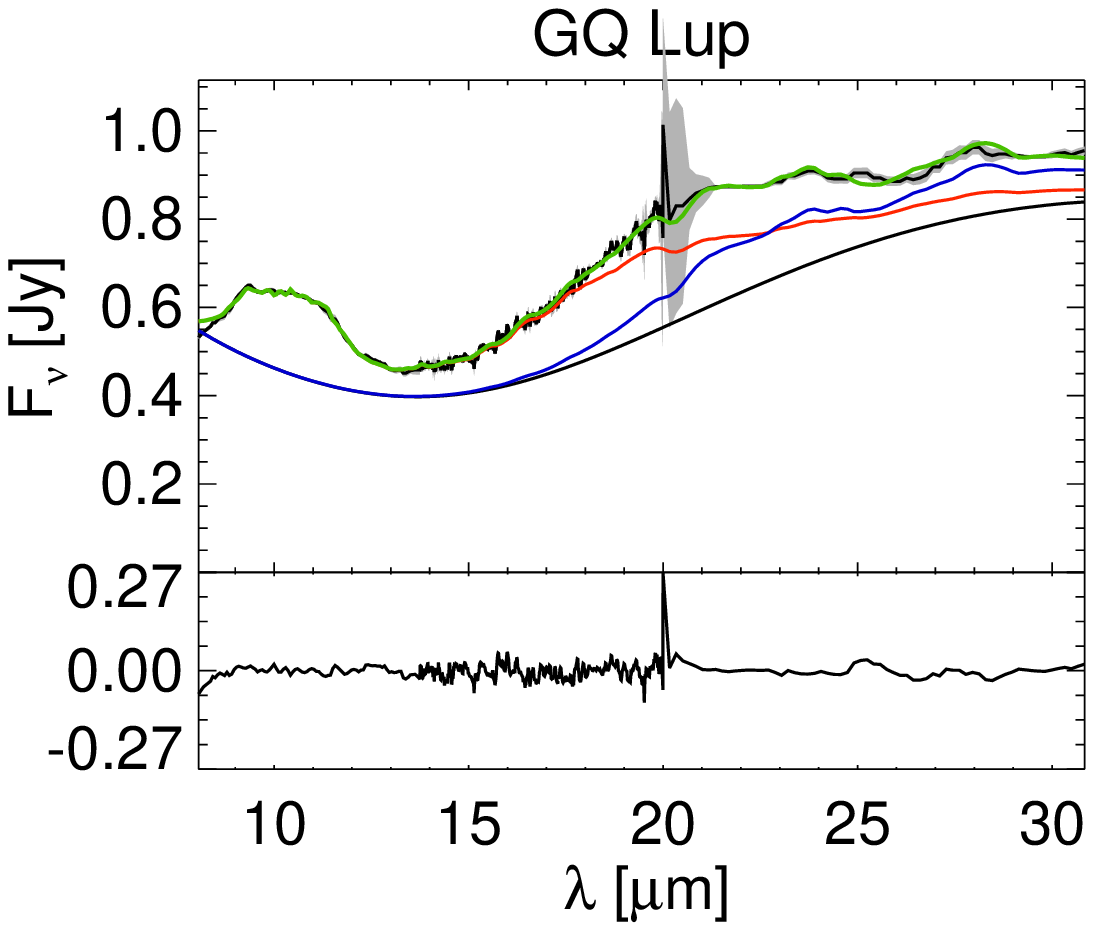}
\includegraphics[width=.2\columnwidth,origin=bl]{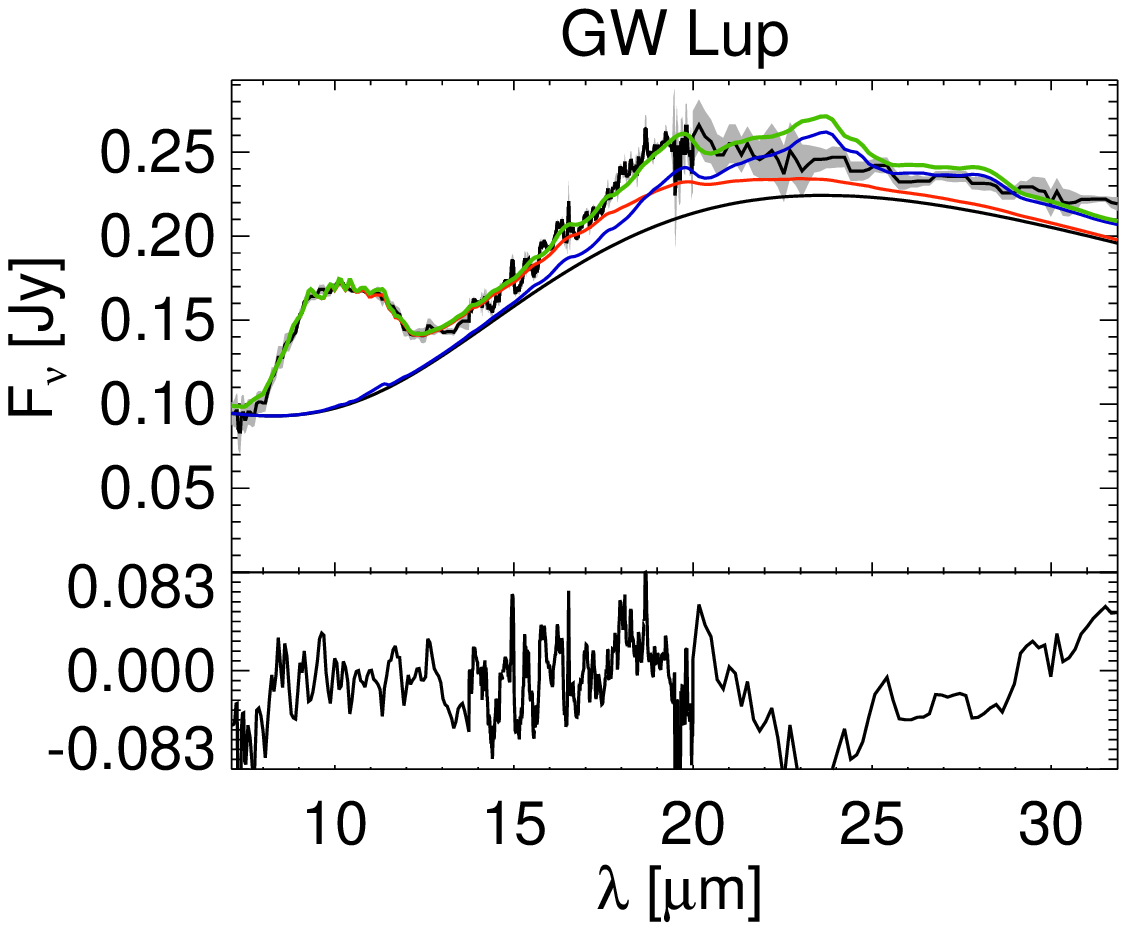}
\includegraphics[width=.2\columnwidth,origin=bl]{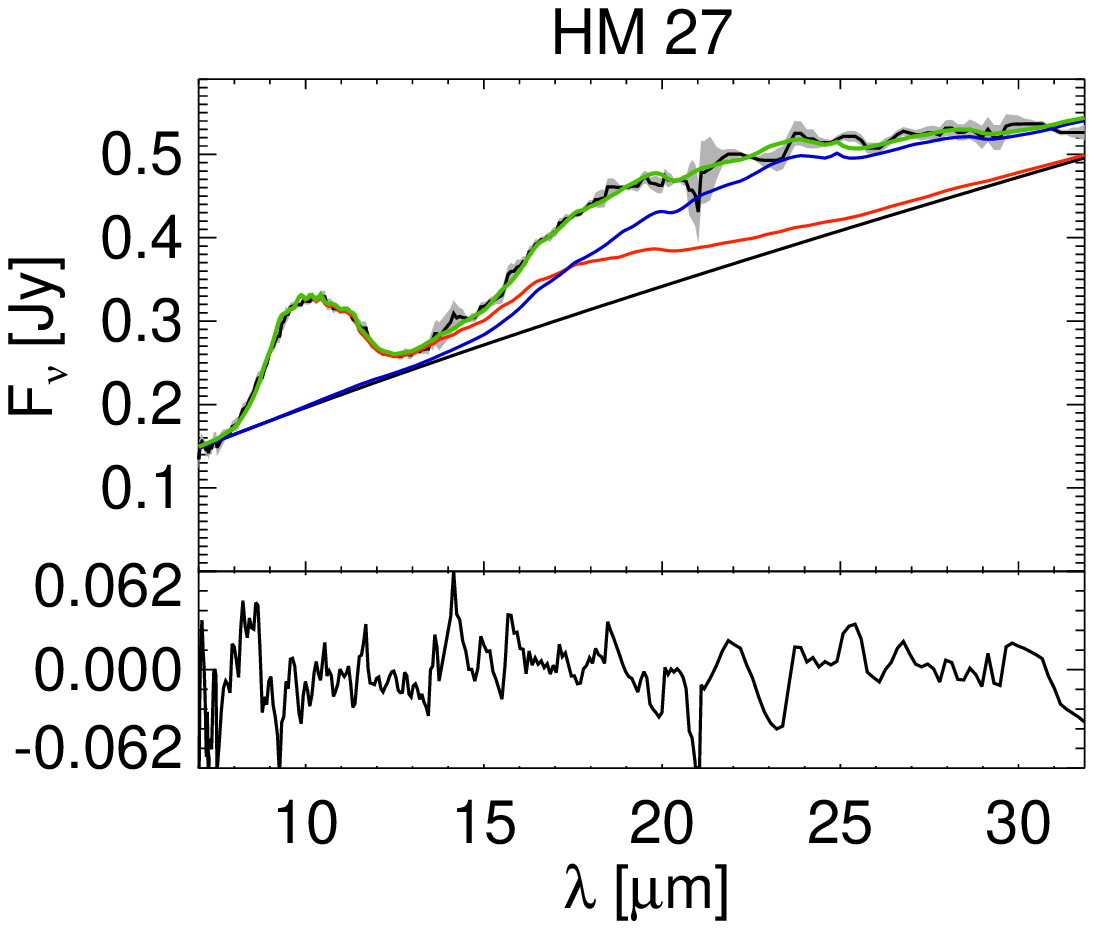}
\includegraphics[width=.2\columnwidth,origin=bl]{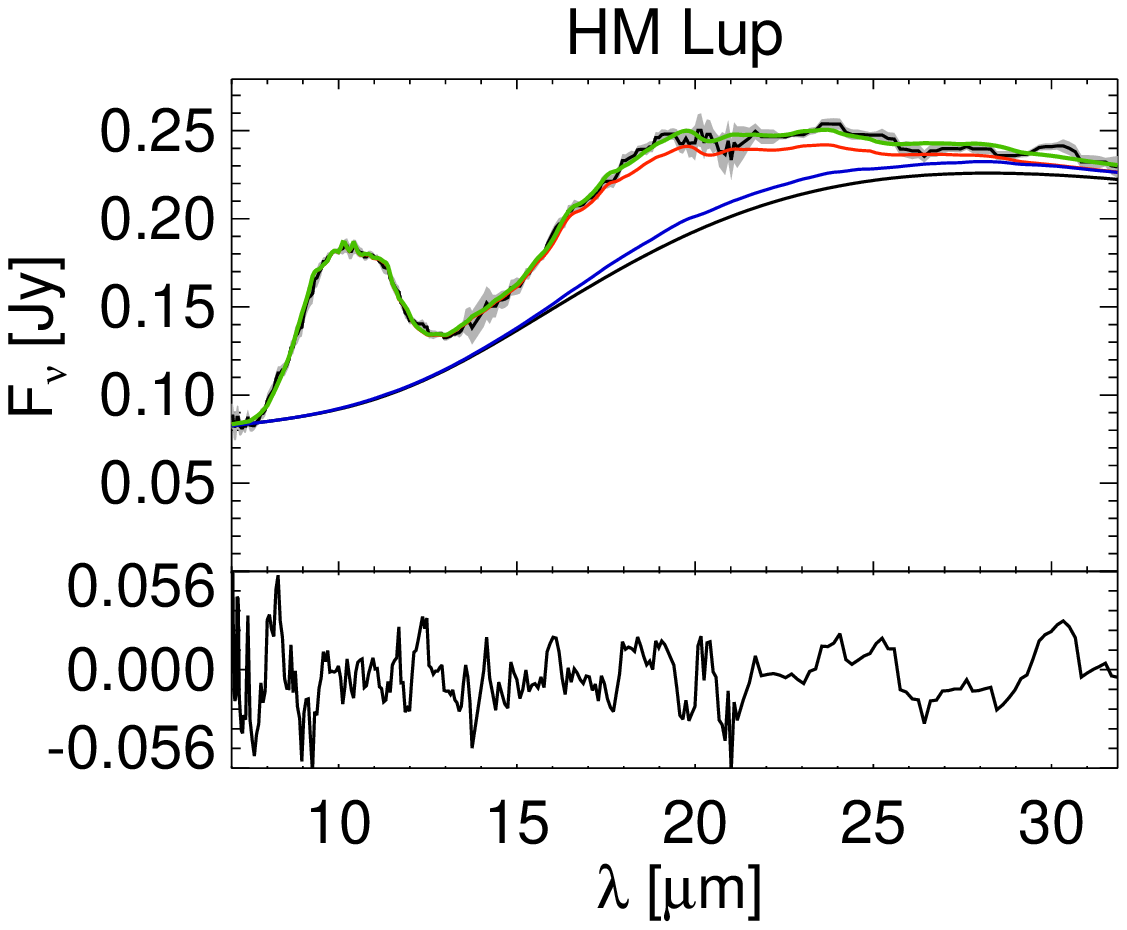}
\includegraphics[width=.2\columnwidth,origin=bl]{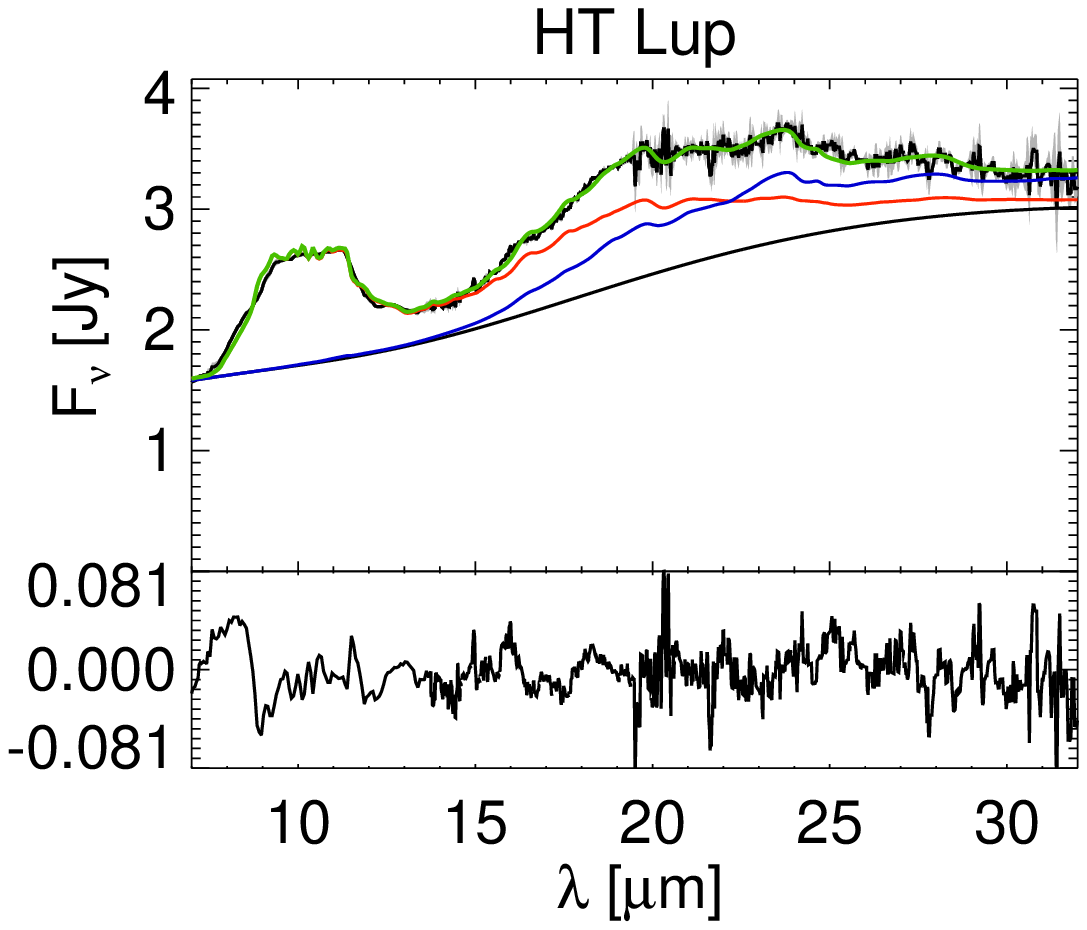}
\includegraphics[width=.2\columnwidth,origin=bl]{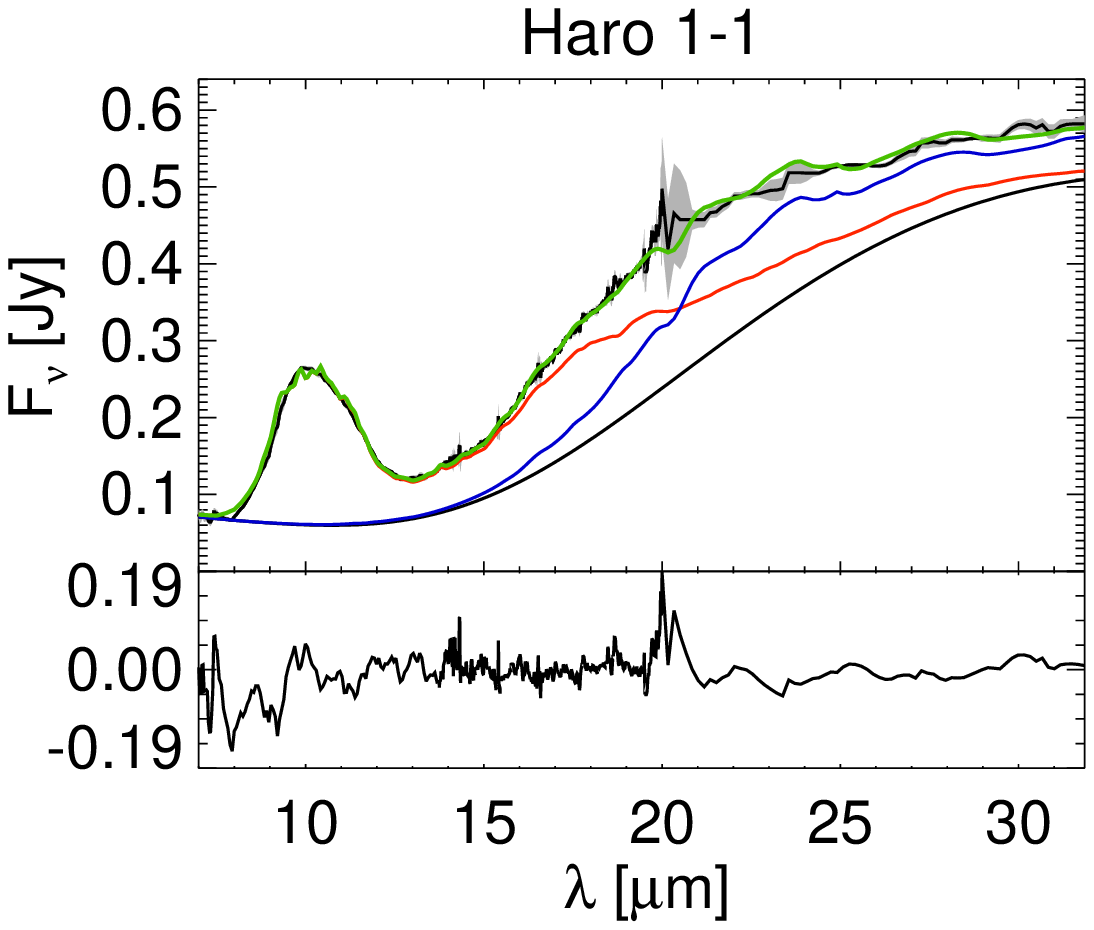}
\includegraphics[width=.2\columnwidth,origin=bl]{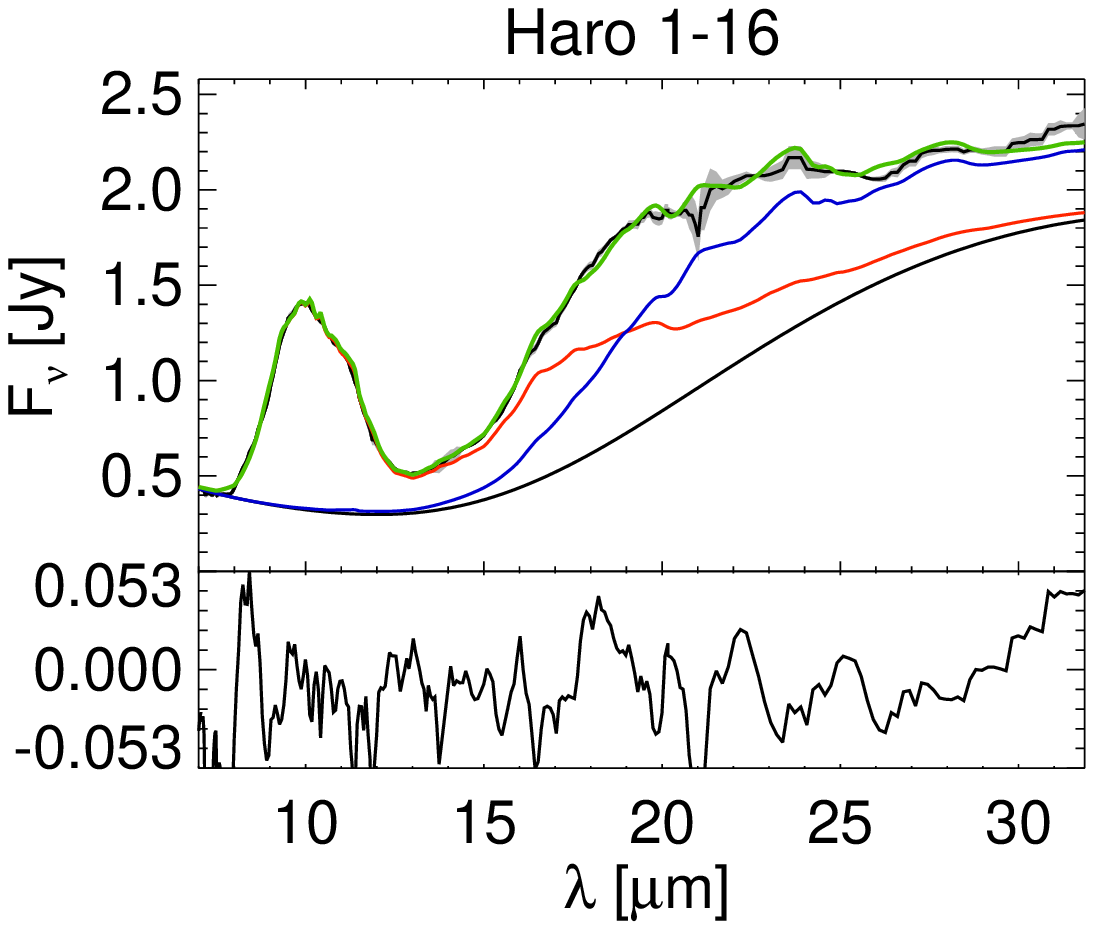}
\includegraphics[width=.2\columnwidth,origin=bl]{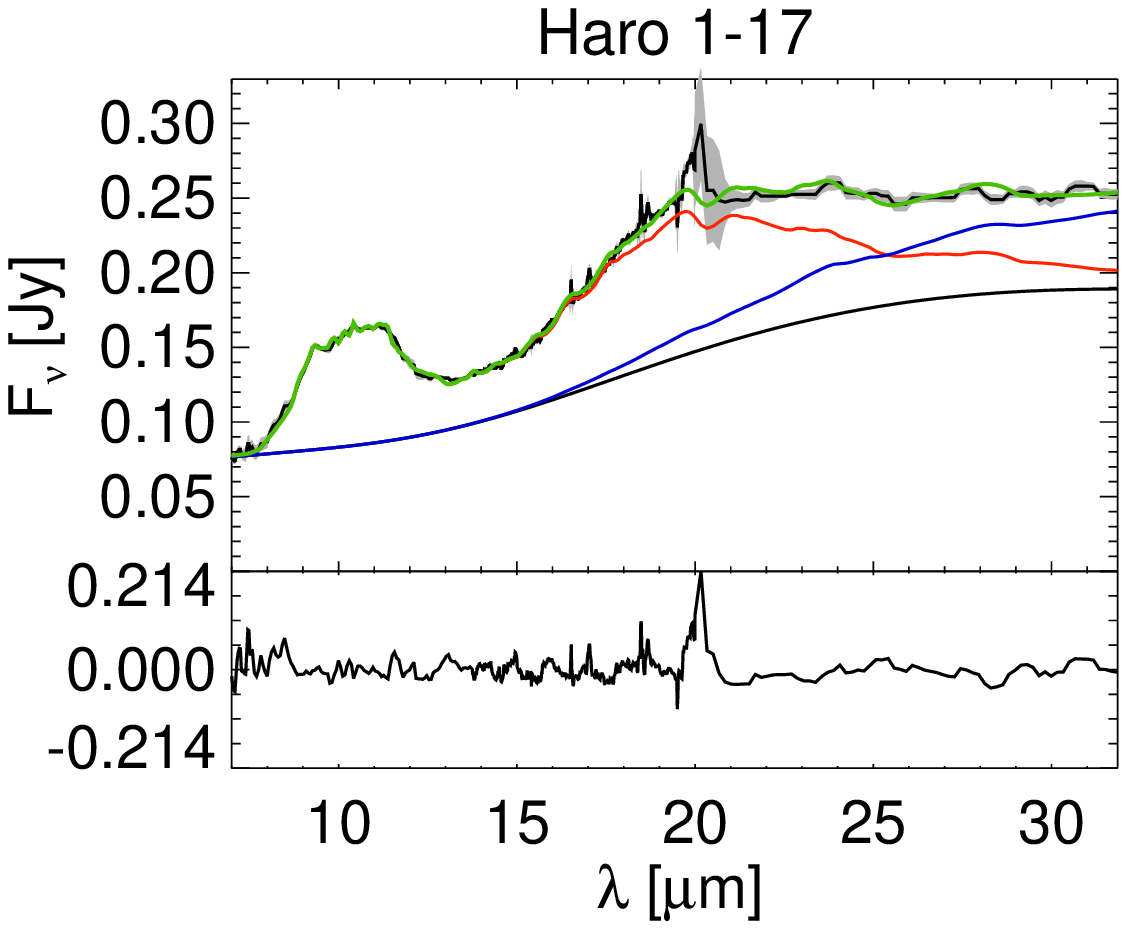}
\includegraphics[width=.2\columnwidth,origin=bl]{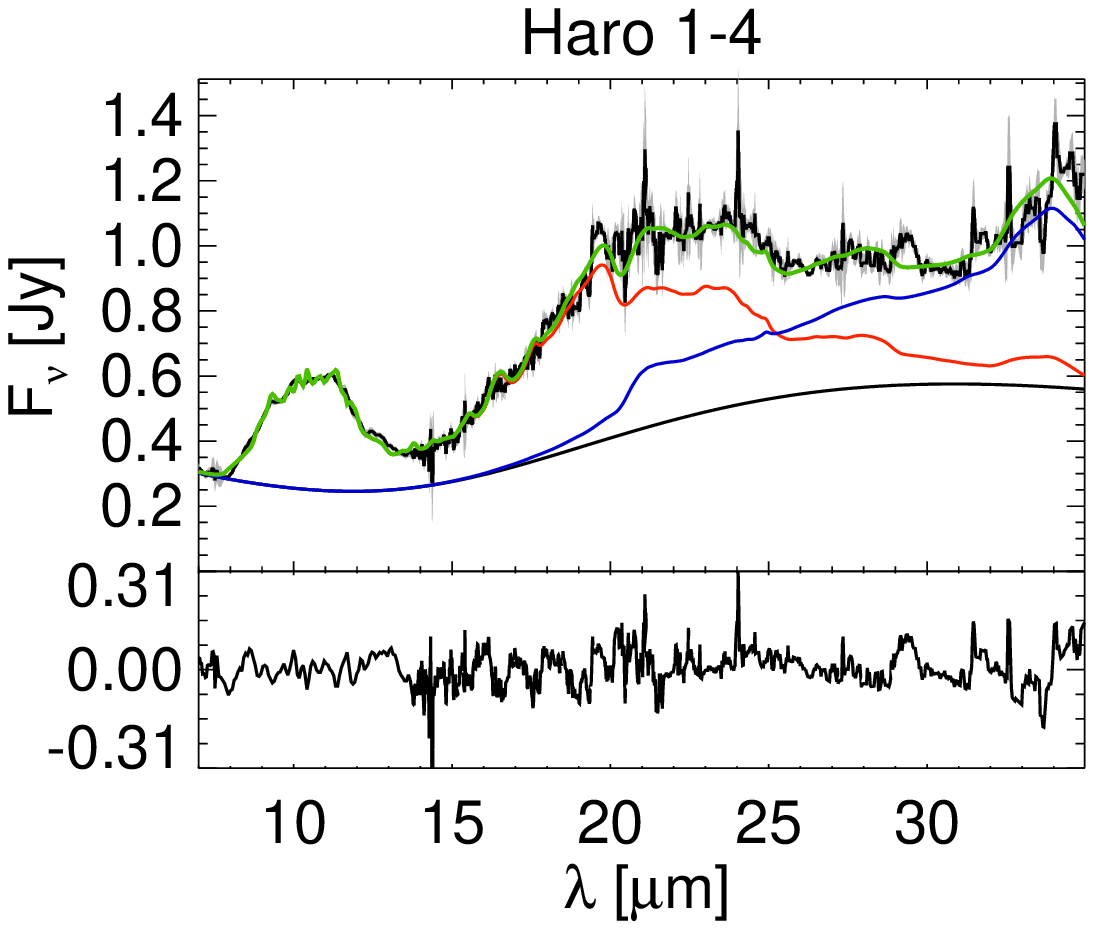}
\includegraphics[width=.2\columnwidth,origin=bl]{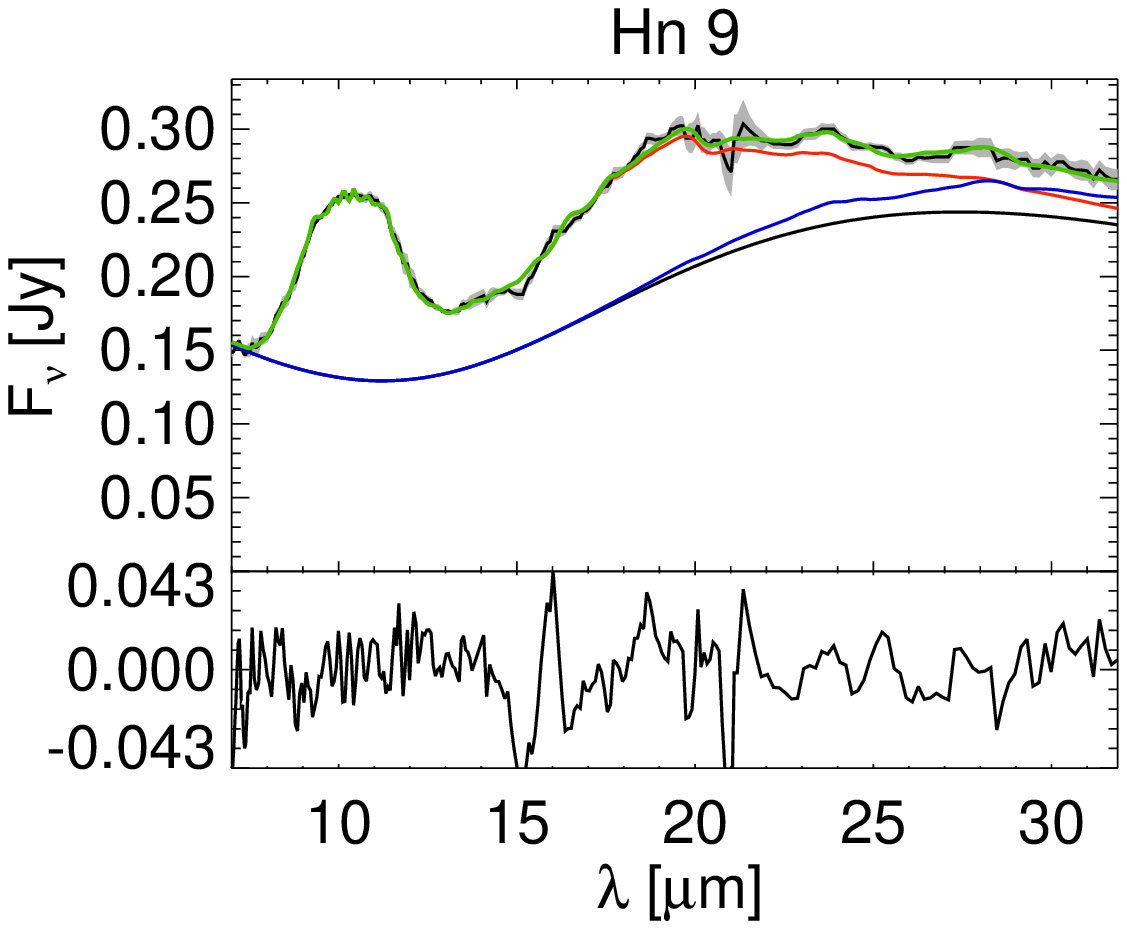}
\includegraphics[width=.2\columnwidth,origin=bl]{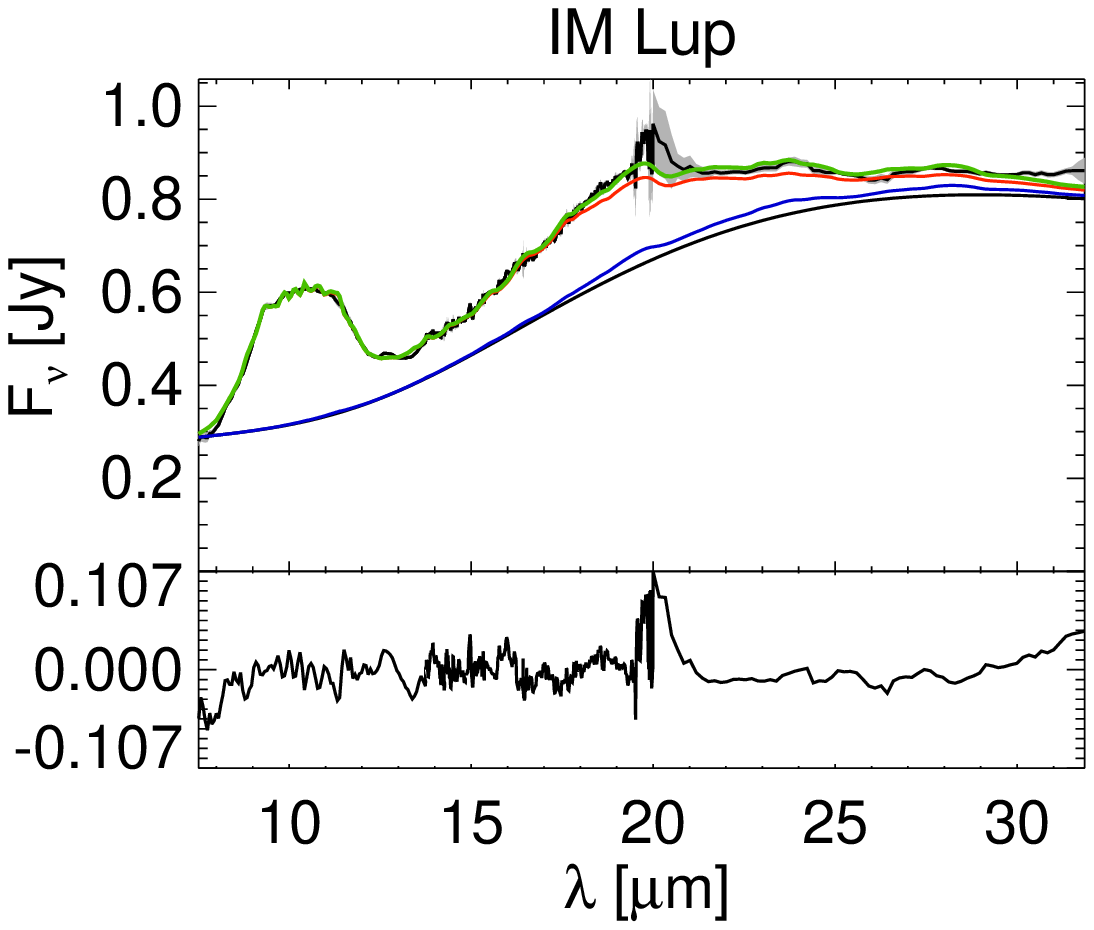}
\includegraphics[width=.2\columnwidth,origin=bl]{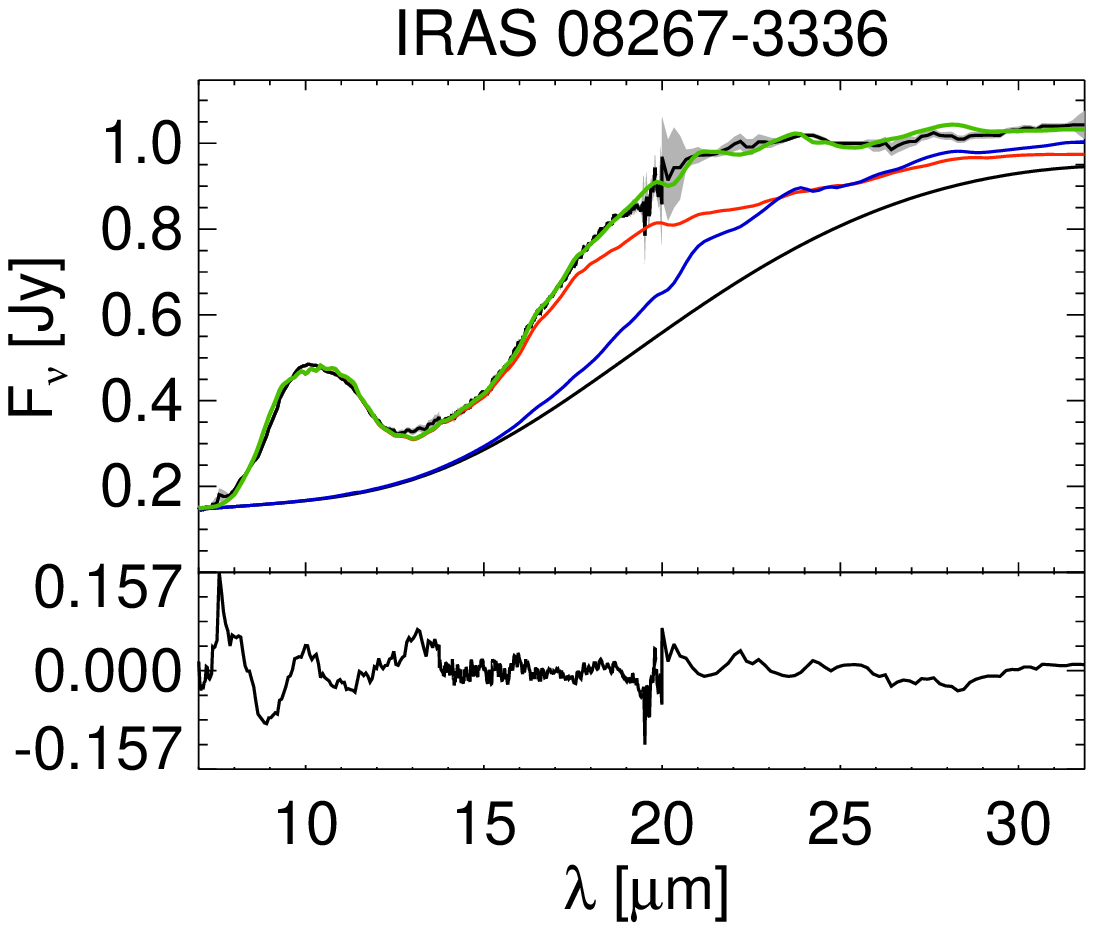}
\includegraphics[width=.2\columnwidth,origin=bl]{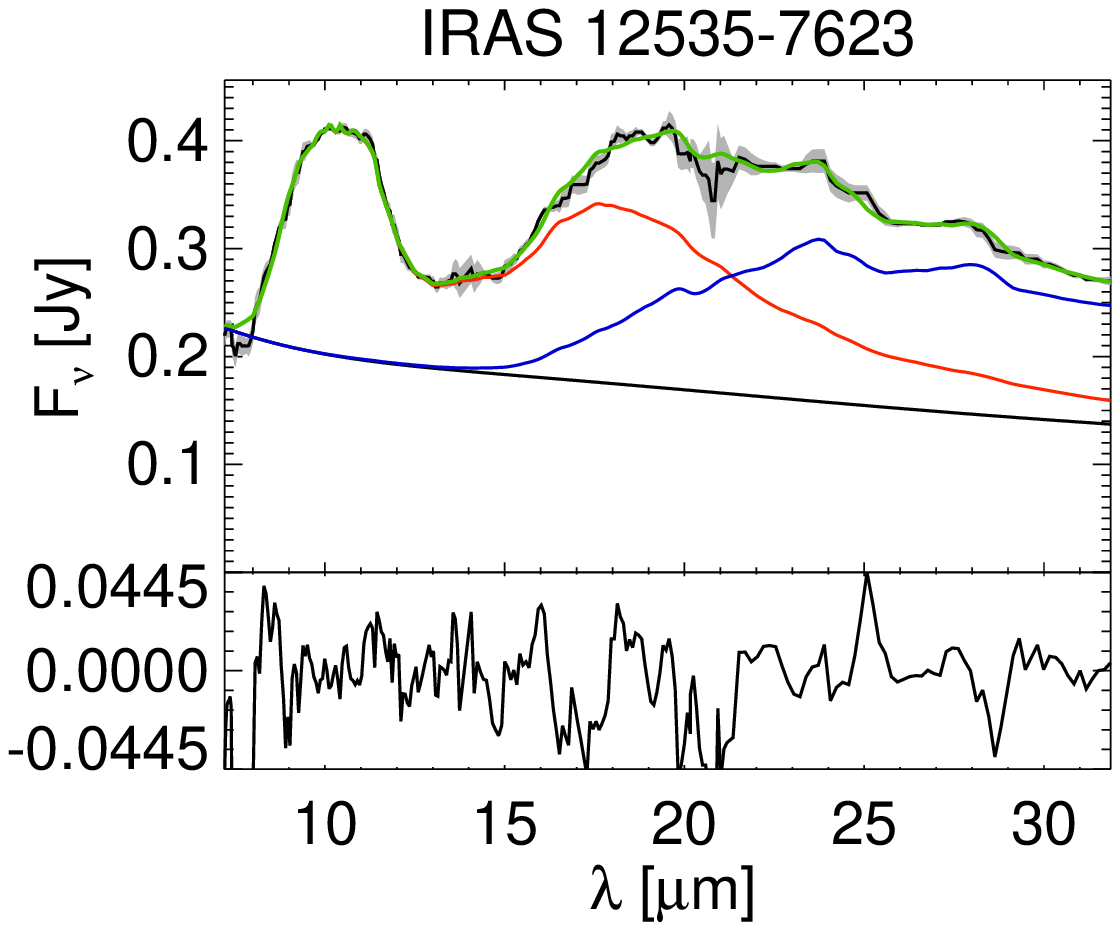}
\includegraphics[width=.2\columnwidth,origin=bl]{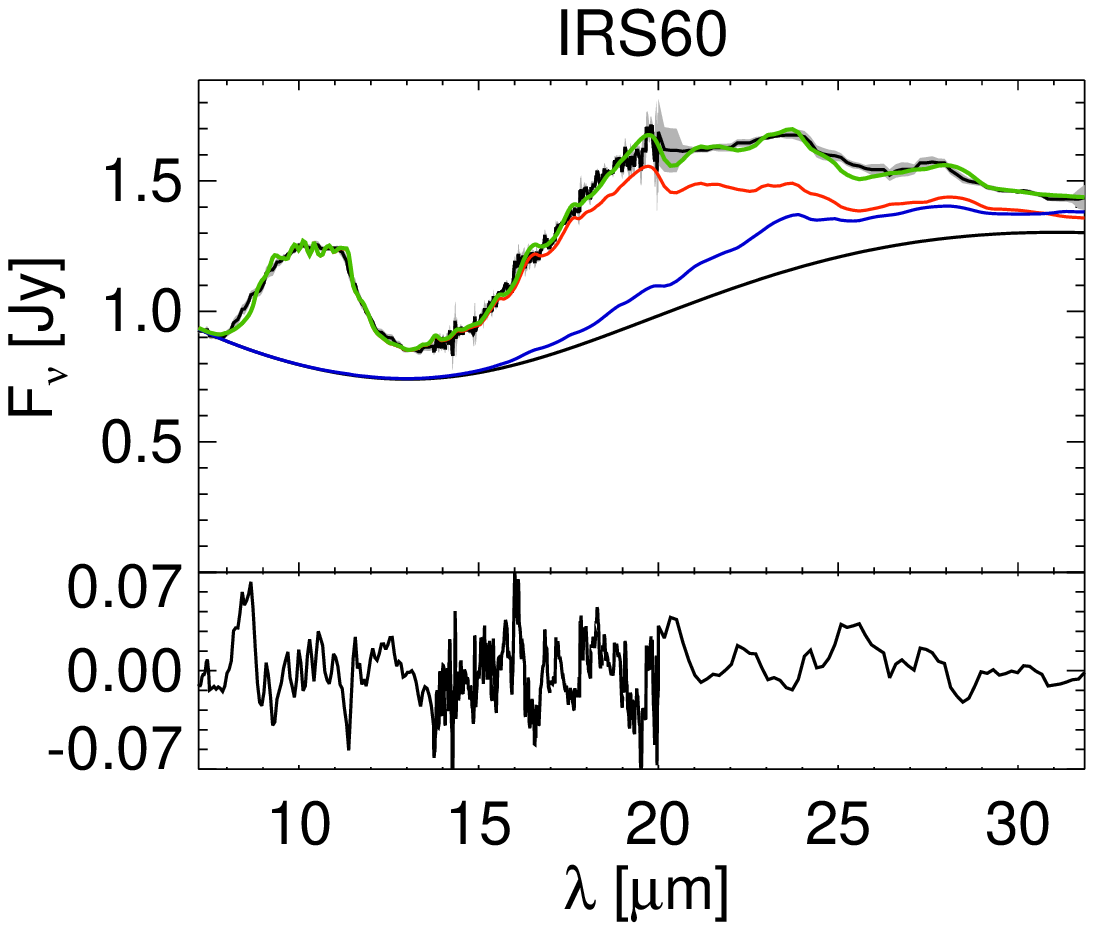}
\includegraphics[width=.2\columnwidth,origin=bl]{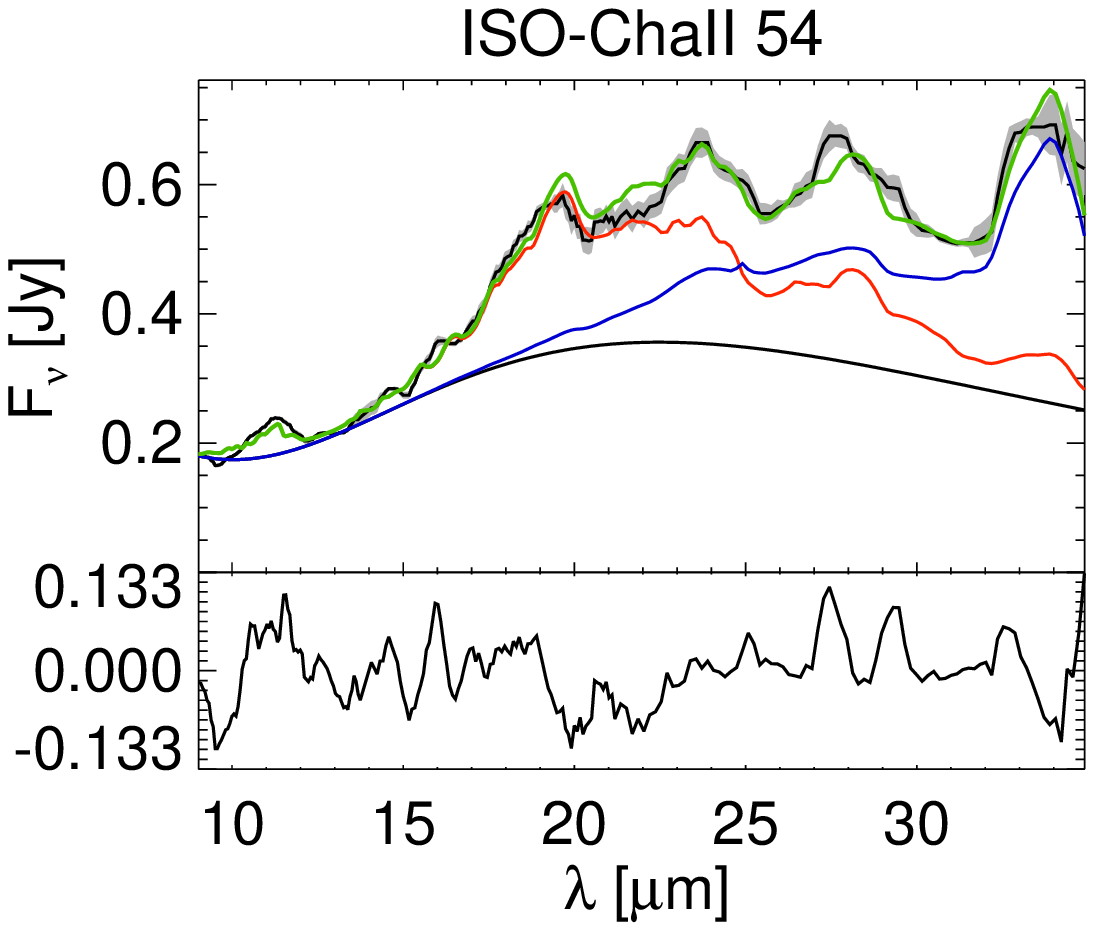}
\includegraphics[width=.2\columnwidth,origin=bl]{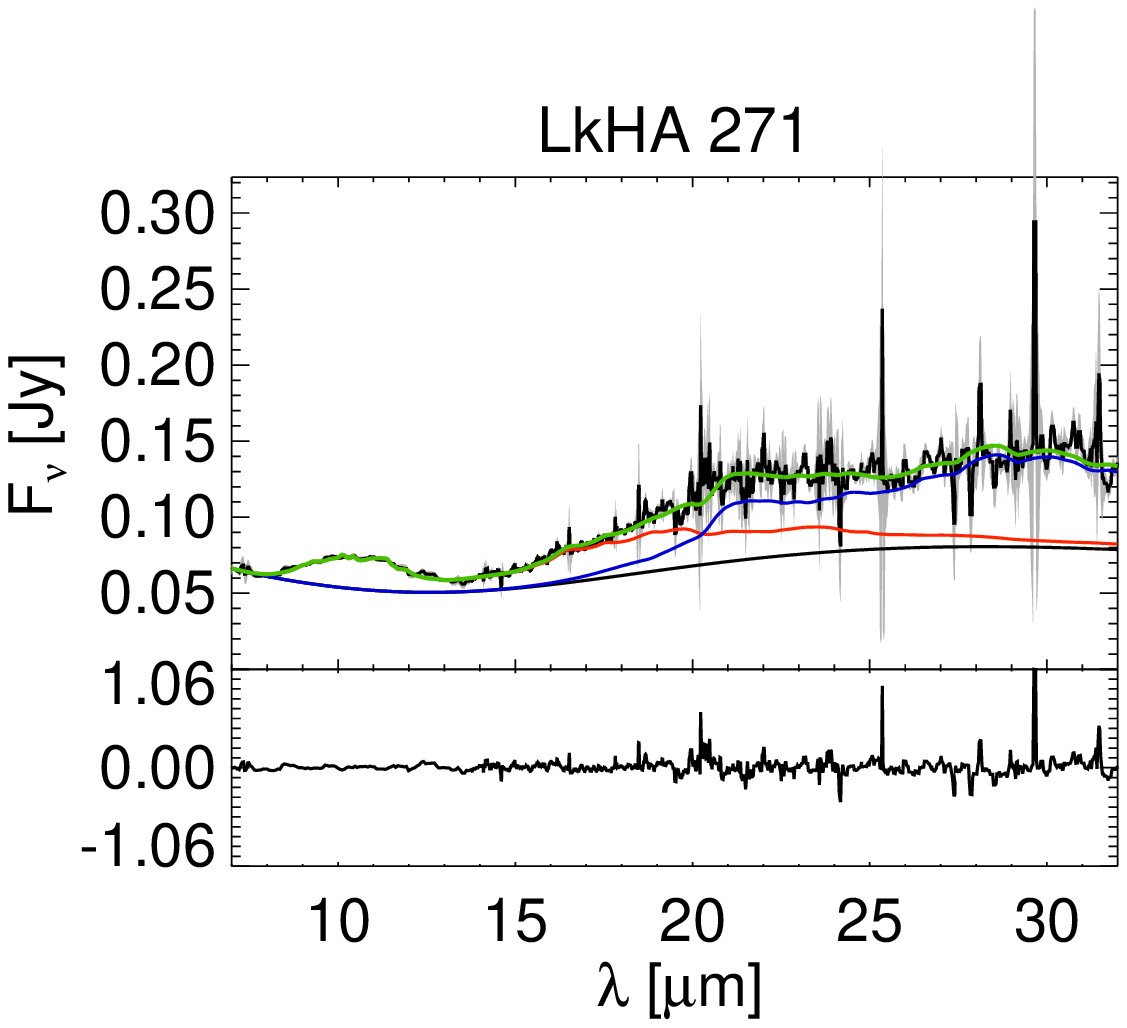}
\includegraphics[width=.2\columnwidth,origin=bl]{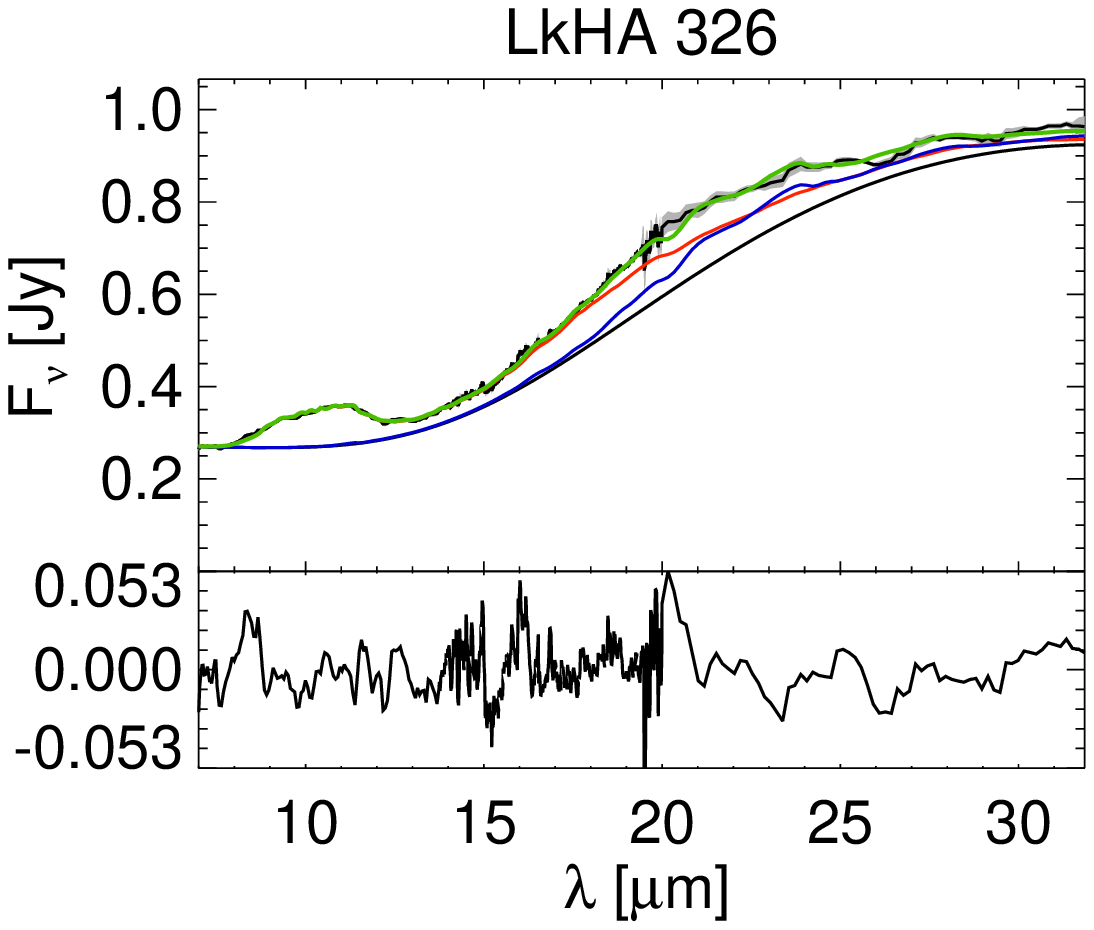}
\includegraphics[width=.2\columnwidth,origin=bl]{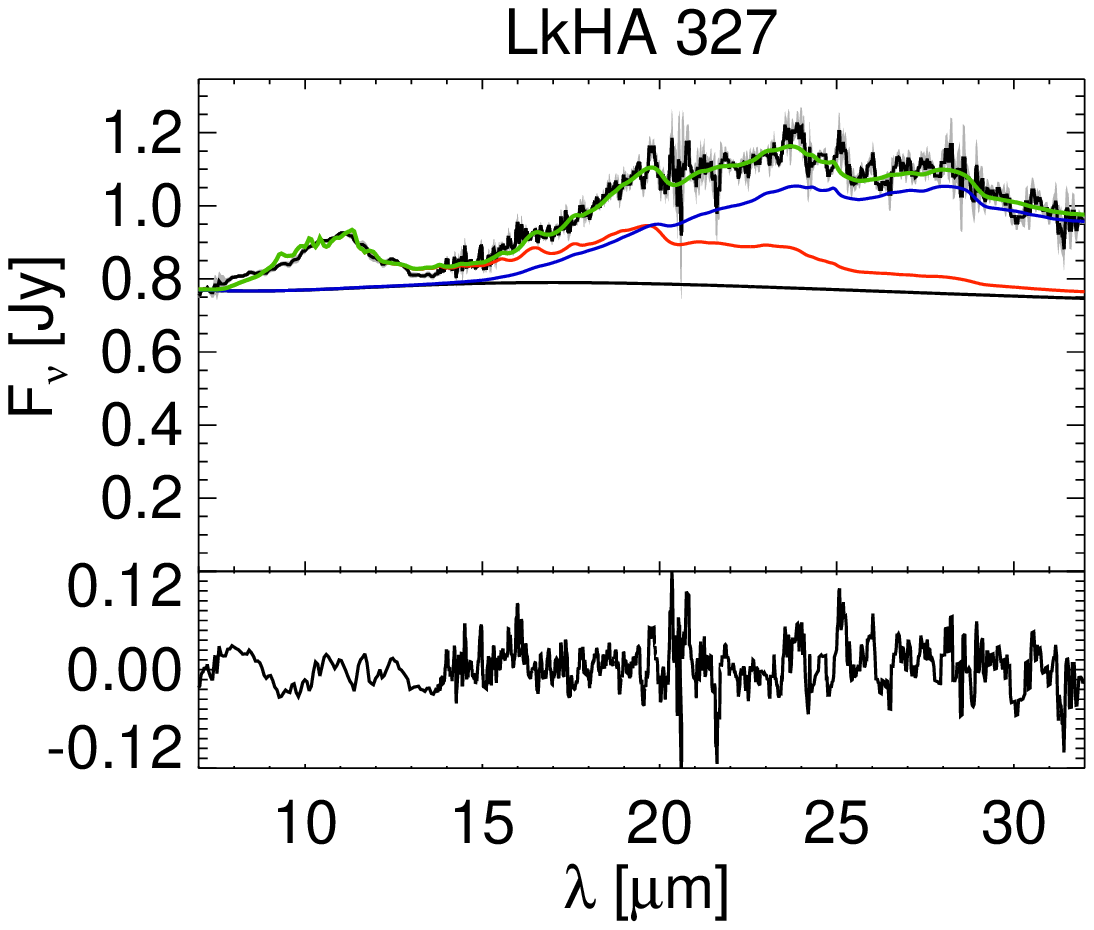}
\includegraphics[width=.2\columnwidth,origin=bl]{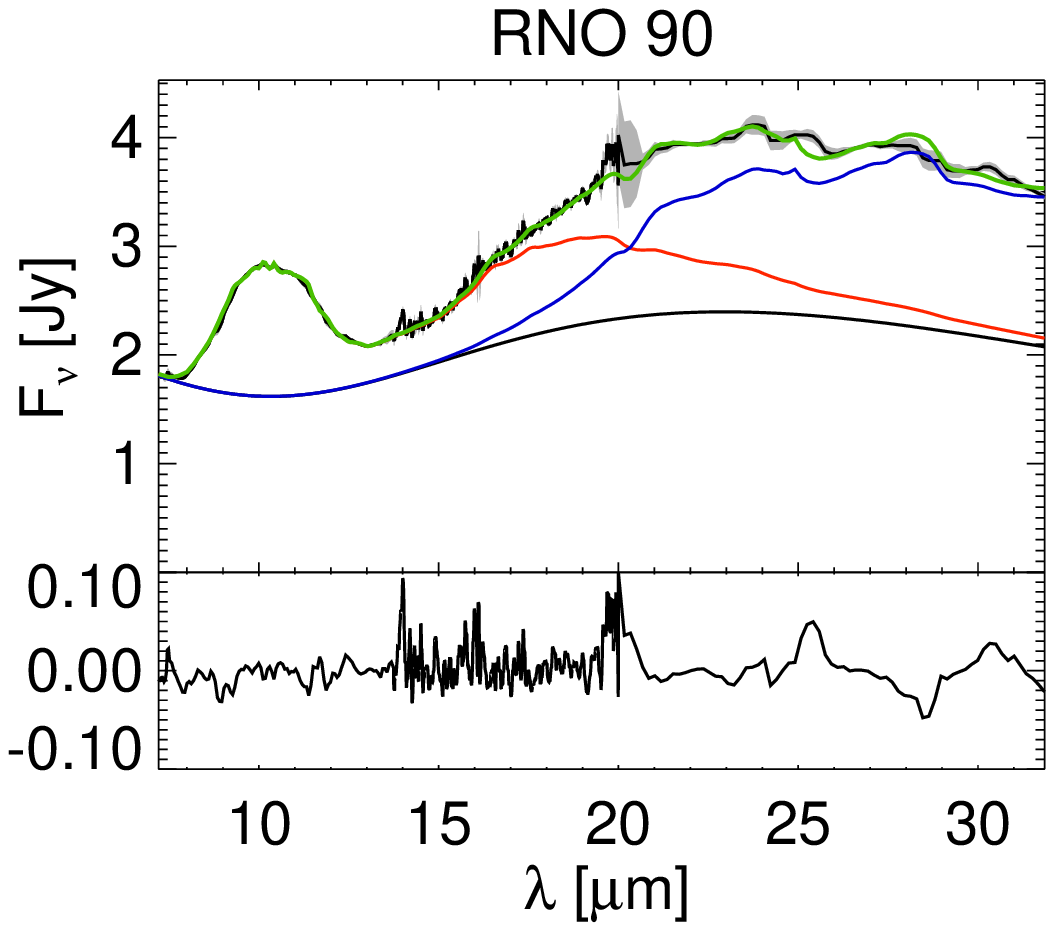}
\includegraphics[width=.2\columnwidth,origin=bl]{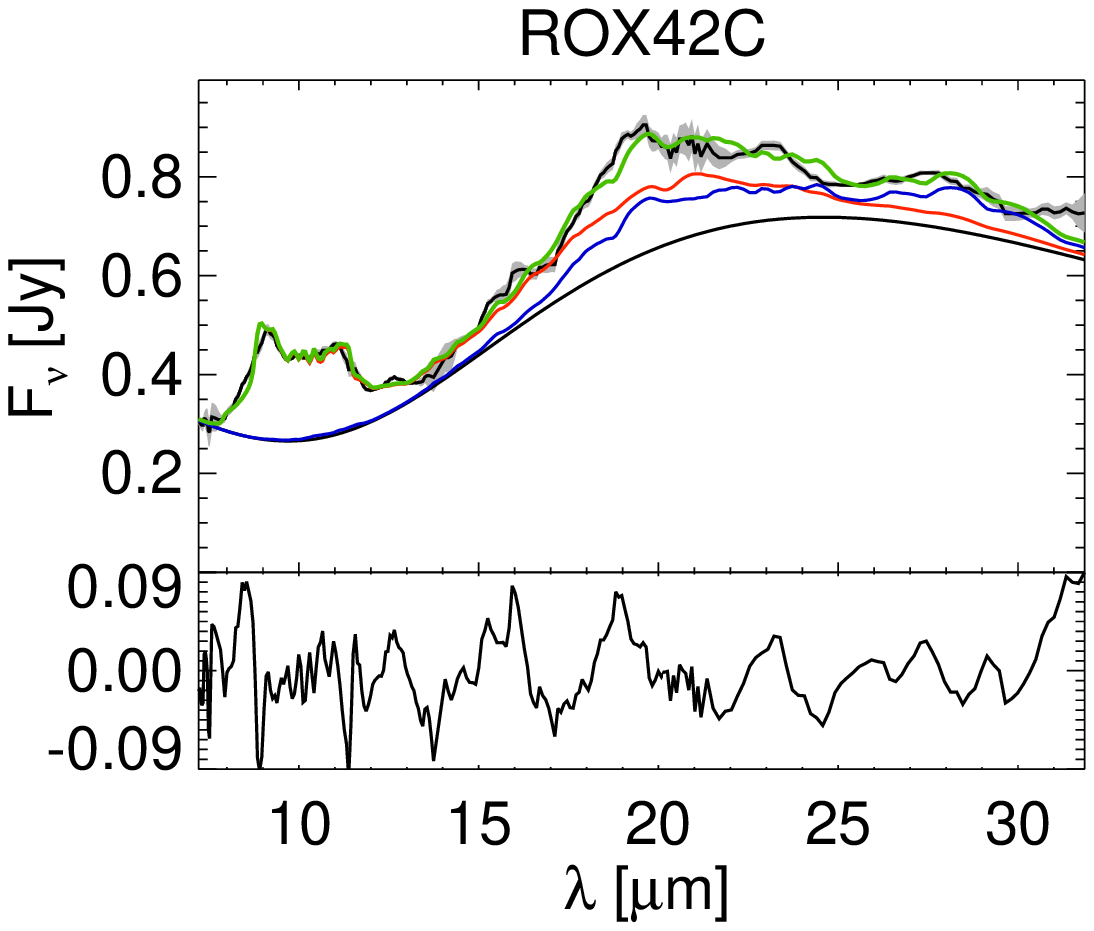}
\includegraphics[width=.2\columnwidth,origin=bl]{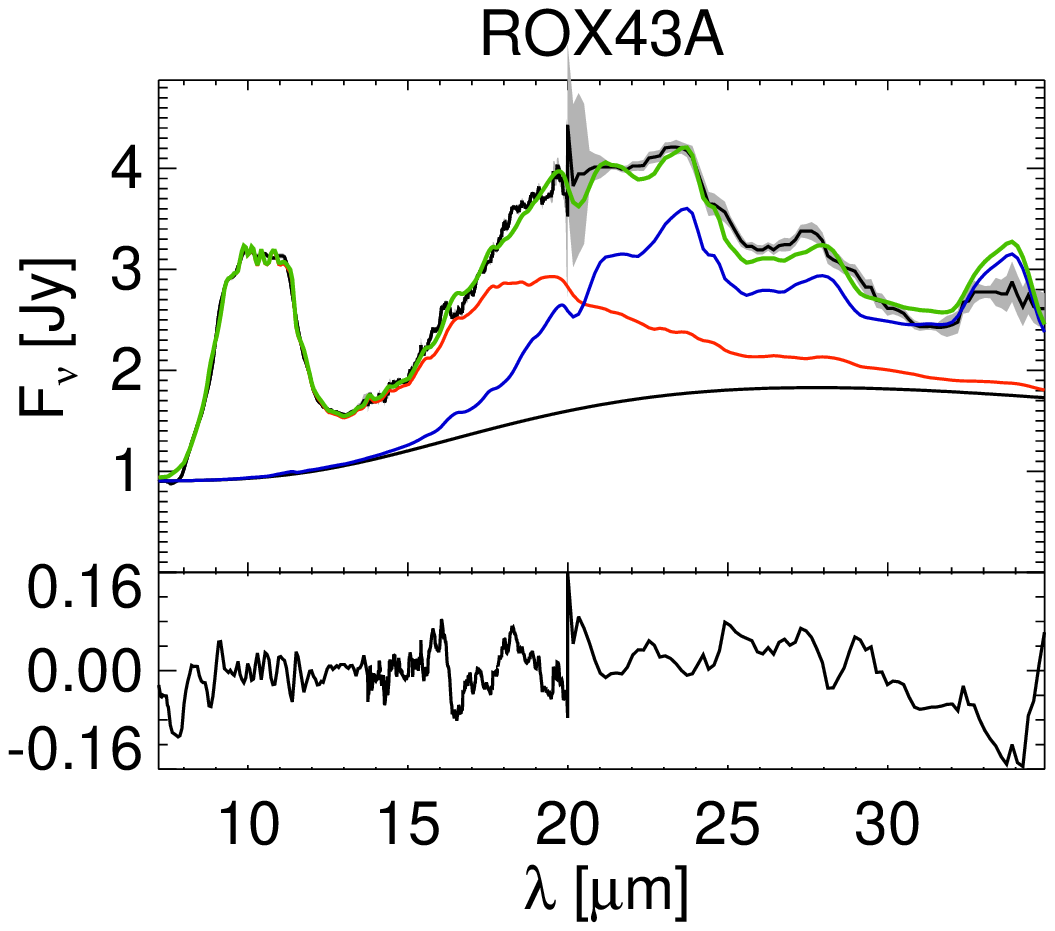}
\includegraphics[width=.2\columnwidth,origin=bl]{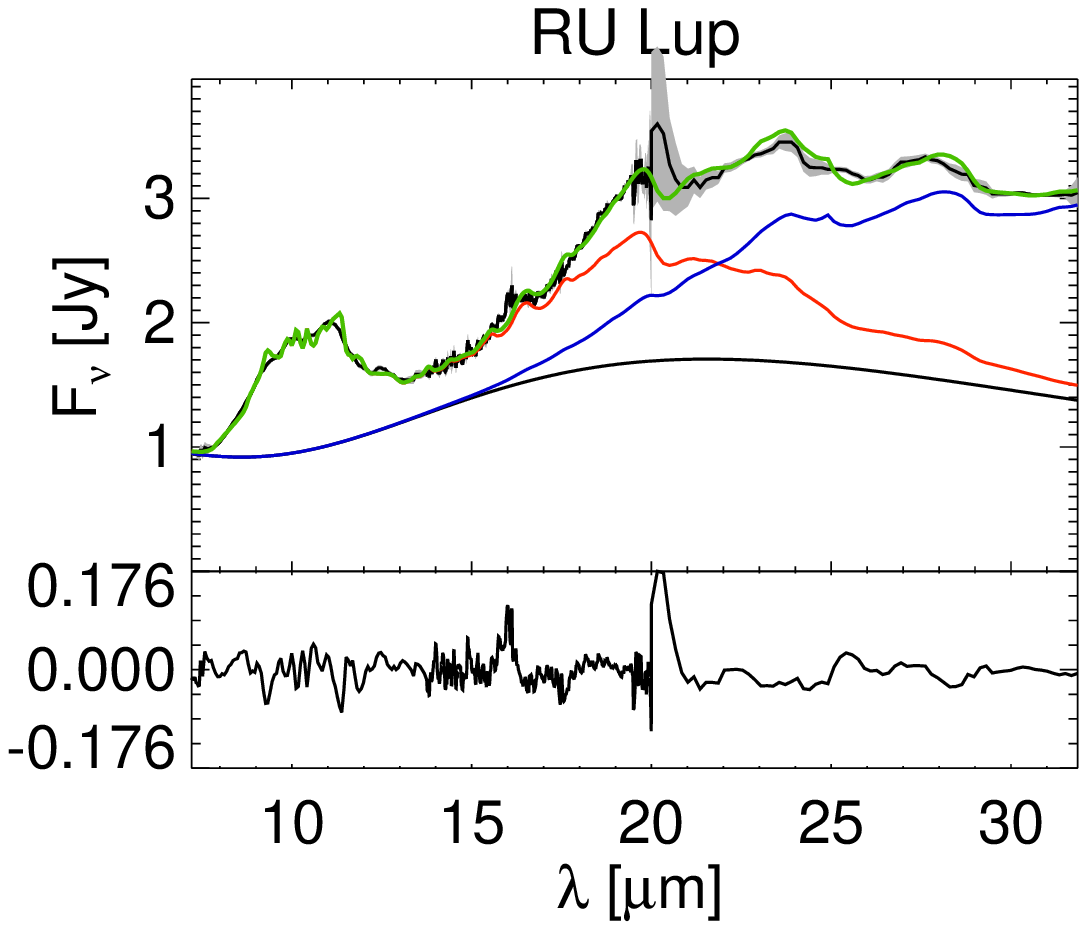}
\end{center}
\end{figure}
\begin{figure}
\begin{center}
\caption{\label{all:fit2}Continued}
\includegraphics[width=.2\columnwidth,origin=bl]{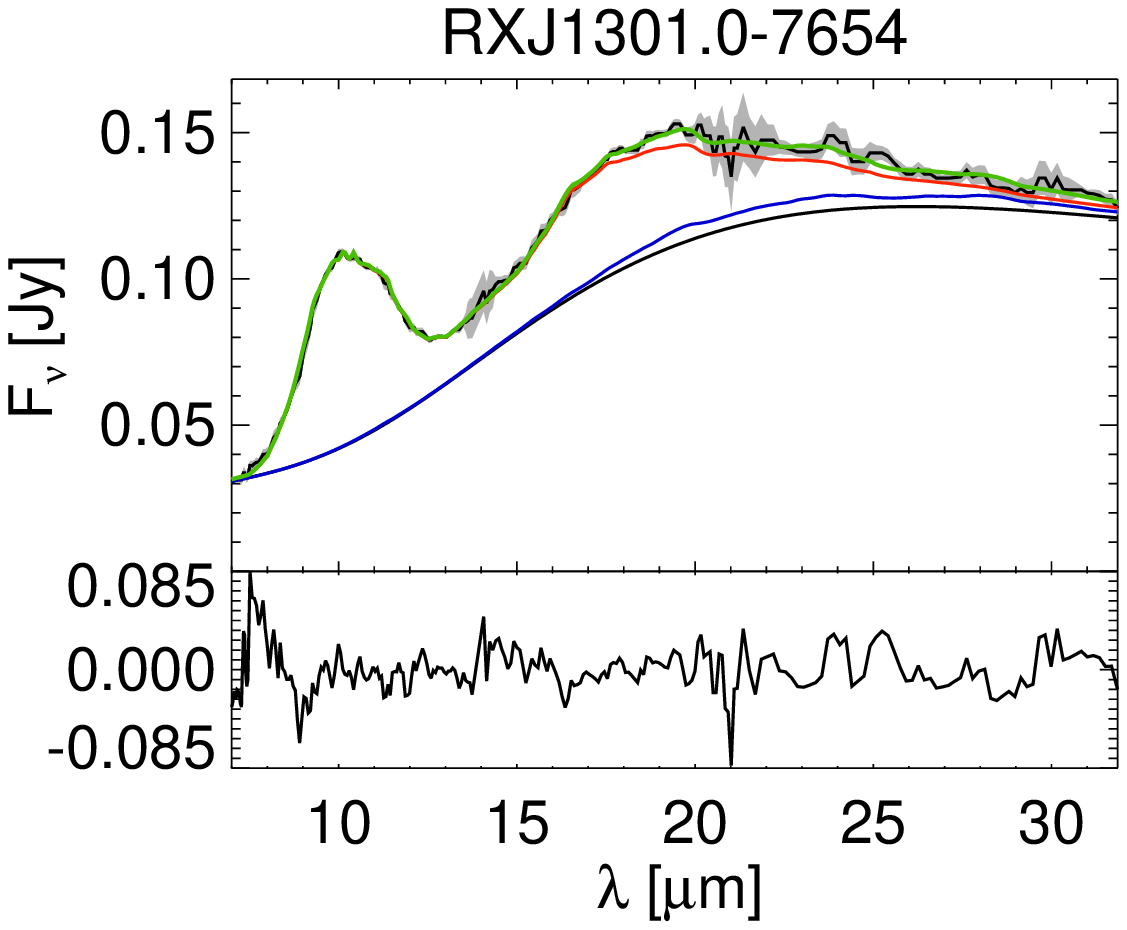}
\includegraphics[width=.2\columnwidth,origin=bl]{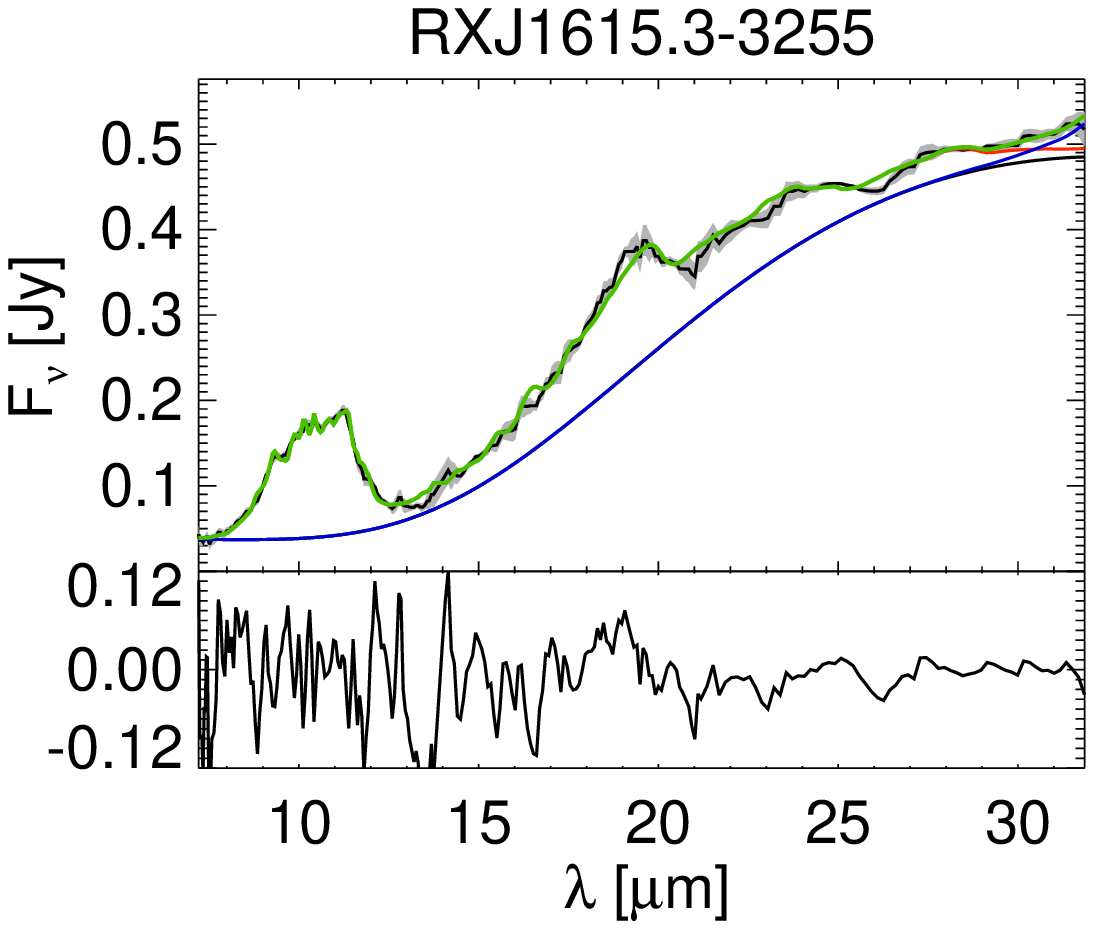}
\includegraphics[width=.2\columnwidth,origin=bl]{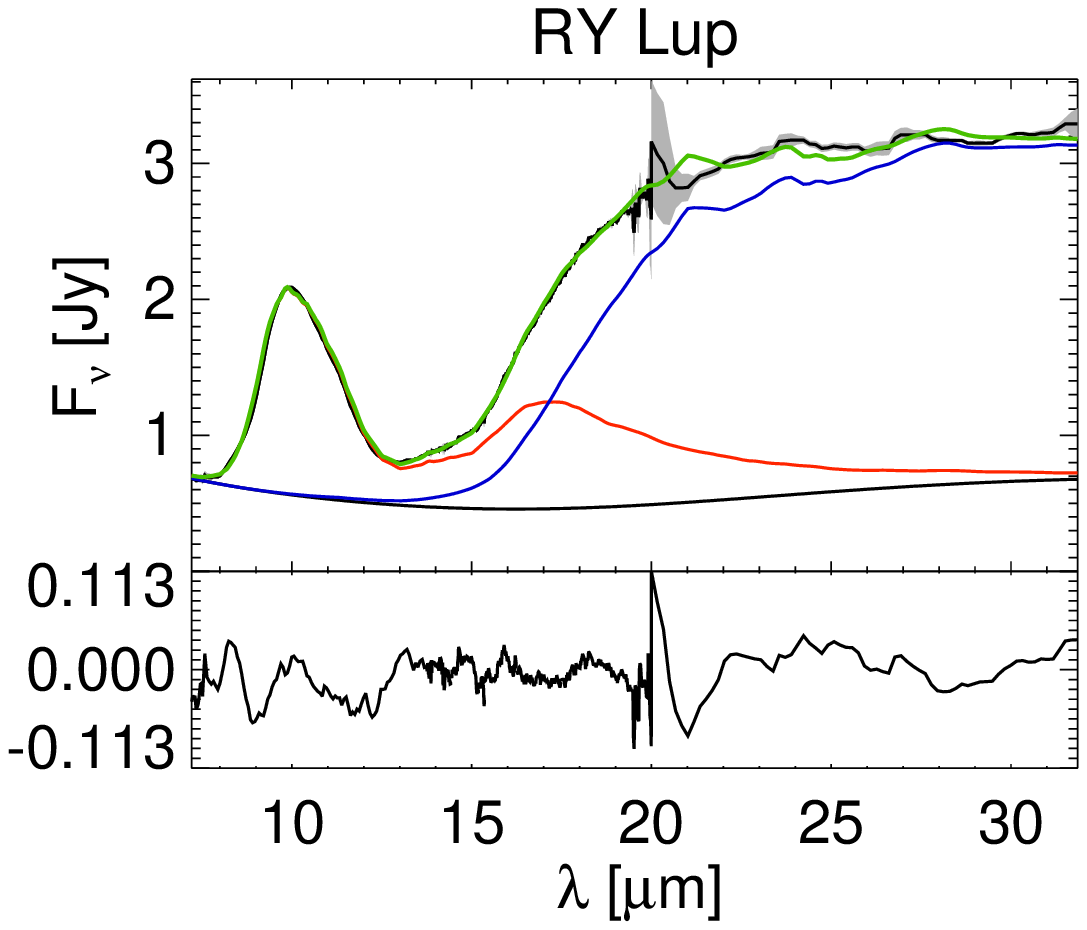}
\includegraphics[width=.2\columnwidth,origin=bl]{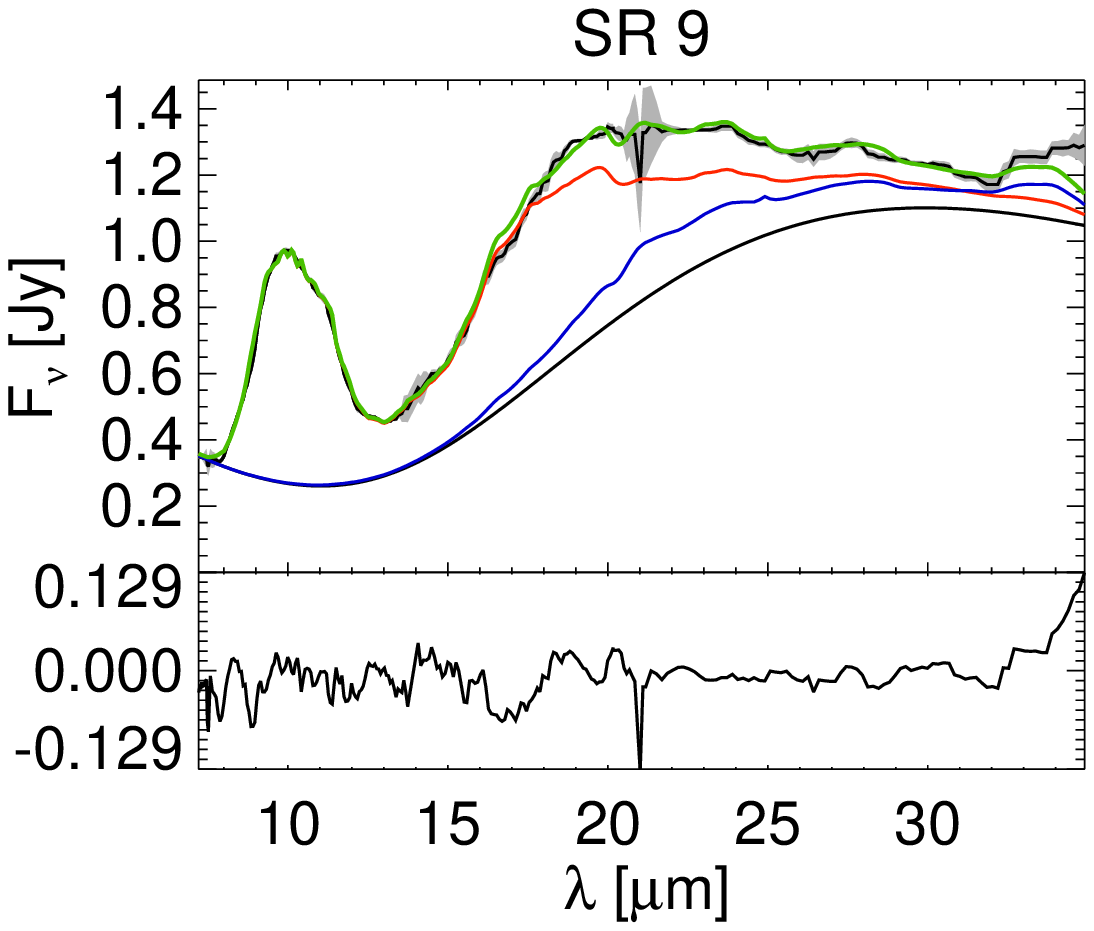}
\includegraphics[width=.2\columnwidth,origin=bl]{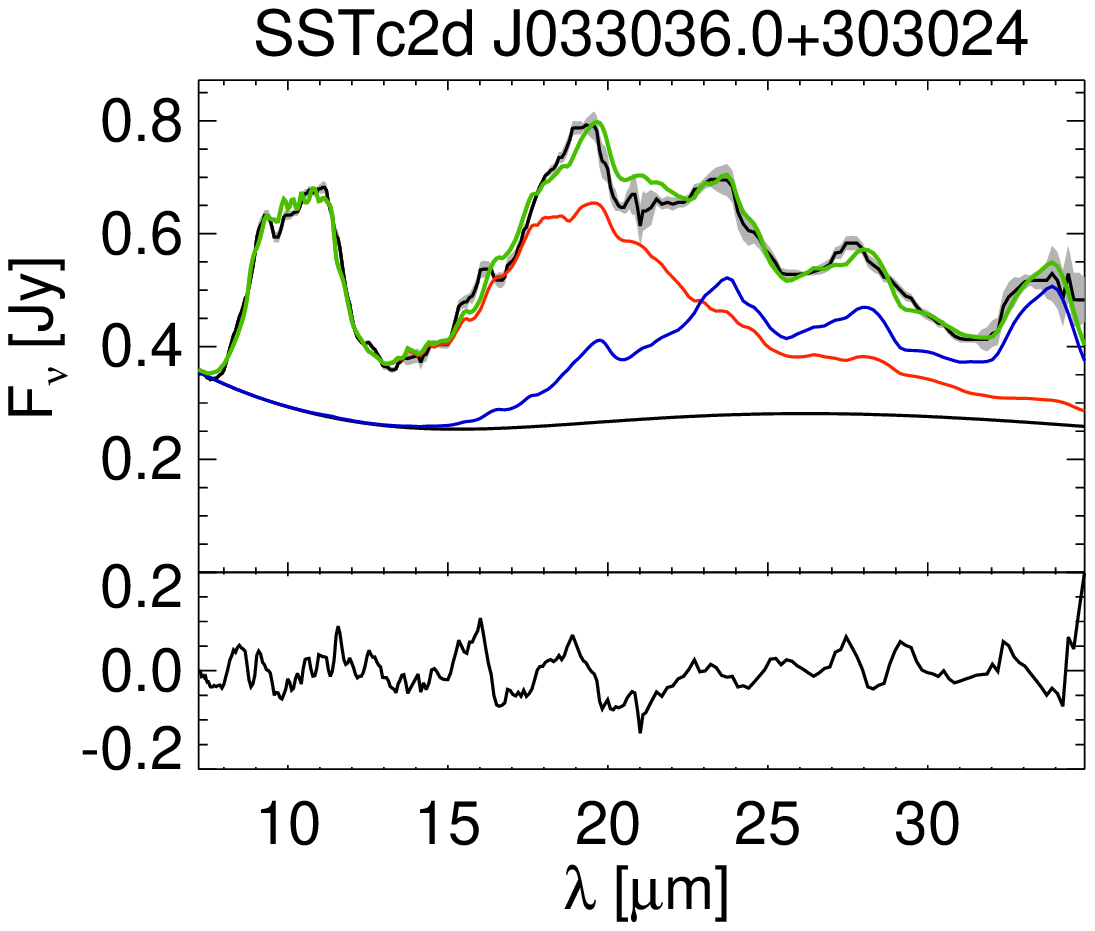}
\includegraphics[width=.2\columnwidth,origin=bl]{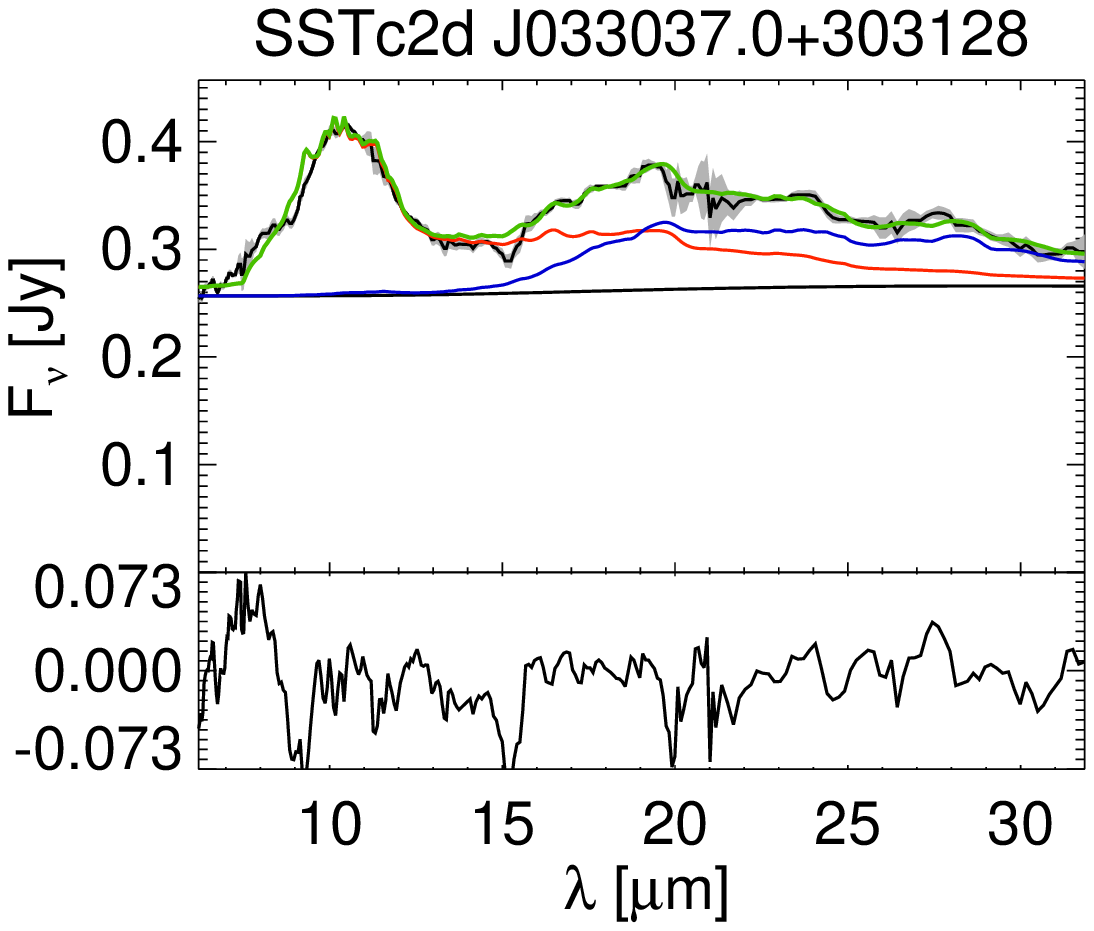}
\includegraphics[width=.2\columnwidth,origin=bl]{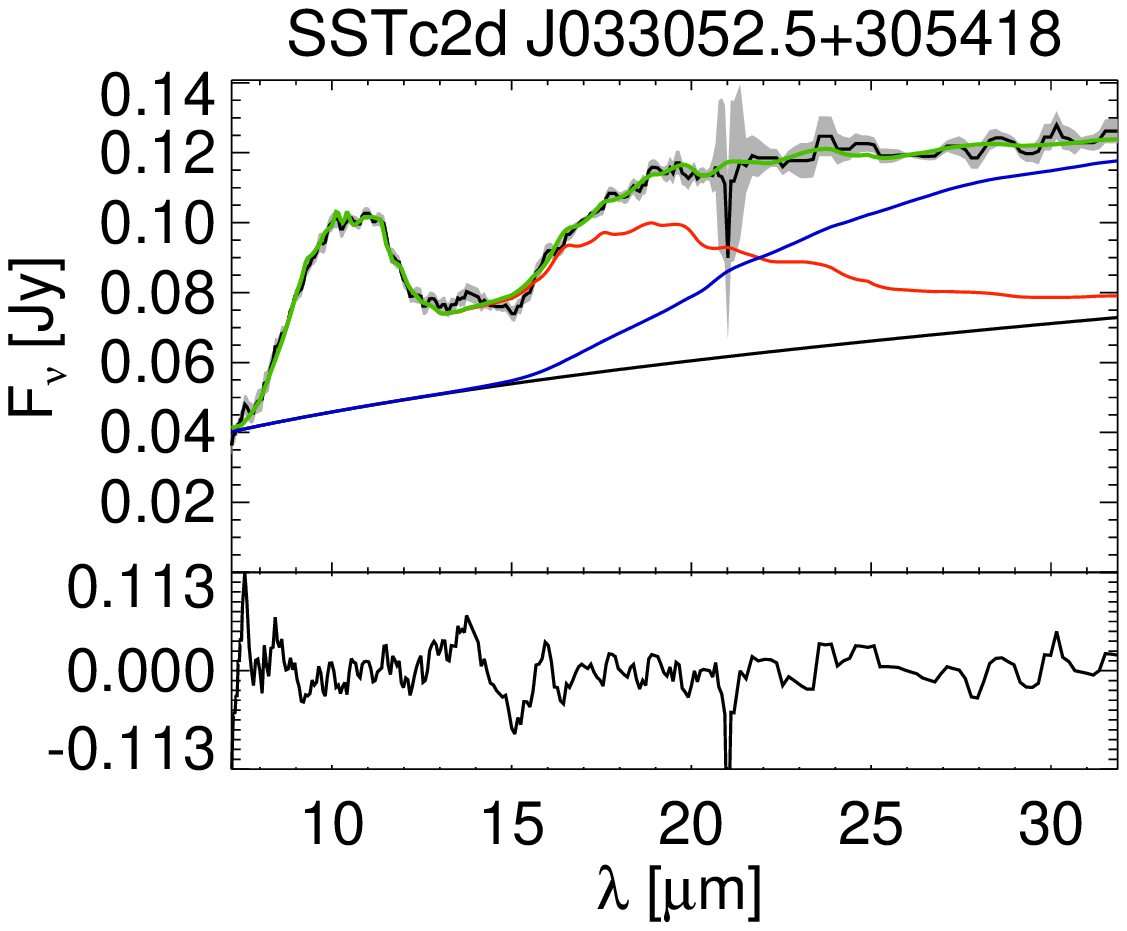}
\includegraphics[width=.2\columnwidth,origin=bl]{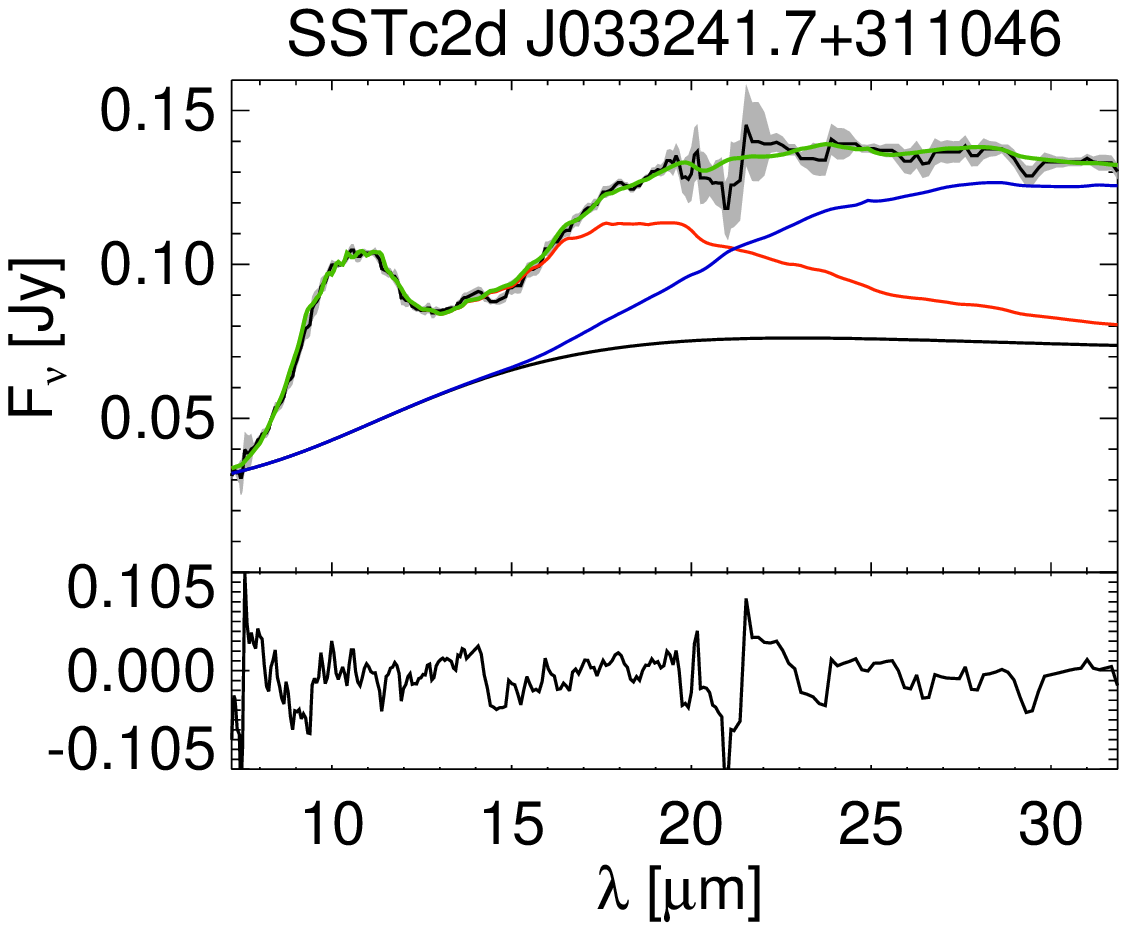}
\includegraphics[width=.2\columnwidth,origin=bl]{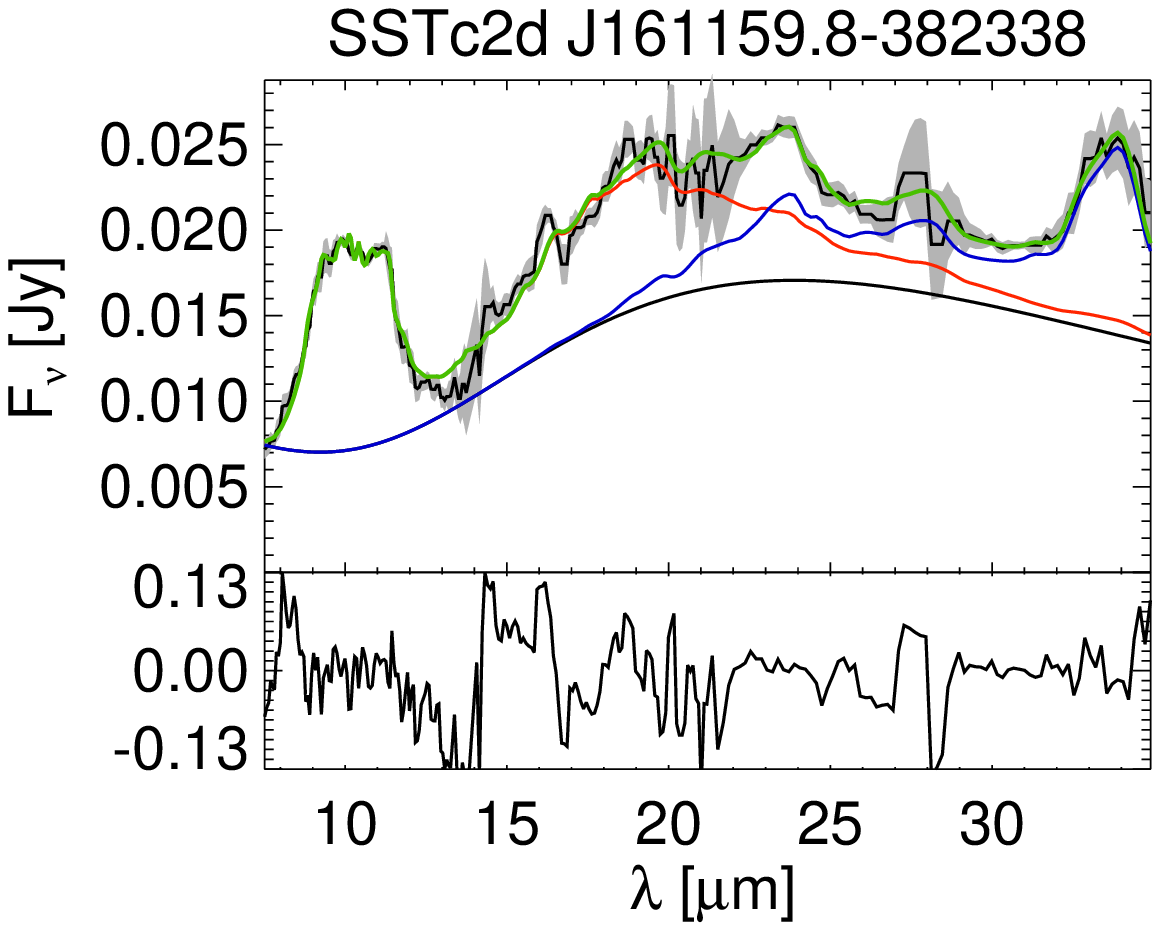}
\includegraphics[width=.2\columnwidth,origin=bl]{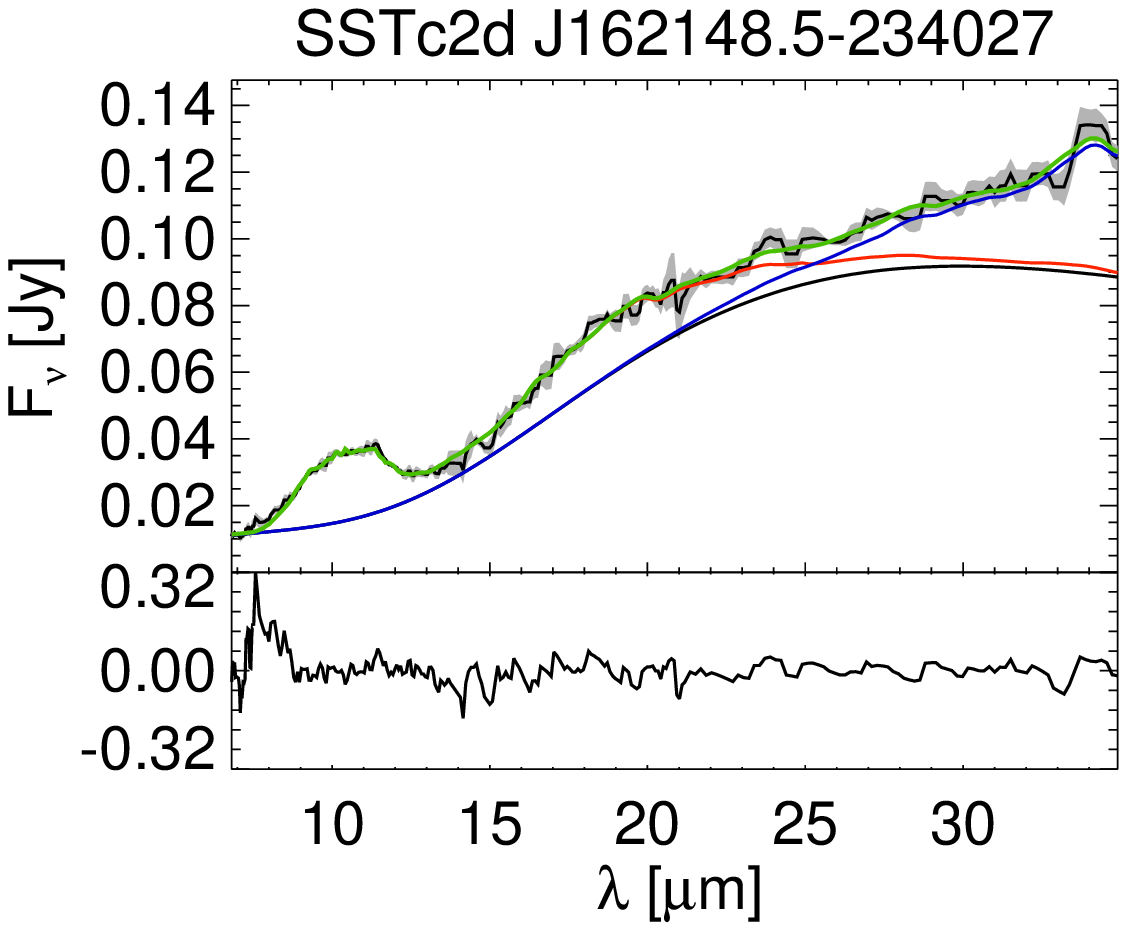}
\includegraphics[width=.2\columnwidth,origin=bl]{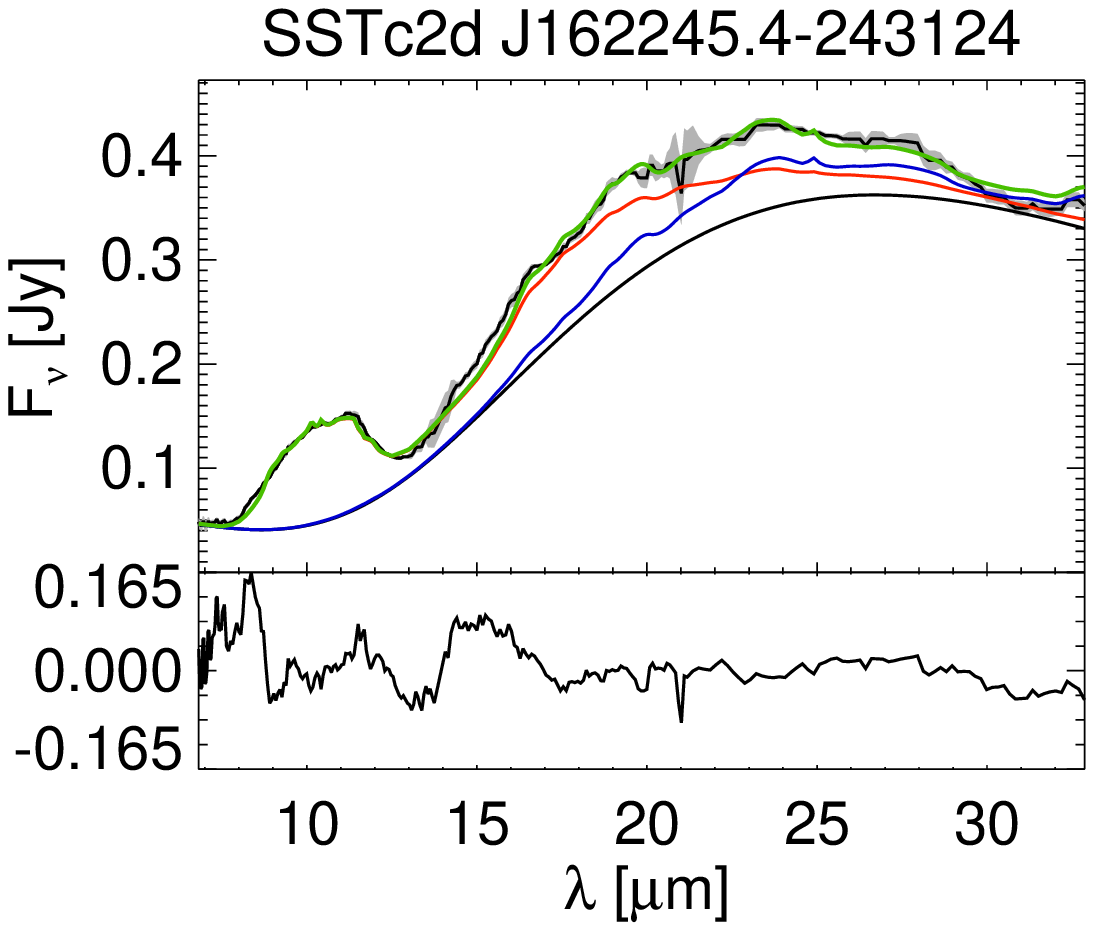}
\includegraphics[width=.2\columnwidth,origin=bl]{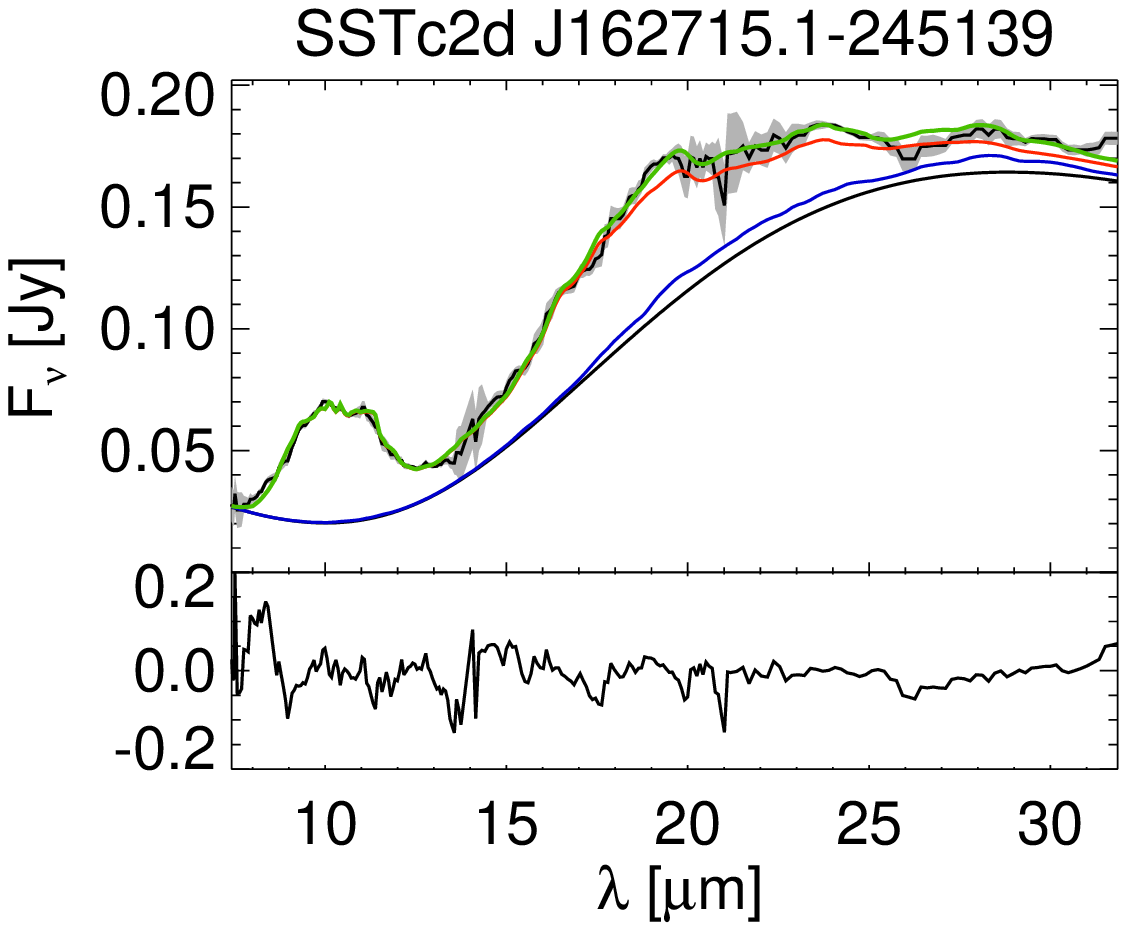}
\includegraphics[width=.2\columnwidth,origin=bl]{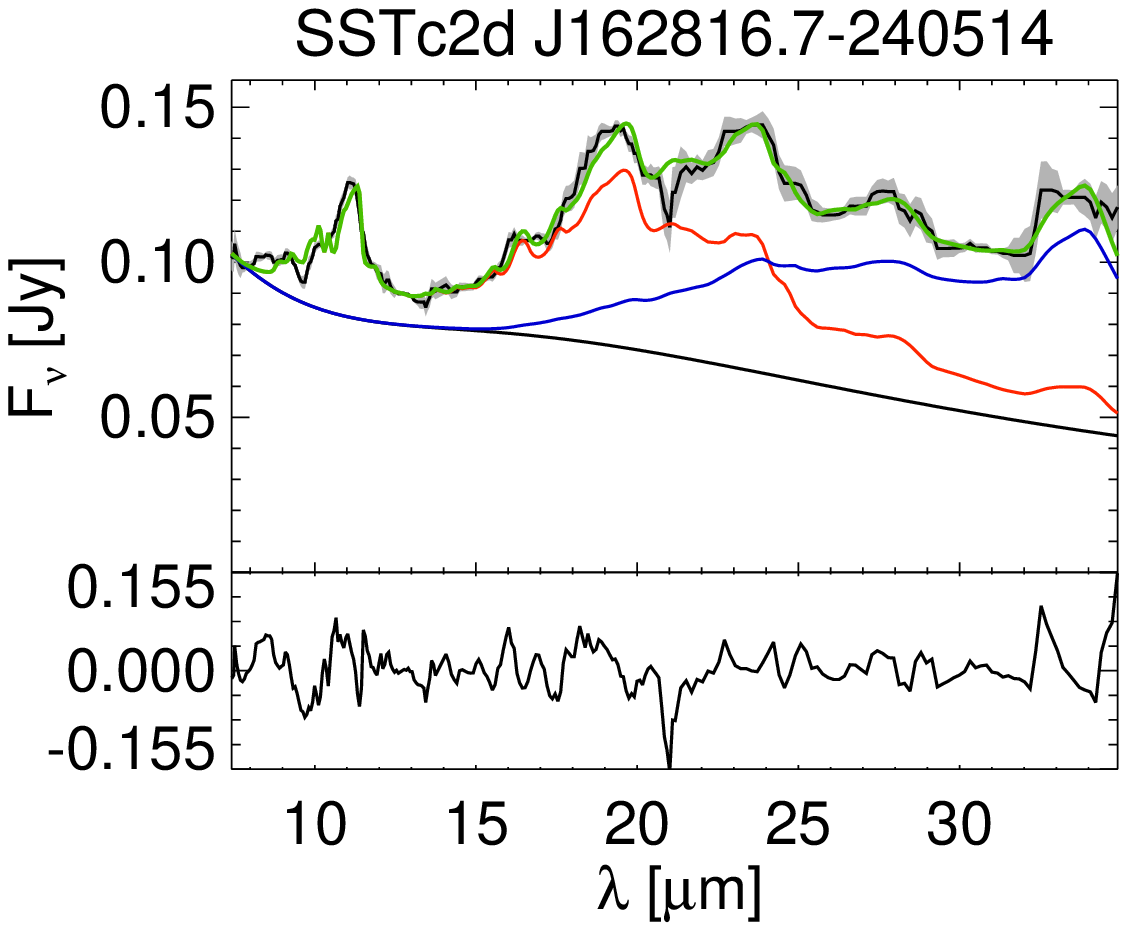}
\includegraphics[width=.2\columnwidth,origin=bl]{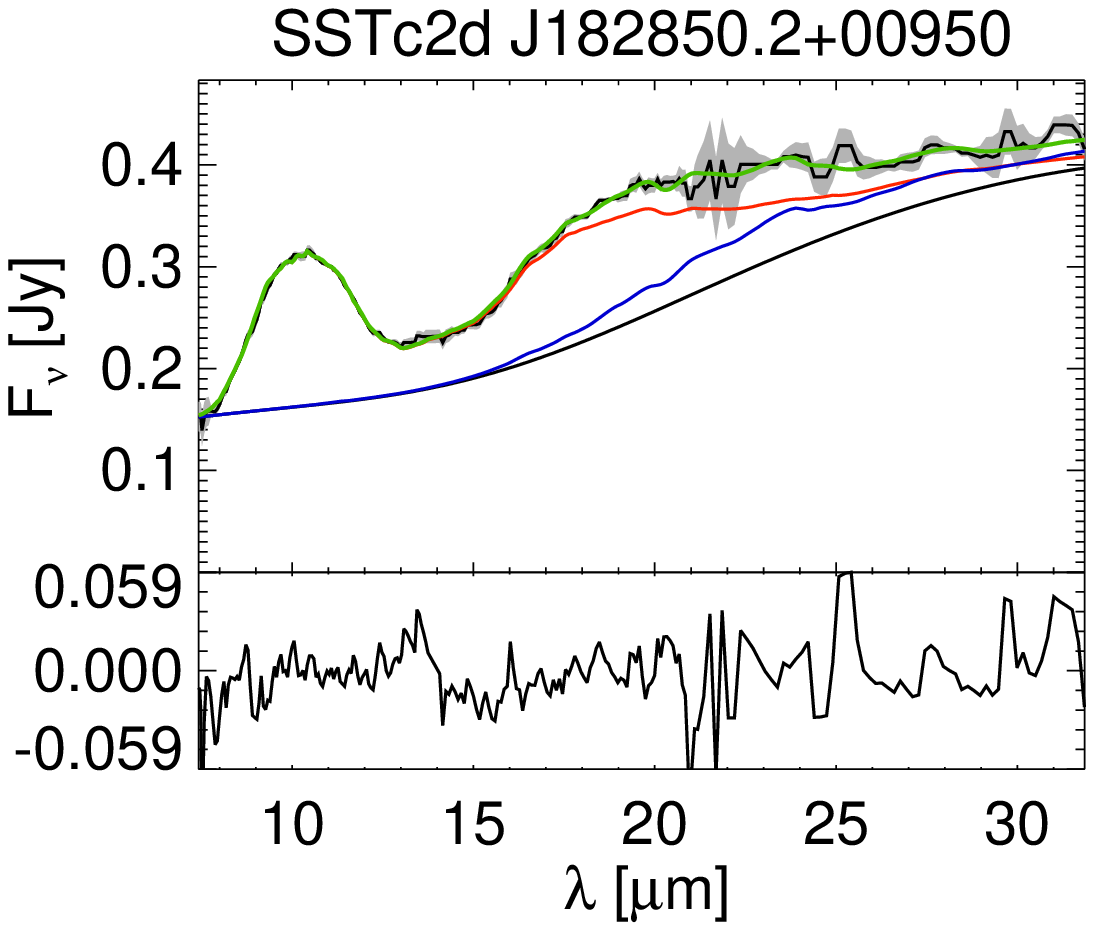}
\includegraphics[width=.2\columnwidth,origin=bl]{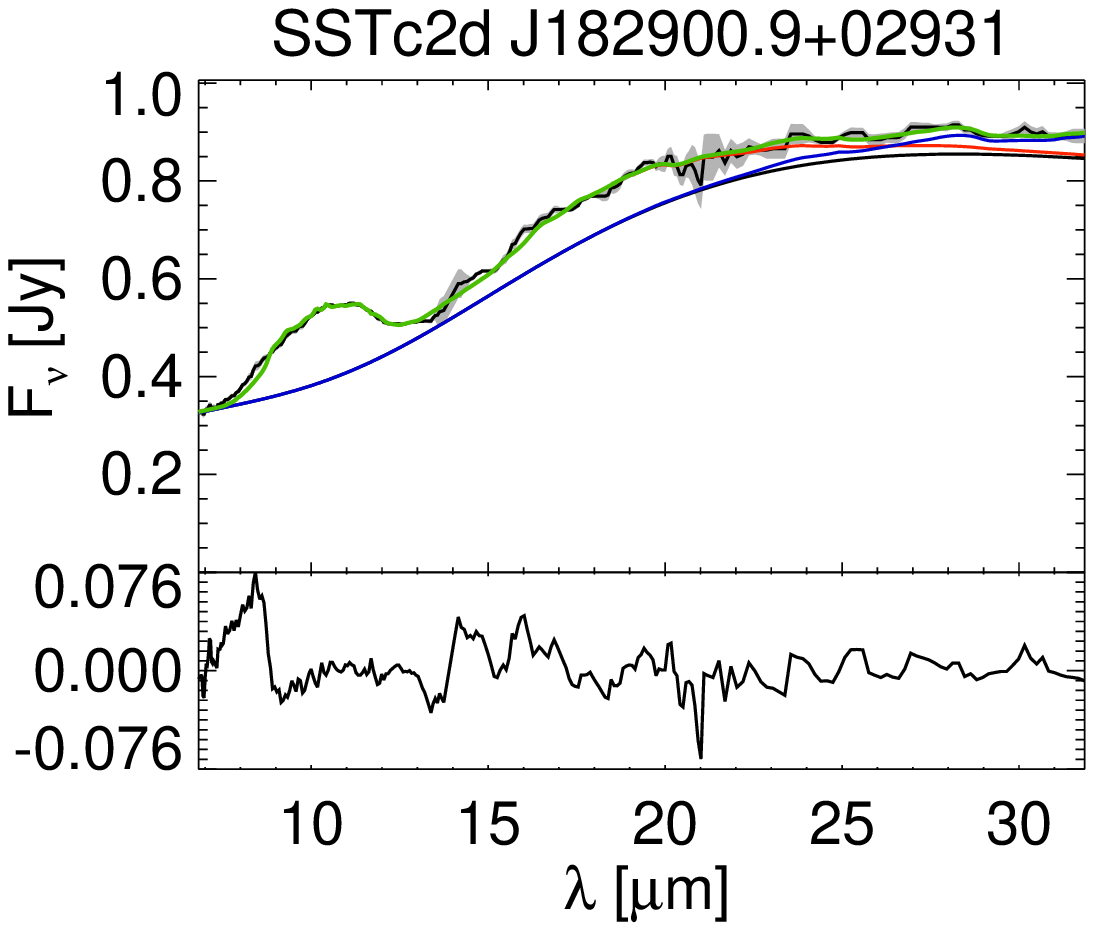}
\includegraphics[width=.2\columnwidth,origin=bl]{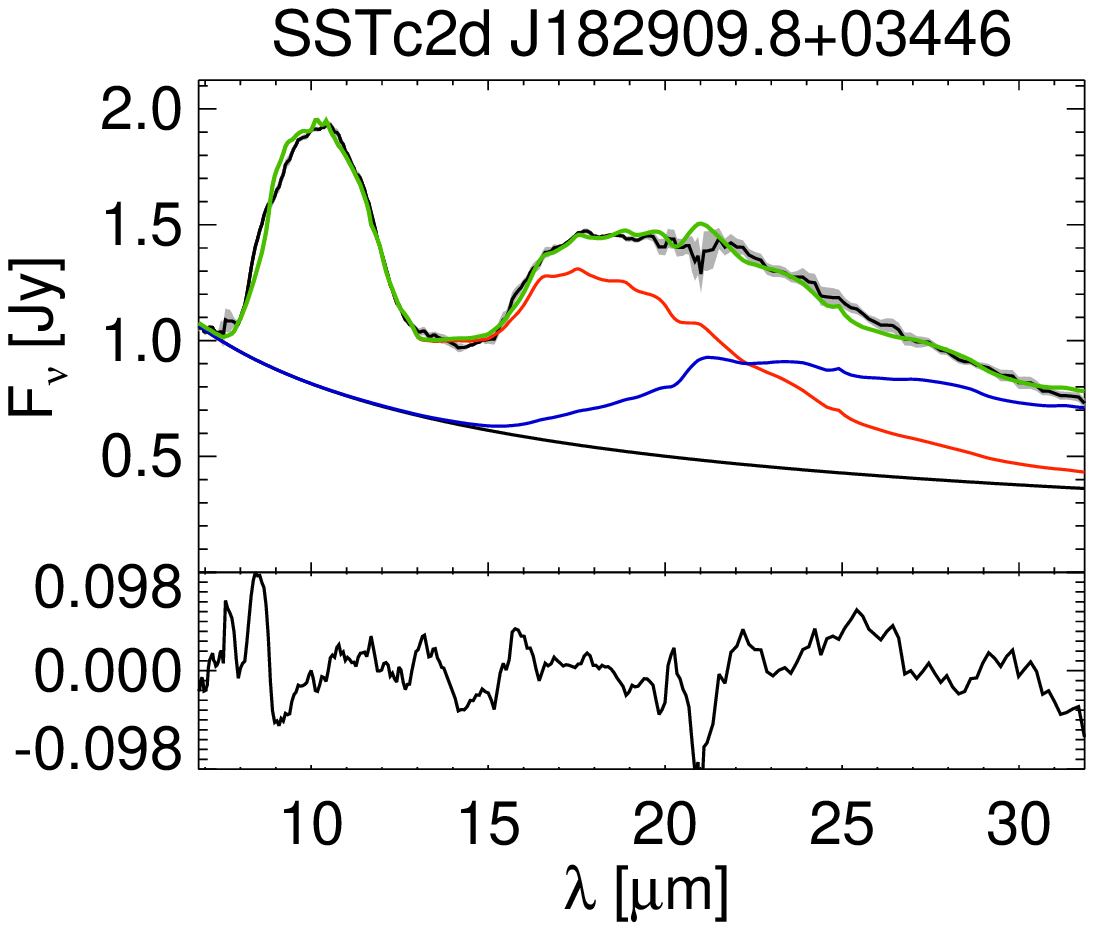}
\includegraphics[width=.2\columnwidth,origin=bl]{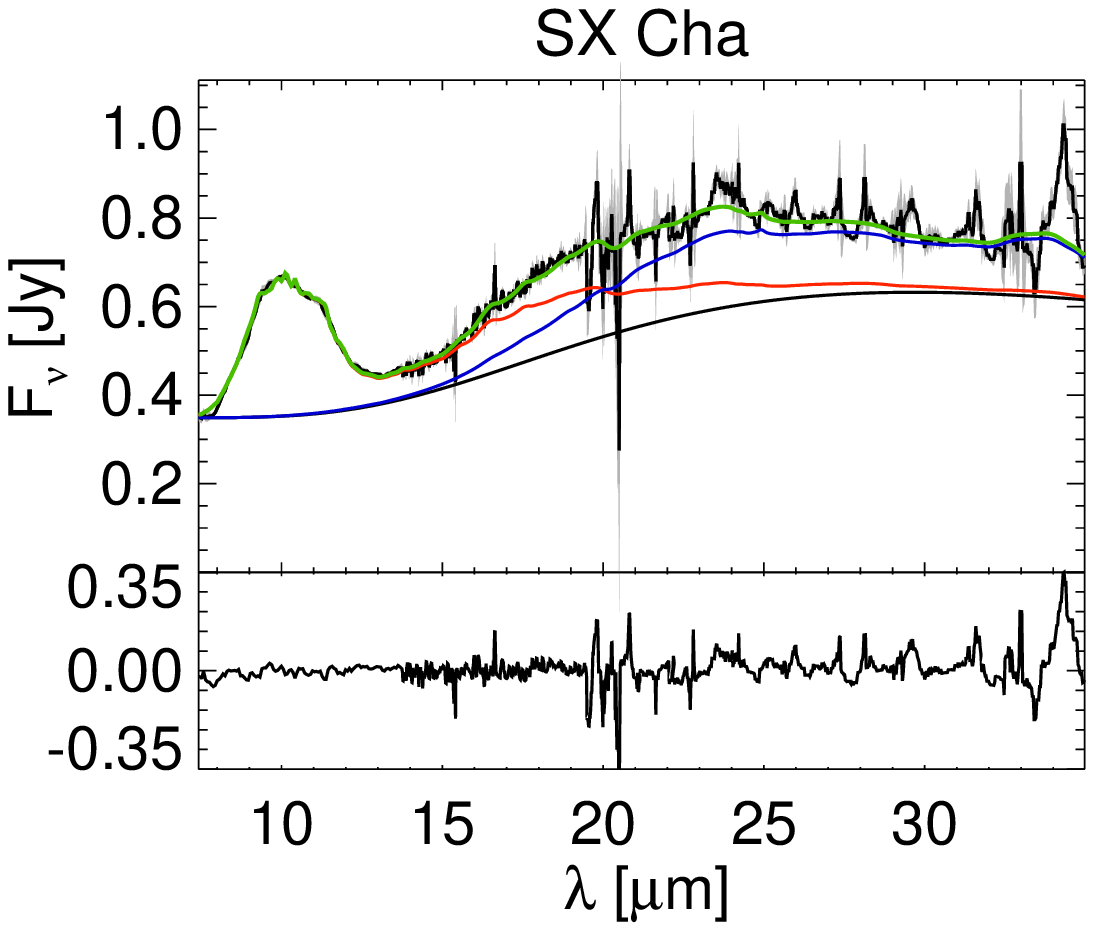}
\includegraphics[width=.2\columnwidth,origin=bl]{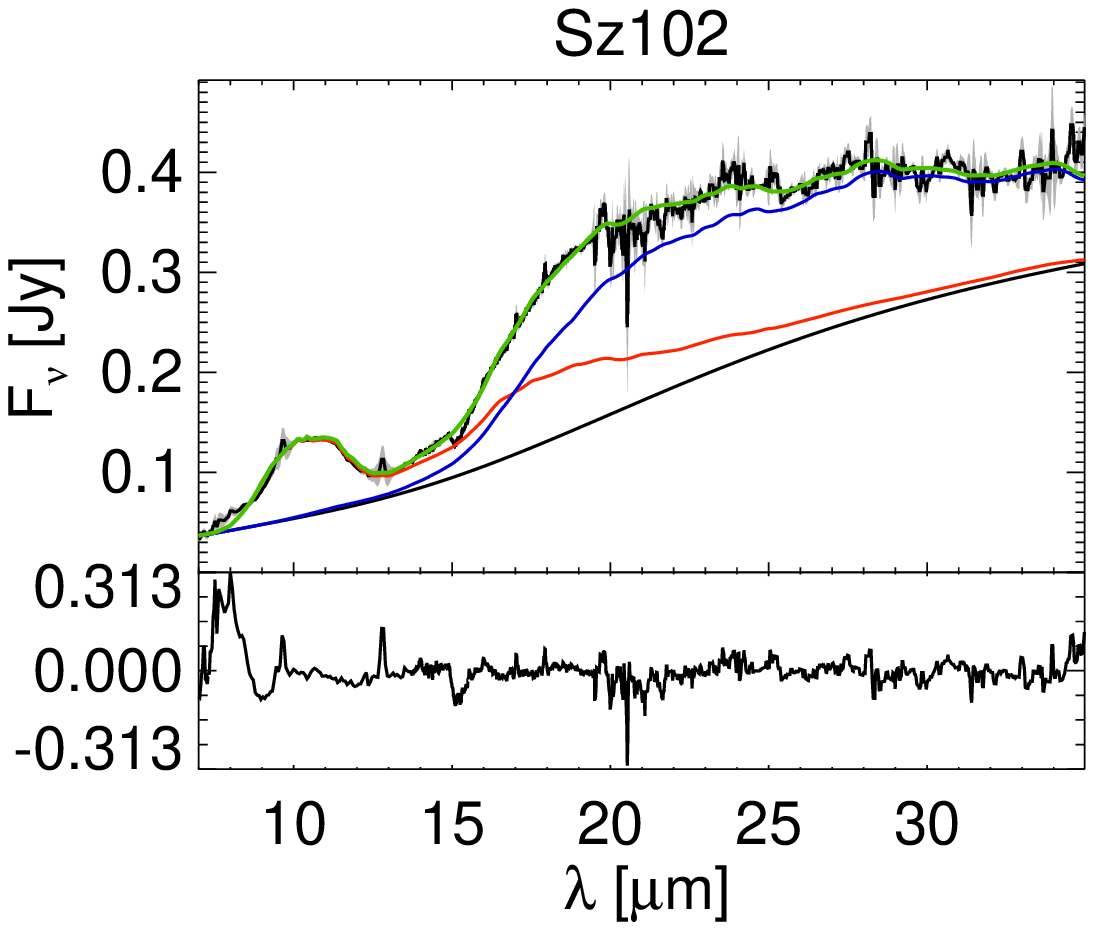}
\includegraphics[width=.2\columnwidth,origin=bl]{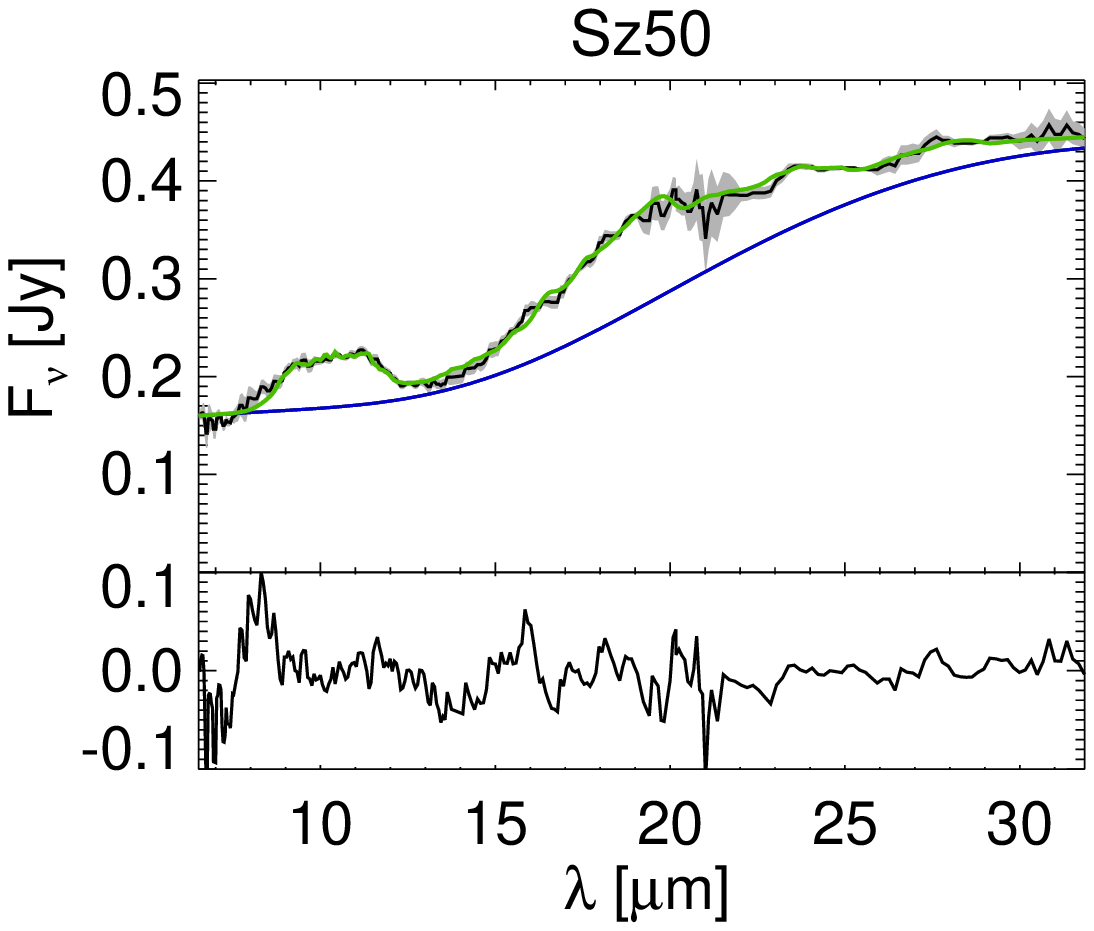}
\includegraphics[width=.2\columnwidth,origin=bl]{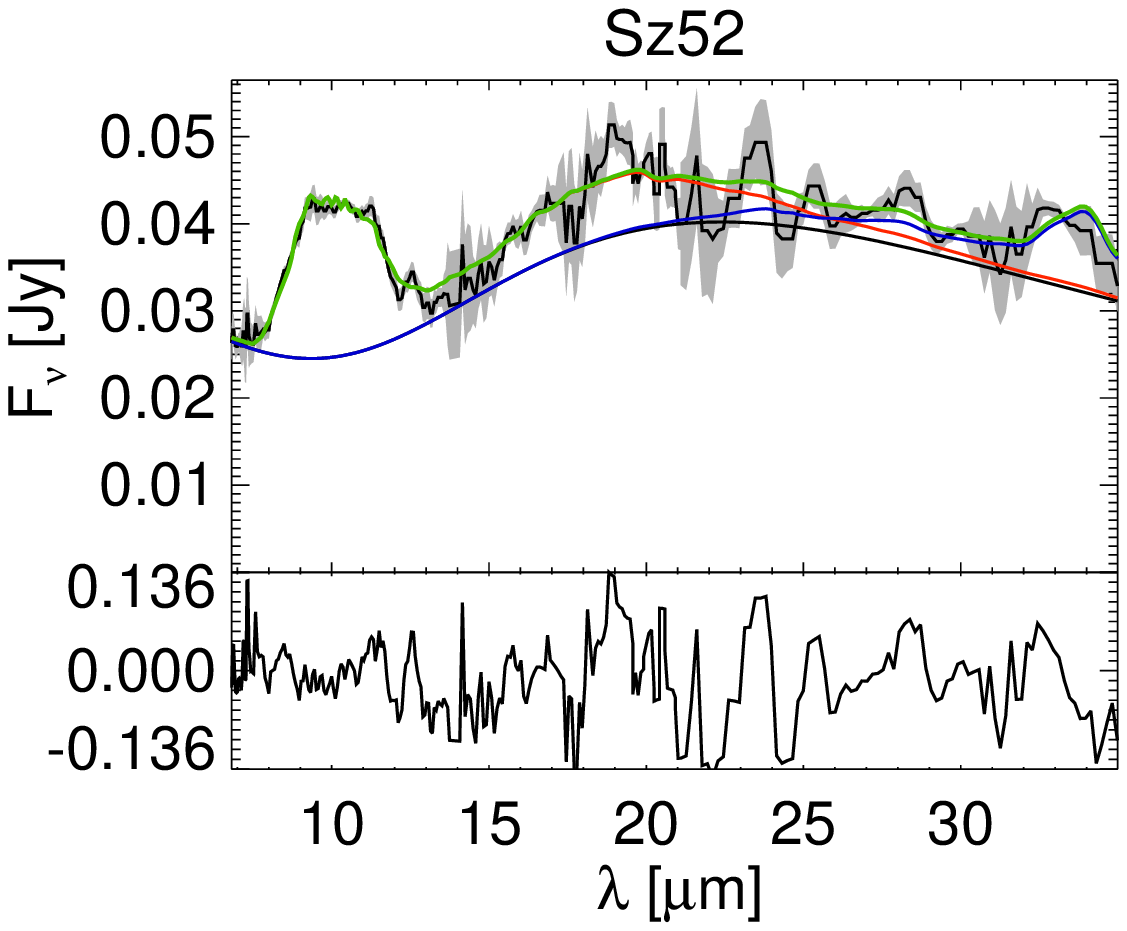}
\includegraphics[width=.2\columnwidth,origin=bl]{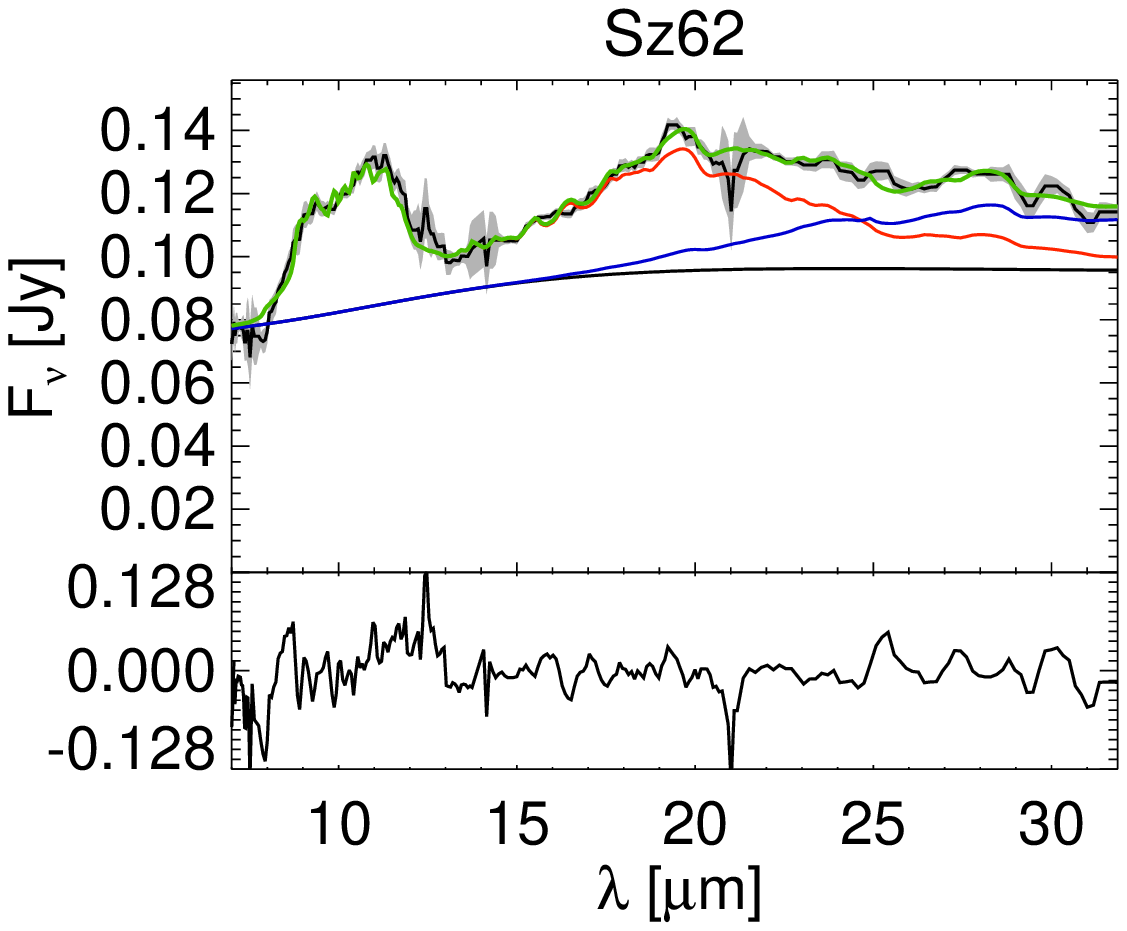}
\includegraphics[width=.2\columnwidth,origin=bl]{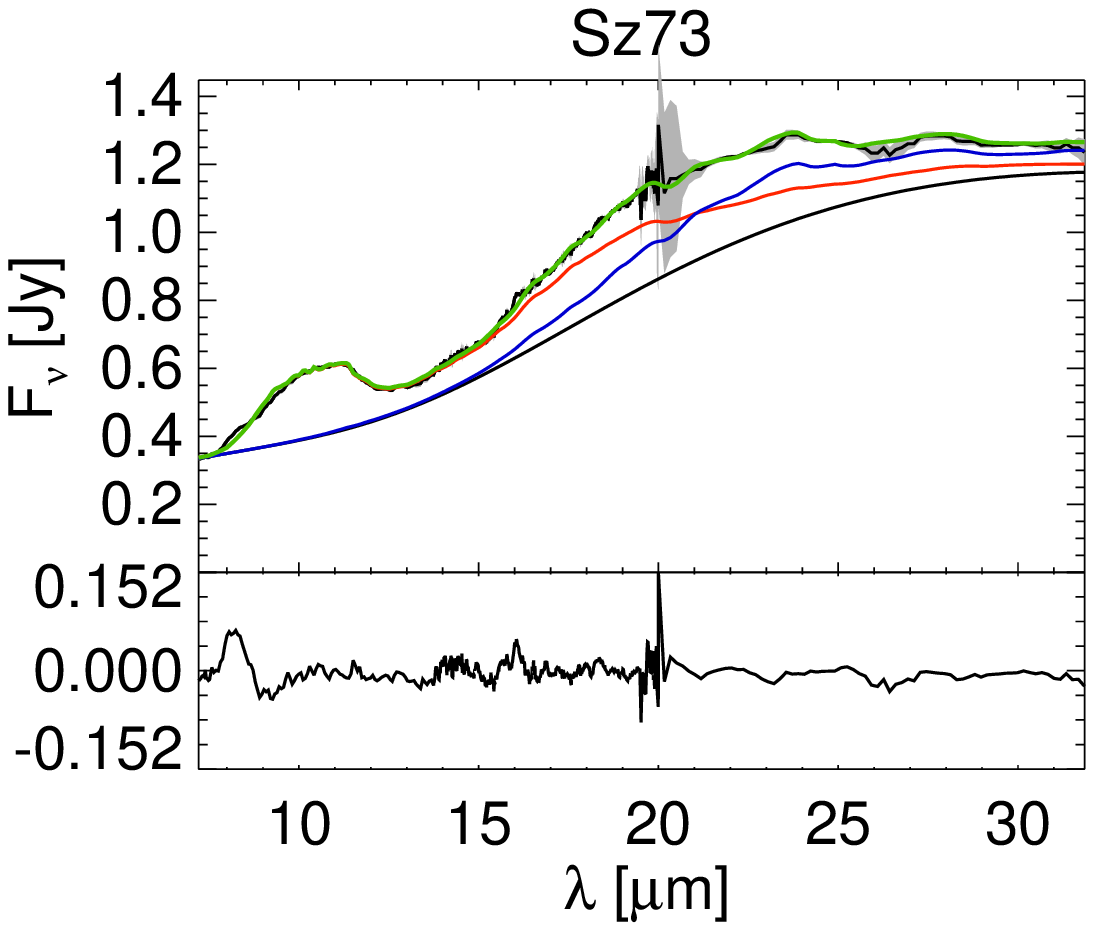}
\includegraphics[width=.2\columnwidth,origin=bl]{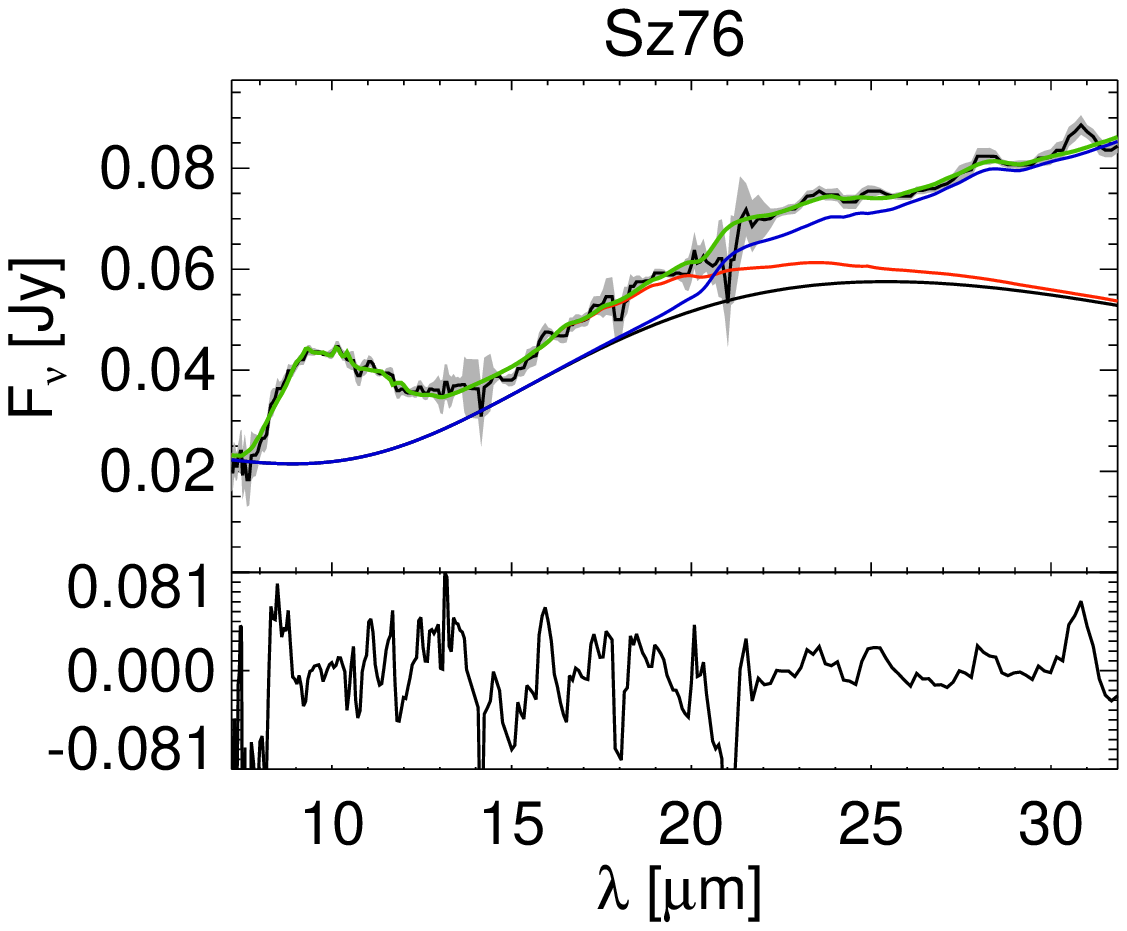}
\includegraphics[width=.2\columnwidth,origin=bl]{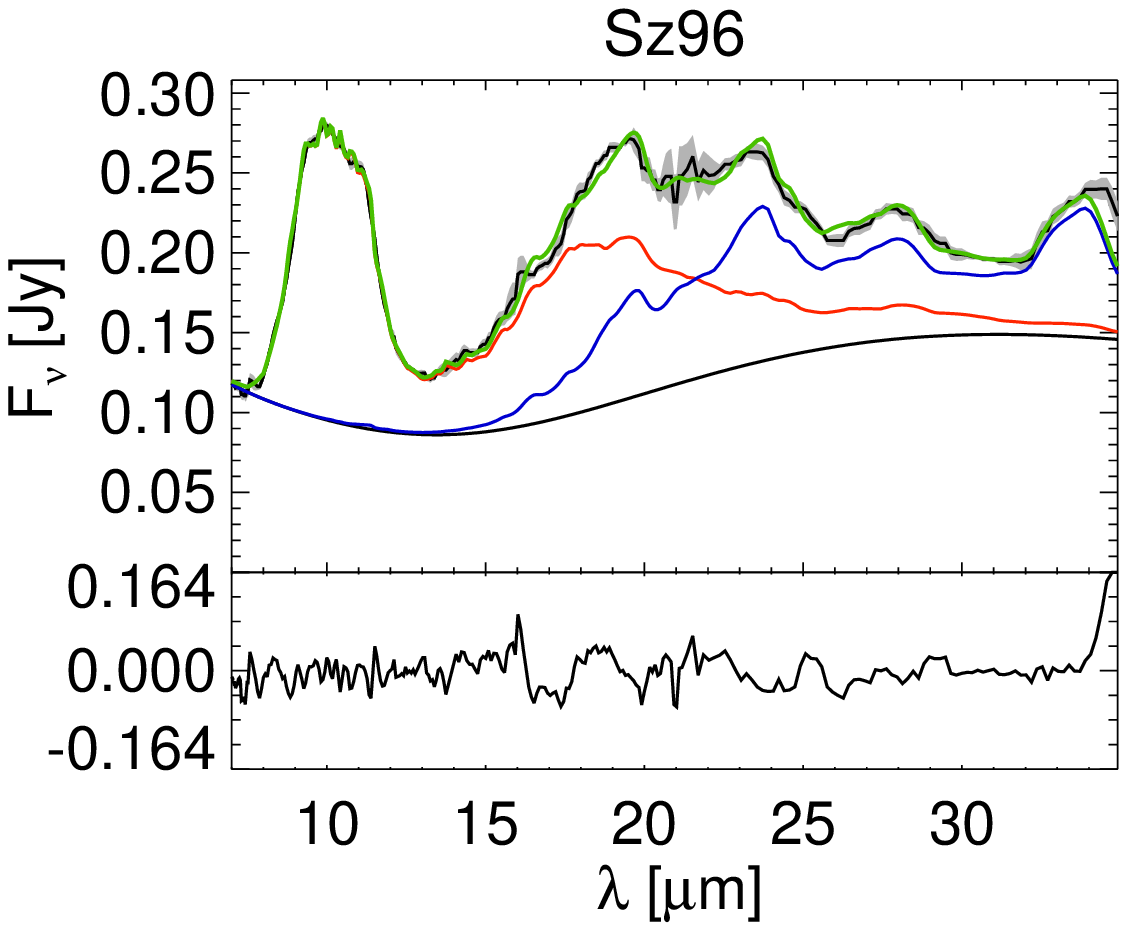}
\includegraphics[width=.2\columnwidth,origin=bl]{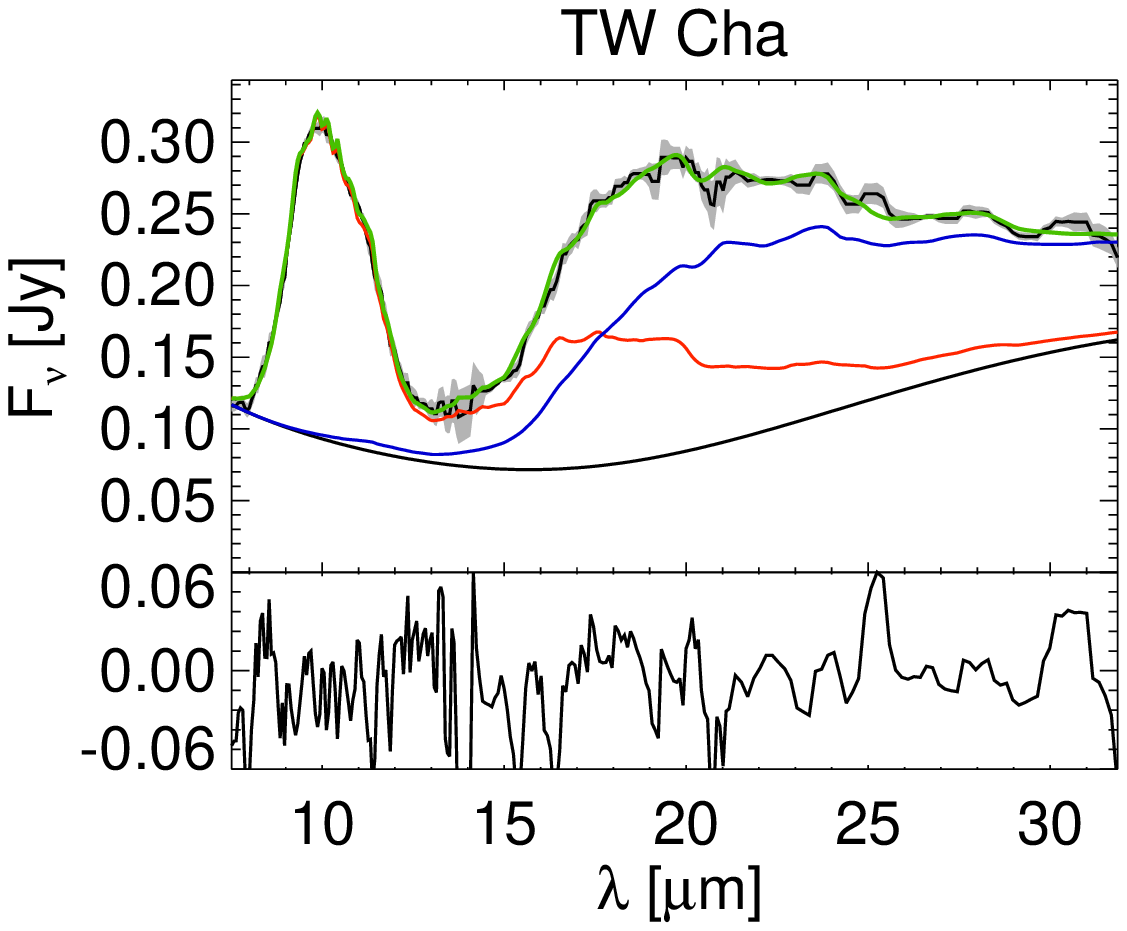}
\includegraphics[width=.2\columnwidth,origin=bl]{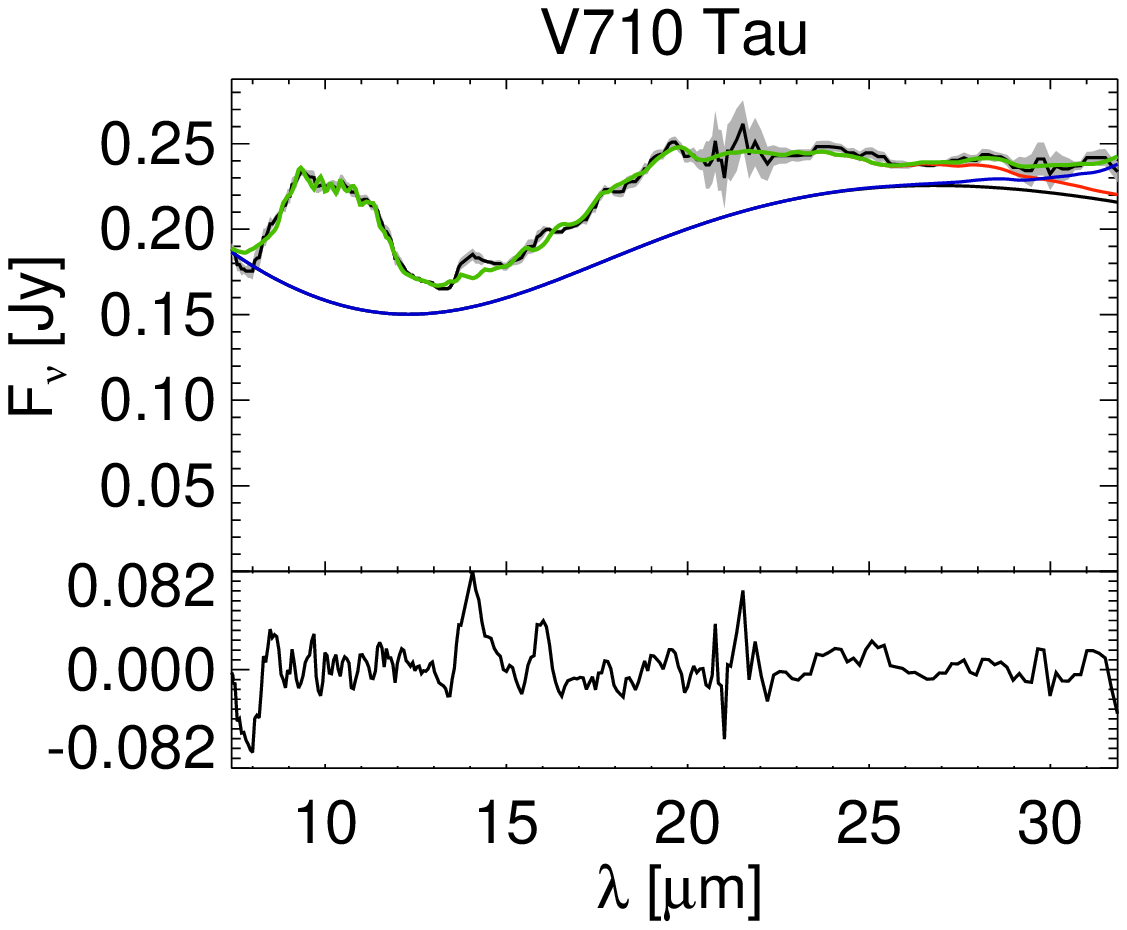}
\includegraphics[width=.2\columnwidth,origin=bl]{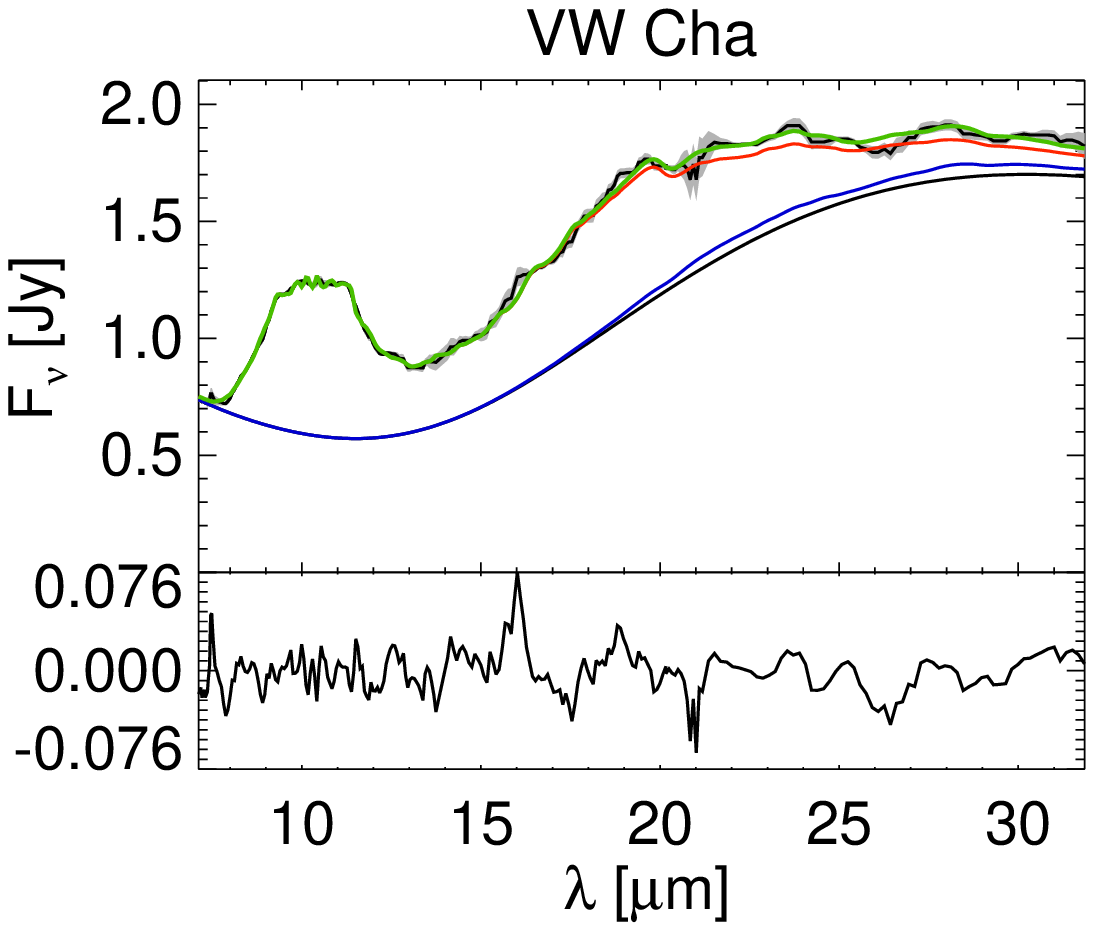}
\includegraphics[width=.2\columnwidth,origin=bl]{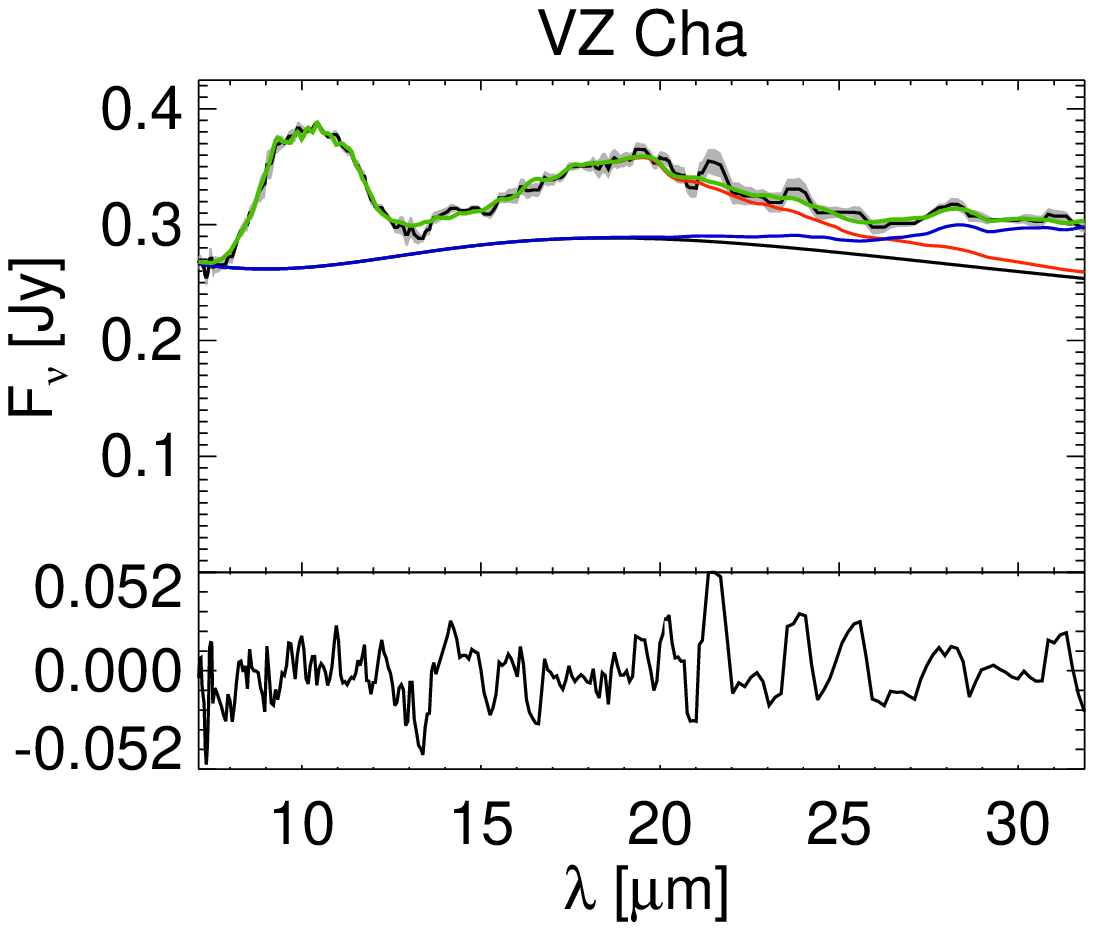}
\end{center}
\end{figure}
\begin{figure}
\begin{center}
\caption{\label{all:fit3}Continued}
\includegraphics[width=.2\columnwidth,origin=bl]{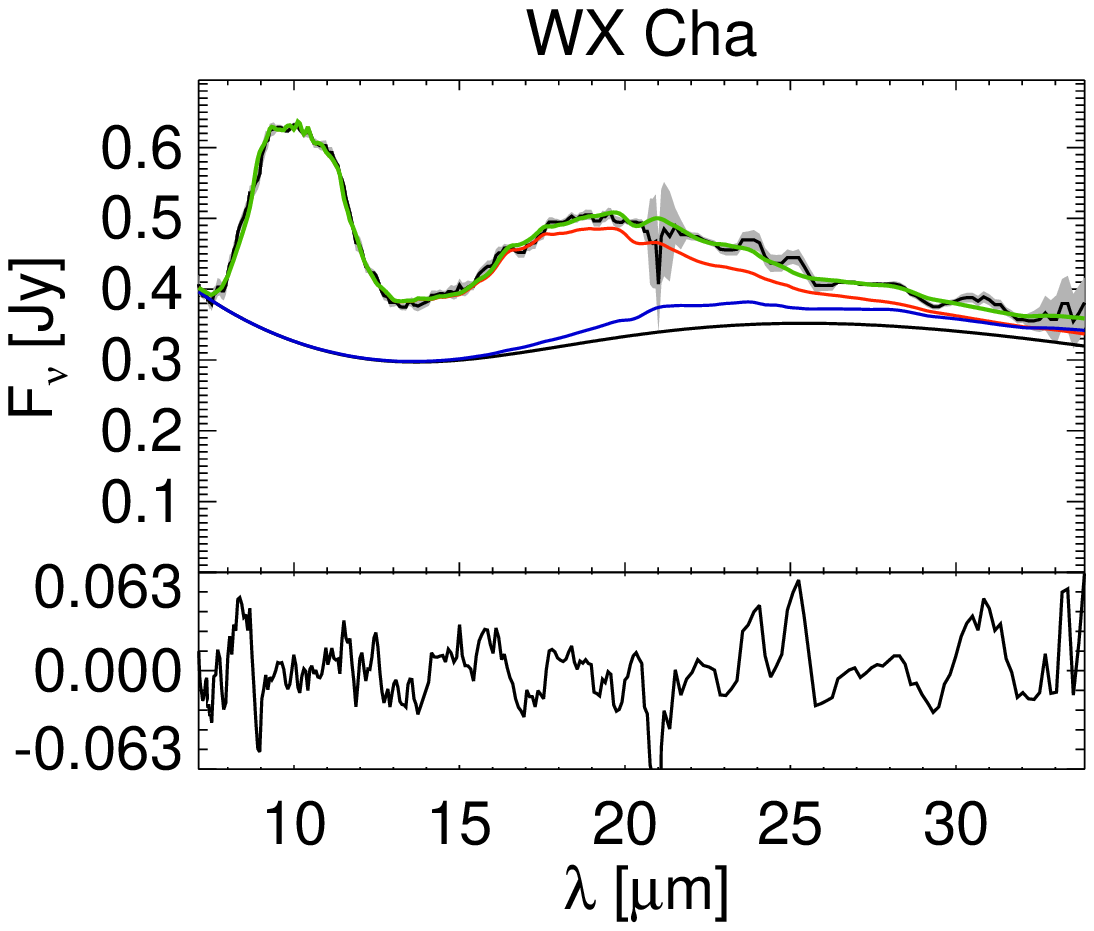}
\includegraphics[width=.2\columnwidth,origin=bl]{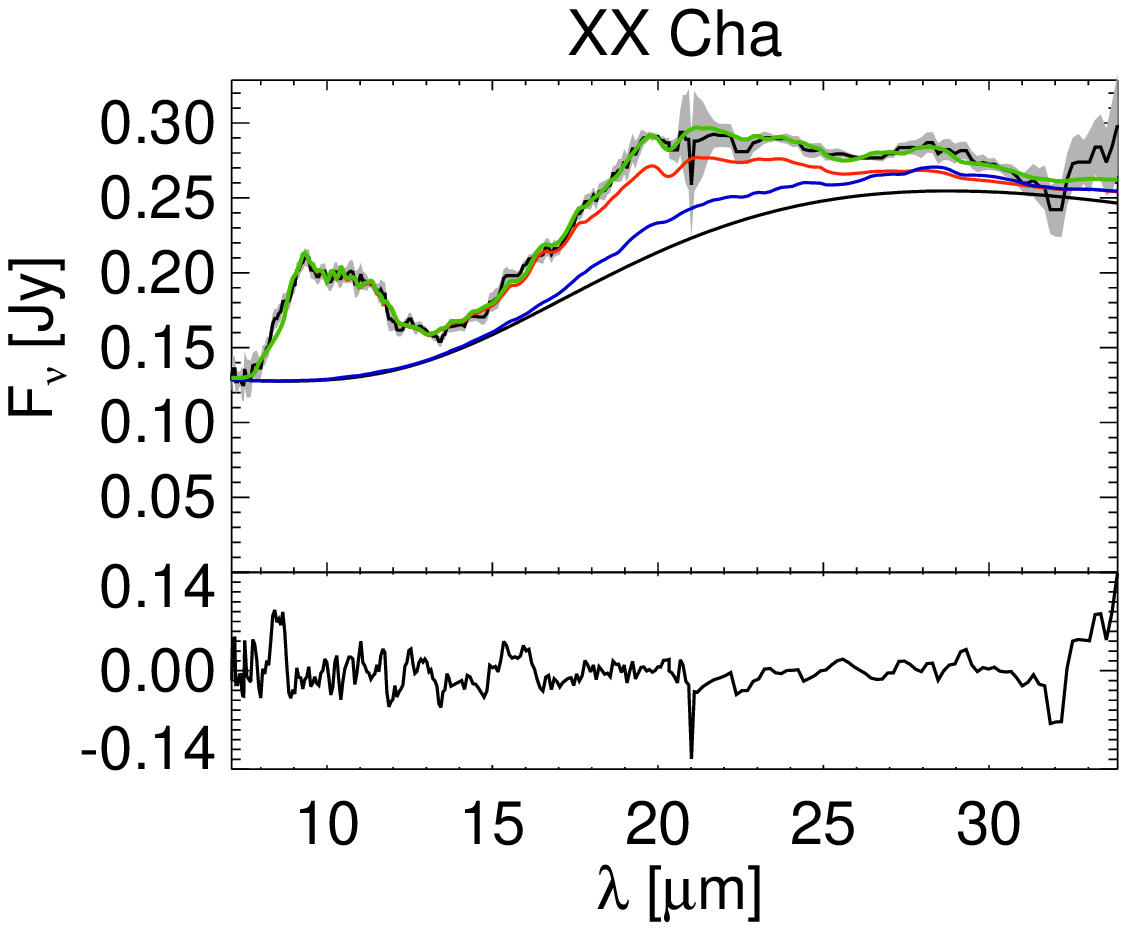}
\end{center}
\end{figure}

\end{document}